\documentclass[11pt]{article}
\usepackage{bm}
\usepackage{amssymb}
\usepackage{amsmath}
\usepackage{amscd}
\usepackage{latexsym,verbatim}
\usepackage[all]{xy}
\xyoption{curve}
\xyoption{line}
\usepackage{amsthm}
\usepackage{epsf}
\usepackage{graphicx}
\usepackage{graphics}
\usepackage{color}
\usepackage[vcentermath,enableskew]{youngtab}
\usepackage{upgreek}

\input{epsf}

\catcode`@=11
\renewcommand{\section}
{\@startsection{section}{1}{0pt}{\medskipamount}{\medskipamount}{\large\bf}}
\makeatletter\renewcommand{\subsection}
{\@startsection{subsection}{2}{\z@}{-3.25ex plus -1ex minus -.2ex}
{1.5ex plus .2ex}{\it }}

\numberwithin{equation}{section}
\catcode`@=12

\def\a{\alpha}
\def\b{\beta}
\def\g{\gamma}
\def\de{\delta}
\def\e{\epsilon}

\def\ve{\varepsilon}

\def\p{\phi}
\def\s{\sigma}

\def\sfrac#1#2{{\textstyle\frac{#1}{#2}}}

\def\pa{\partial}

\def\Dots{\cdot\cdot}

\def\Dirac{{D\!\!\!\!/\,}} 
\def\Diraccal{{{\cal D}\!\!\!\!/\,}} 
 
\def\DiracYtilde{{\widetilde{\cal Y}\!\!\!\!/\,}} 
\def\H{{H\!\!\!\!/ \,}}
\def\tgamma{{\widetilde\gamma}}
\def\psid{{\breve\psi}}

\def\pa{{\partial\!\!\!/}}
\def\cp2star{\star}

\def\beq{\begin{equation}}
\def\eeq{\end{equation}}
\def\bea{\begin{eqnarray}}
\def\eea{\end{eqnarray}}

\newcommand{\Qbar}{{\overline{Q}}}

\def\+{\dagger}
\def\und{\qquad\textrm{and}\qquad}
\def\and{\quad\textrm{and}\quad}

\def\with{\quad\textrm{with}\quad}

\renewcommand{\e}{\,\mathrm{e}\,}

\newcommand{\im}{\,\mathrm{i}\,}
\newcommand{\diff}{\mathrm{d}}

\newcommand{\R}{{\mathbb{R}}}

\newcommand{\C}{{\mathbb{C}}}
\newcommand{\Z}{{\mathbb{Z}}}

\newcommand{\F}{{\mathbb{F}}}

\newcommand{\Idd}{\mathbf{1}}
\newcommand{\Hcal}{{\cal H}}
\newcommand{\Ecal}{{\cal E}}

\newcommand{\Ncal}{{\cal N}}

\newcommand{\Lcal}{{\cal L}}
\newcommand{\Mcal}{{\cal M}}
\newcommand{\Ocal}{{\cal O}}

\newcommand{\yb}{{\bar{y}}}

\newcommand{\zb}{{\bar{z}}}
\newcommand{\zeb}{\bar{\zeta}}
\newcommand{\ab}{{\bar{\alpha}}}
\newcommand{\bb}{{\bar{\beta}}}

\newcommand{\ca}{{\cal{A}}}
\newcommand{\cf}{{\cal{F}}}

\newcommand{\man}{{\cal M}}

\newcommand{\HQ}{{\rm H}}

\newcommand{\ch}{{\mathrm{ch}}}

\newcommand{\Tr}{{\rm Tr}}

\newcommand{\su}{{{\rm SU}(2)}}

\newcommand{\sut}{{{\rm SU}(3)}}
\newcommand{\uo}{{{\rm U}(1)}}
\newcommand{\uoL}{{{\rm u}(1)}}

\newcommand{\urm}{{{\rm U}}}
\newcommand{\urmL}{{{\rm u}}}

\newcommand{\utwo}{{{\rm U}(2)}}

\newcommand{\sutL}{{{\rm su}(3)}}

\newcommand{\SU}{{\rm SU}}

\newcommand{\sltc}{{{\rm SL}(3,\C)}}
\newcommand{\sltcL}{{{\rm sl}(3,\C)}}

\newcommand{\spin}{{\rm Spin}}

\newcommand{\sfg}{{\sf g}}

\newcommand{\mbf}[1]{{\boldsymbol {#1} }}

\def\Hom{{\rm Hom}}

\def\ind{{\rm index}}

\def\>{\rangle}
\def\<{\langle}
\def\+{\dagger}
\def\={\ =\ }

\newcommand{\Acal}{{\cal A}}
\newcommand{\Fcal}{{\cal F}}
\newcommand{\J}{{\cal J}}

\newcommand{\sfR}{{\sf R}}

\newcommand{\Rscal}{{\sf Scal}}
\newcommand{\deepsigh}{\uppsi}

\def\a{\alpha}
\def\b{\beta}
\def\g{\gamma}

\def\de{\delta}
\def\ve{\varepsilon}

\def\t{\theta}
\def\Th{\Theta}

\def\s{\varsigma}
\def\p{\phi}

\def\ome{\omega}
\def\Om{\Omega}
\def\La{\Lambda}

\def\1{{\bar 1}}
\def\2{{\bar 2}}
\def\3{{\bar 3}}
\def\4{{\bar 4}}

\def\hra{\,\hookrightarrow\,}

\begin{document}

\begin{titlepage}
\setcounter{page}{0}
\begin{flushright}
DIAS--STP--12--05\\
HWM--12--10\\
EMPG--12--13\\
NI--12057--BSM
\end{flushright}

\vskip 1.8cm

\begin{center}

{\Large\bf Solitons and Yukawa Couplings\\[10pt] in Nearly K\"ahler Flux Compactifications}

\vspace{15mm}

{\large Brian P. Dolan${}^{1,2}$}
\ \ and \ \ {\large Richard J. Szabo${}^{3,4}$}
\\[5mm]
\noindent ${}^1${\em Department of Mathematical Physics, National
  University of Ireland\\ Maynooth, Co. Kildare, Ireland}
\\[5mm]
\noindent ${}^2${\em School of Theoretical Physics, Dublin Institute
  of Advanced Studies \\ 10 Burlington Road, Dublin 4, Ireland}
\\[5mm]
\noindent ${}^3${\em Department of Mathematics, Heriot-Watt
  University\\ Colin Maclaurin Building, Riccarton, Edinburgh EH14
  4AS, U.K.}
\\[5mm]
\noindent ${}^4${\em Maxwell Institute for Mathematical Sciences\\
  Edinburgh, U.K.}
\\[5mm]
{Email: {\tt bdolan@thphys.nuim.ie , R.J.Szabo@hw.ac.uk}}

\vspace{15mm}

\begin{abstract}
\noindent
We study vacuum states and symmetric fermions in equivariant
dimensional reduction of Yang--Mills--Dirac theory over the
six-dimensional homogeneous space ${\rm SU(3)/U(1)}\times{\rm U(1)}$
endowed with a family of SU(3)-structures including a nearly
K\"ahler structure. We derive the fixed tree-level scalar potentials of the induced Yang--Mills--Higgs theory, and compute the
dynamically generated gauge and Higgs boson masses as functions of the metric moduli of the
coset space. We find an integrable subsector of the Higgs
field theory which is governed by a sine-Gordon type model whose
topological soliton solutions are determined non-perturbatively by the
gauge coupling and which tunnel between families of infinitely
degenerate vacua. The reduction of the Dirac action for
symmetric fermions yields exactly massless chiral fermions, containing
subsectors which have fixed tree-level Yukawa
interactions. We compute dynamical fermion mass matrices explicitly and compare them at different points of the moduli space, some of
which support consistent heterotic flux vacua.
\end{abstract}
\end{center}
\end{titlepage}

\newpage

{\baselineskip=12pt
\tableofcontents
}

\bigskip

\section{Introduction\label{Intro}}

\noindent
Superstring compactifications with fluxes along the internal
manifold are believed to provide a means of connecting superstring
theory to observable low-energy physics while evading the unfavourable
features of the more common Calabi--Yau compactifications. The
presence of fluxes deforms the compactification manifold and requires
the introduction of non-K\"ahler geometries in six
dimensions~\cite{LopesCardoso:2002hd,Strominger:1986uh}. In
particular, in
heterotic string theory the relevant flux is the Neveu--Schwarz
three-form background $H$ which is usually taken as a source for
torsion on six-dimensional manifolds with $\sut$-structure. One of the
main goals in the study of heterotic string compactifications is to
understand how the dimensional reduction of the ten-dimensional $\Ncal=1$
supersymmetric Yang--Mills gauge sector can be used to fix the
multitude of free parameters present in the Higgs and Yukawa sectors
of the Standard Model, and its extensions.

The pioneering dimensional reduction schemes are the Scherk--Schwarz
reduction~\cite{Scherk:1979zr} and coset space dimensional reduction
(see~\cite{Kapetanakis:1992hf} for a review). Coset space dimensional reduction of heterotic supergravity over nearly K\"ahler homogeneous
spaces is considered in~\cite{Chatzistavrakidis:2009mh}. Of the four known compact six-dimensional nearly K\"ahler manifolds,
only the
flag manifold $\F_3:=\sut/\uo\times\uo$ seems to produce interesting
and non-trivial consistent heterotic string
vacua, see e.g.~\cite{Klaput:2011mz,Lust,LNP}. Coset space dimensional
reduction of the supersymmetric Yang--Mills gauge sector over $\F_3$ is considered
in e.g.~\cite{Irges:2011de}.
An alternative dimensional reduction scheme over homogeneous spaces is provided by equivariant
dimensional reduction (see~\cite{Lechtenfeld:2007st} for reviews);
this scheme was applied to the internal space $\F_3$
in~\cite{LPS3,Popov:2010rf}.

In this paper we will study the equivariant dimensional reduction of
Yang--Mills--Dirac theory over the coset space $\F_3$, focusing
attention on the Higgs and Yukawa sectors of the induced field
theory. Our field theory should be regarded as a toy model which is
the first step in describing the
full gauge sector of heterotic string theory, in the sense that it
involves two important omissions. Firstly, we start with unitary gauge
groups, rather than the desired ${\rm E}_8$ gauge group of the
$\Ncal=1$ supersymmetric gauge theory. Secondly, we do not demand
supersymmetry of our initial lagrangean. Incorporating both of these
restrictions would be an important test of the viability of
equivariant dimensional reduction in producing realistic physical
relatives of the Standard Model; our preliminary analysis in this
paper demonstrates that indeed an interesting vacuum structure and
physical masses are induced by this scheme. However, our constructions
and results are interesting in their own right, without any reference
to heterotic string compactifications, as we now explain.

We consider the most general family of quasi-K\"ahler
$\sut$-structures on $\F_3$, one member of which is its standard
nearly K\"ahler structure. This extends the analysis
of~\cite{Dolan:2009ie,Dolan:2009nz}, which demonstrates how
equivariant dimensional reduction over K\"ahler coset spaces can yield
physical particle spectra that are qualitatively analogous to that of
the Standard Model, to non-K\"ahler compactification
manifolds; it extends the considerations of~\cite{Popov:2010rf} to incorporate symmetric
fermions. We will study the vacuum structure of the induced Higgs
sector, and compute the Higgs and gauge boson masses, induced by
dynamical symmetry breaking, as functions on the moduli space of
$\sut$-structures. We will also describe in detail the structure of
the Yukawa couplings at various points of the moduli space, and
compute induced fermion mass matrices explicitly after dynamical symmetry
breaking. We will see that the nearly K\"ahler
member of the family of $\sut$-structures is naturally singled out, as
has been observed previously from considerations based on supersymmetry.

The present paper is structured as follows. In \S\ref{dimred} we
review some standard facts about the geometry of the coset space
$\F_3$, including the construction of homogeneous gauge fields, a
three-parameter family of $\sut$-structures, Dirac operators
associated to the torsional connections of the $\sut$-structures
twisted by homogeneous background fields, and the structure of their
harmonic spinors. In \S\ref{dimredeq} we review the construction of
$\sut$-invariant gauge fields on product manifolds $M\times\F_3$
associated to irreducible representations of $\sut$, and
extend the construction to $\sut$-symmetric fermion fields. In
\S\ref{VacYMH} we study the vacuum structure of the induced
Yang--Mills--Higgs theory on $M$, and compute the tree-level Higgs
potential and the induced boson masses after spontaneous symmetry
breaking. In particular, we identify a subsector of the Higgs field
theory which contains infinitely degenerate vacua which are connected
by sine-Gordon type soliton field configurations, and we completely
classify the physical field content associated to an arbitrary
irreducible $\sut$-module. In \S\ref{IndYukawa} we classify those
representations of $\sut$ that allow for non-trivial tree-level Yukawa
interactions between symmetric fermions in the reduced
Yang--Mills--Higgs--Dirac theory on $M$. We show that at the nearly K\"ahler
locus of the moduli space a natural class of Yukawa couplings can be
obtained via reduction using constant harmonic spinors on $\F_3$, and
that non-zero fermion masses are induced by dynamical symmetry
breaking.
We also compare the Yukawa couplings at other points of the
quasi-K\"ahler moduli space and with those associated to the standard
K\"ahler geometry of the homogeneous space $\F_3$. Our results are summarised
in the concluding section \S\ref{Conclusions} and finally two appendices at
the end of the paper contain some technical details of the
constructions that are used in the main text: In Appendix~A we
summarise the relevant data for the $\sut$ representations that we
use, while in Appendix~B we list the $\sut$-invariant field strengths
for arbitrary irreducible $\sut$-modules.

\bigskip

\section{Geometry of the homogeneous space $\mbf{\sut/\uo\times\uo}$\label{dimred}}

\noindent
The coset space $\F_3:=\sut/ \uo\times \uo$ is a reductive but not symmetric
homogeneous space. In this section we describe the quasi-K\"ahler geometry of $\F_3$,
referring to~\cite{LPS3} and~\cite{Popov:2010rf} for further
details. We will also describe the spin geometry of $\F_3$ and the
construction of $\sut$-invariant spinor fields.

\subsection{Bimonopole fields\label{HombunCP2}}

The projective plane
$\C P^2$ and the complete flag manifold
$\F_3$ on $\C^3$ are related through the fibrations
\beq
\xymatrix{
 & \sut \ar[dl]_{\pi_3} \ar[dr]^{\pi_1} & \\
\F_3 \ \ar[rr]_{\pi_2}  & & \ \C P^2
}
\label{2.1}\eeq
with fibres $\uo\times\su$,
$\su/\uo$ and $T:= \uo\times\uo$ for the bundle projections
$\pi_1, \pi_2$ and $\pi_3$, respectively. The $\C P^1$-bundle $\pi_2$
has structure group $\su$ and
describes $\F_3$ as the twistor space of $\C P^2$. In the following we
will exploit this
description to construct natural gauge potentials on $\F_3$. We could
also consider non-maximal embeddings $\uo\times\uo\to \sut$ which are
parametrized by a pair of integers $(r,s)$ and lead to quotients
$\F_3/\Z_r\times \Z_s$ by freely acting cyclic groups corresponding to
inclusion of discrete Wilson line fluxes; the case $(r,s)=(3,1)$ is considered in~\cite{Irges:2011de}.

With $y^1, y^2$ local complex
coordinates on the base space $\C P^2$, we define
\begin{equation}\label{2.3}
T:=\begin{pmatrix}\yb^{\2}\, \\y^1\end{pmatrix} \ , \qquad
W:=\g\, {\bf 1}_2 - \frac{1}{\g +1}\, T\, T^{\+}\und
\g = \sqrt{1+ T^{\+}\, T}=\sqrt{1+y^\a\, \yb^{\ab}} \ .
\end{equation}
Let us
introduce one-forms on $\C P^2$ given by
\begin{equation}\label{2.6}
 b=\frac{1}{4\g^2}\,\big(T^{\+}\, \diff T - \diff T^{\+}\, T\big)\und
B=\frac{1}{\g^2}\, \big(W\, \diff W + T\, \diff T^{\+}-
\mbox{$\frac{1}{2}$}\, \diff T^{\+}\, T -
\mbox{$\frac{1}{2}$}\,T^{\+}\, \diff T \big)\ ,
\end{equation}
together with
\begin{equation}\label{2.7}
\t= \begin{pmatrix}\t^{\2}\, \\
\t^1\end{pmatrix} :=\frac{2\La}{\g^2}\, W\,\diff T=\frac{2\La}{\g}\,
\begin{pmatrix}\diff\yb^{\2}\, \\ \diff y^1\end{pmatrix}-
\frac{2\La}{\g^2\, (\g +1)}\, \begin{pmatrix}\yb^{\2}\, \\
y^1\end{pmatrix}\,\big(\yb^{\1}\, \diff y^{1} + y^2\, \diff\yb^{\2}\,
\big)\ , 
\end{equation}
where the real parameter $\La$ characterizes the ``size'' of the base
$\C P^2$. Here $\t^1$ and $\t^2$ form a local $\sut$-equivariant orthonormal basis of $(1,0)$-forms on
$\C P^2$, with respect to the natural right isometric action of
$G=\sut$ on the coset. In this frame, the one-form $b$ is an
anti-self-dual $\uoL$-connection (monopole potential) on a complex
line bundle over $\C P^2$, while the gauge potential $B$ is the (canonical)
$\urmL(2)$-valued Levi--Civita connection on the tangent bundle of the
coset space~$\C P^2$.
The corresponding field strengths are given by
\begin{equation}\label{2.8}
 f^-:=\diff b=\frac{1}{8\La^2}\,\t^{\+}\wedge\t = -\frac{1}{8\La^2}\,
\big (\t^{1}\wedge\t^{\1}-\t^{2}\wedge\t^{\2}\, \big)
\end{equation}
and
\begin{equation}\label{2.9}
 F^{+}:=\diff B^{+} + B^{+}\wedge B^{+}=-\frac{1}{8\La^2}\,
\begin{pmatrix}\t^{1}\wedge\t^{\1}+\t^2\wedge\t^{\2}&2\,\t^{\1}
\wedge\t^{\2}\\-2\, \t^1\wedge\t^{2}&-\t^{1}\wedge\t^{\1}
-\t^2\wedge\t^{\2}\end{pmatrix}\ ,
\end{equation}
where
\begin{equation}\label{2.10}
 B^{+}= \begin{pmatrix}a_+&-\overline b_+\\
b_+&-a_+\end{pmatrix}:=B  + b\, {\bf 1}_2
\end{equation}
and
\begin{equation}\label{2.11}
 F:=\diff B + B\wedge B=\frac{1}{4\La^2}\,\t\wedge\t^{\+}=
F^{+}-f^-\, {\bf 1}_2\ .
\end{equation}

A representative element of the fibre space $\C P^1\cong\su/\uo\cong S^2$ is a local section of the Hopf fibration $S^3\to S^2$
given by the matrix
\begin{equation}\label{2.12}
h=\frac{1}{\sqrt{1+\zeta\, \overline\zeta}}\,
\begin{pmatrix}1&-\overline\zeta \ \\ \zeta & 1\end{pmatrix}
\ \in \ {\rm SU}(2)\cong S^3\ ,
\end{equation}
where $\zeta$ is a local complex coordinate on $\C P^1$.
We may then define one-forms on $\F_3$ by the fibrewise gauge transformations
\begin{equation}\label{2.15}
 \widehat\t = h^{\+}\, \t =\frac{1}{\sqrt{1+\zeta\, \zeb}}\, 
\begin{pmatrix}
\t^{\2}+\overline\zeta\, \t^1\\ \t^1-\zeta\, \t^{\2}
\end{pmatrix}
=:
\begin{pmatrix}
 \widehat\t\,^{\2}\, \\ \widehat\t\,^1
\end{pmatrix}
\end{equation}
and
\begin{equation}\label{2.16}
 \widehat B = h^{\+}\, B\,h +h^{\+}\,\diff h=\widehat B^+-b\, {\bf 1}_2=:
\begin{pmatrix}
 \widehat a_+& -\sfrac{1}{2R}\, \widehat\t\,^{\3}\, \\ \, \sfrac{1}{2R}\,\widehat\t\,^{3}&-\widehat a_+
\end{pmatrix} -b\, {\bf 1}_2 \ ,
\end{equation}
with
\begin{equation}\label{2.17}
\widehat a_+ =\frac{1}{1+\zeta\, \zeb}\,\Big( \big(1-\zeta\, \zeb\,\big)\,
a_+ + \zeb\, b_+ -
\zeta\, \overline b_+ + \mbox{$\frac{1}{2}$}\, \big(\,\zeb\, \diff\zeta
-\zeta\, \diff\zeb\, \big)\Big) 
\end{equation}
and
\begin{equation}\label{2.18}
\widehat\t\,^{3} =\frac{2R}{1+\zeta\, \zeb}\, \left(\diff\zeta + b_+ - 2\zeta\, a_+ +
\zeta^2\, \overline b_+\right)\ .
\end{equation}
Here $b$, $a_+$ and $b_+$ are given by (\ref{2.6}) and (\ref{2.10}), while $R$ is the
radius of the fibre two-sphere $S^2\cong\C P^1$. Note that the
restriction of the one-form (\ref{2.17}) to the fibre is the usual
Dirac monopole potential on $\C P^1$.
The corresponding field strengths are given by
\begin{equation}\label{2.20}
 \widehat f^-=f^-=\diff b=-\frac{1}{8\La^2}\,\big(\, \widehat\t\,^1\wedge\widehat\t\,^{\1}-
 \widehat\t\,^2\wedge\widehat\t\,^{\2}\, \big)=
-\frac{1}{8\La^2}\,\big(\t^1\wedge\t^{\1}-\t^2\wedge\t^{\2}\, \big)
\end{equation}
and
\begin{equation}\label{2.21}
\widehat F^+ =\diff\widehat B^++\widehat B^+\wedge \widehat B^+ = -\frac{1}{8\La^2}\,
 \begin{pmatrix}
\widehat\t\,^1\wedge\widehat\t\,^{\1}+\widehat\t\,^2\wedge\widehat\t\,^{\2}&
2\, \widehat\t\,^{\1}\wedge\widehat\t\,^{\2}\\
-2\, \widehat\t\,^{1}\wedge\widehat\t\,^{2}&-\widehat\t\,^1\wedge\widehat\t\,^{\1}-
\widehat\t\,^2\wedge\widehat\t\,^{\2}
 \end{pmatrix} \ ,
\end{equation}
together with the Cartan--Maurer equation
\begin{equation}\label{2.22}
\diff\widehat\t + \big(\widehat B - 2\, b\, {\bf 1}_2 \big)\wedge\widehat\t =0\ .
\end{equation}
The gauge fields $\widehat f^-=\pi^\ast_2f^-$ and $\widehat F^+=\pi^\ast_2F^+$
are pull-backs of the monopole and instanton gauge fields $f^-$ and $F^+$ on $\C P^2$ to the
flag manifold $\F_3$ by the
twistor fibration $\pi_2$ from~(\ref{2.1}); in this setting $\F_3=\C
P(E)$ is the split manifold of the tangent bundle of $\C
P^2$~\cite{BottTu}, regarded as a complex vector bundle $E$ of rank two over $\C P^2$ with
structure group $\utwo$. We call the pair $(-\widehat a_+,b)$ of
$\uoL$-valued gauge potentials on $\F_3$ a \emph{bimonopole potential}; it
will play an instrumental role throughout this paper.

\subsection{$\sut$-structures\label{SU3struct}}

The metric and a corresponding almost K\"ahler
structure on $\F_3$ read
\begin{equation}\label{2.23}
\widehat \sfg = \widehat\t\,^1\otimes\widehat\t\,^{\1} + \widehat\t\,^2\otimes\widehat\t\,^{\2} + \widehat\t\,^3\otimes\widehat\t\,^{\3}\und
\widehat\ome =\mbox{$\frac{\im}{2}$}\, \big(\, \widehat\t\,^1\wedge\widehat\t\,^{\1}+\widehat\t\,^2\wedge\widehat\t\,^{\2}
+ \widehat\t\,^3\wedge\widehat\t\,^{\3}\, \big)\ ,
\end{equation}
where $\widehat\t\,^\a$ with $\a =1,2,3$ are given in (\ref{2.15}) and (\ref{2.18}).
The $\sut$-invariant one-forms $\widehat\t\,^\a$ define a compatible
integrable almost complex structure $\J_+$ on $\F_3$, i.e. a complex structure, such that
$\J_+\widehat\t\,^\a =\im  \widehat\t\,^\a$ so that $\widehat\t\,^\a$ are $(1,0)$-forms with respect to $\J_+$.
{}From (\ref{2.20})--(\ref{2.22}) we obtain the Cartan structure equations
\begin{equation}\label{2.26}
\diff\widehat\t\,^\a + \widehat\Gamma^\a_\b\wedge\widehat\t\,^\b =0\ ,
\end{equation}
which define the Levi--Civita connection 
\begin{equation}\label{2.25}
\widehat\Gamma =\big(\, \widehat\Gamma^\a_\b\, \big) = \begin{pmatrix}-\widehat a_+-3\, b&0&-\sfrac{1}{2R}\, \widehat\t\,^{\2}\, \\[1mm]
0&-\widehat a_++3\, b&\sfrac{1}{2R}\, \widehat\t\,^{\1}\\[1mm]
\sfrac{R}{4\La^2}\,\widehat\t\,^2& -\sfrac{R}{4\La^2}\,\widehat\t\,^1&
-2\, \widehat a_+
\end{pmatrix}
\end{equation}
on
the tangent bundle of $\F_3$. {}From (\ref{2.26})--(\ref{2.25}) it follows that $\widehat \ome$ is K\"ahler, i.e. $\diff\widehat \ome =0$,
if and only if
\begin{equation}\label{2.27}
R^2=2\La^2\ .
\end{equation}
Then the connection matrix $\widehat\Gamma$ in (\ref{2.25}) takes values in
the Lie algebra $\sutL$, i.e. the holonomy group is
$\sut$.

Let us now introduce the forms
\begin{equation}\label{2.33}
 \Theta^1:=\widehat\t\,^1\ ,\qquad \Theta^2:=\widehat\t\,^2\und\Theta^3:=\widehat\t\,^{\3}\ ,
\end{equation}
which are of type $(1,0)$ with respect to an almost complex structure
$\J_-$, i.e. $\J_-\,\Theta^\a = \im \Theta^\a$. 
The almost complex structure $\J_-$ is obtained
from $\J_+$ by changing its sign along the $\C P^1$-fibres of the twistor bundle $\pi_2$, i.e.
$\J_{\pm}\Th^{1,2}=\im  \Th^{1,2}, \
\J_{\pm}\Th^{3}=\mp \im \Th^{3}$. It is never integrable, i.e. the
corresponding Nijenhuis tensor is non-vanishing.
Using the redefinition (\ref{2.33}), we obtain from
(\ref{2.26})--(\ref{2.25}) the Cartan structure
equations
\begin{equation}\label{2.38}
\diff\Th^\a+\Gamma^\a_\b\wedge\Th^\b=H^\a \qquad \mbox{and} \qquad
\diff\Th^\ab+\Gamma^\ab_\bb \wedge\Th^\bb=H^\ab \ ,
\end{equation}
where the left-hand sides define the (torsional) metric connection
\begin{equation}\label{2.37}
\Gamma =\big(\Gamma^\a_\b\big) = \begin{pmatrix}-\widehat a_+-3\, b&0&0\\
0&-\widehat a_++3\, b&0\\
0& 0& 2\, \widehat a_+
\end{pmatrix}
\qquad \mbox{and} \qquad \Gamma^\ab_\bb=-\Gamma^\a_\b
\end{equation}
with holonomy $\uo\times \uo\subset \sut$, while the right-hand sides define the Nijenhuis tensor (torsion) with components
$H^\a_{\bb\bar\g}$ given by
\begin{equation}\label{2.39}
\big(H^\a\big) = \big(\mbox{$\frac{1}{2}$}\, H^\a_{\bb\bar\g}\,
\Th^{\bb}\wedge\Th^{\bar\g} \big) = \frac{1}{2R}\, 
\begin{pmatrix}\Theta^{\2}\wedge\Theta^{\3}\\ \Theta^{\3}\wedge\Theta^{\1}
\\ \frac{R^2}{\La^2}\, \Theta^{\1}\wedge\Theta^{\2}\end{pmatrix}\ .
\end{equation}
We also have
\begin{equation}\label{2.39a}
\diff{b}=-\frac{1}{8\La^2}\, \big(\Th^1\wedge\Th^{\1} -\Th^2\wedge\Th^{\2}\, \big)
\end{equation}
and
\begin{equation}\label{2.39b}
\diff \widehat a_+=-\frac{1}{8\La^2}\, \big(\Th^1\wedge\Th^{\1} +\Th^2\wedge\Th^{\2}\, \big)
+\frac{1}{4R^2}\,\Th^3\wedge\Th^{\3}
\end{equation}
for the abelian gauge fields on $\F_3$.

The pair of forms $(\omega,\Omega)$ given by
\beq
\ome=\mbox{$\frac{\im}{2}$} \,\big( \Th^1\wedge\Th^{\1} +
\Th^2\wedge\Th^{\2} + \Th^3\wedge\Th^{\3} \big) \und
\Om=\Th^1\wedge\Th^2 \wedge \Th^3
\label{2.31}\eeq
defines a one-parameter family of invariant
$\sut$-structures on $\F_3$, parametrized by the ratio~$\frac{R^2}{\Lambda^2}$. From (\ref{2.38})--(\ref{2.39}) it follows that the homogeneous
manifold $\F_3$ is nearly K\"ahler, i.e. $\diff\ome =
\frac{3}{2}\,W_1\, {\rm Im}\,\Om$ and $\diff\Om =
W_1\,\ome\wedge\ome$ with $W_1\in \R$, if and only if
\begin{equation}\label{2.41}
R^2=\La^2 \ ,
\end{equation}
in which case $W_1=\frac1R$.
In this instance we will fix the scales of the fibre $\C P^1$ and the
base $\C P^2$ in $\F_3$ as
\begin{equation}\label{2.53}
R=\La =\sqrt{3} \ ,
\end{equation}
in order that the connection $\Gamma$ in (\ref{2.37}) coincides with the canonical
connection on the principal torus bundle~$\pi_3$ from (\ref{2.1}).
The $(1,1)$-form $\omega$ is almost K\"ahler for the metric
\begin{equation}\label{2.43}
\sfg=\Th^1\otimes \Th^{\1}+\Th^2\otimes \Th^{\2}+\Th^3\otimes\Th^{\3} \ .
\end{equation}

We obtain from the nearly K\"ahler structure a
three-parameter family of invariant quasi-K\"ahler SU(3)-structures by
rescaling the one-forms $\Th^\a$ by constant metric moduli $\s_\alpha\in\R$ as
\begin{equation}\label{2.64}
\Th^\a\ \longmapsto\ \widetilde\Th^\a=\mbox{$\frac1{2\sqrt3}$}\, \s_\a^{-1}\, \Th^\a
\end{equation}
for $\a=1,2,3$. The metric and the fundamental two-form become
\begin{equation}\label{2.66}
\widetilde \sfg=\widetilde\Th^1\otimes \widetilde\Th^{\1}+\widetilde\Th^2\otimes \widetilde\Th^{\2}+
\widetilde\Th^3\otimes \widetilde\Th^{\3}
\und
\widetilde\ome=\sfrac{\im}{2}\, \big(\, \widetilde\Th^1\wedge\widetilde\Th^{\1}+
\widetilde\Th^2\wedge\widetilde\Th^{\2}+\widetilde\Th^3\wedge\widetilde\Th^{\3}\,
\big)\ .
\end{equation}
The three-form $\diff\widetilde\omega$ has only $(3,0)$ and $(0,3)$ components
with respect to the almost complex structure $\J_-$,
i.e. $\widetilde\omega$ is quasi-K\"ahler.
The associated family of connections $\widetilde\Gamma$ on the
tangent bundle $T\F_3$ corresponds
to different regularization schemes which are related by field
redefinitions of the underlying worldsheet sigma-model in heterotic
string theory.
In particular, at the nearly K\"ahler locus
$\s_1=\s_2=\s_3=\frac1{2R}$ of the moduli space we can restore the
fibre $\C P^1$ radius $R$, while for $\s_1=\s_2=\frac1{2\La}$, $\s_3=\frac1{2R}$ we
can restore both of our original base and fibre size
parameters $\La$ and~$R$, with the K\"ahler locus in the
moduli space given by (\ref{2.27}). Notice that the K\"ahler
and nearly K\"ahler loci correspond not only to different choices of
almost complex structures $\J_+$ and $\J_-$ on $\F_3$,
but also to metrics $\widehat\sfg$ and $\sfg$
which differ by a factor of 2 along the fibre direction $\C
P^1\hra\F_3$. 

The generic case of an SU(3)-structure
is classified by intrinsic torsion~\cite{CS} which can be characterized by
the decomposition of the torsion $H$ into irreducible SU(3)-modules
(see Appendix~A);
they are referred to as the five torsion classes
$W_1,\dots,W_5$. For an almost K\"ahler manifold all five intrinsic
torsion classes can be generically non-vanishing. For a K\"ahler manifold $H\in W_5= \widehat{V}^{1,0}\oplus
\widehat{V}^{0,1}$, while in the nearly K\"ahler case $H\in W_1= \widehat{V}^{0,0}\oplus
\widehat{V}^{0,0}$. Quasi-K\"ahler structures have $H\in
W_1\oplus W_2$. Calabi--Yau manifolds correspond to the vanishing of all
five intrinsic
torsion classes.

\subsection{Dirac operators with torsion}

The homogeneous space $\F_3$ is a spin manifold with an
$\sut$-invariant spin structure. Its main feature as a six-dimensional $\sut$-structure manifold
is that its first Chern class vanishes and the canonical bundle is
trivial. This implies the existence of a metric connection (the
canonical connection) with totally skew-symmetric
torsion and holonomy contained in $G= \sut$ which admits a covariantly constant spinor without
coupling to gauge fields. We will describe invariant fermion fields
via the index theorem for the twisted Dirac operator on $\F_3$
corresponding to these torsional connections, which gives the chiral
asymmetry of zero modes of the Dirac operator.
Torsion does not affect the principal
symbol of the Dirac operator, hence the index is the same at every
point in the moduli space; the torsion part can be regarded as a
continuous deformation of the Dirac operator constructed from the
Levi--Civita spin connection, and the index is
invariant under compact perturbations. Nevertheless, the presence of
intrinsic torsion can affect the fermion counting,
as fermions of different chirality can become intertwined and the
usual chirality operator cannot be used to define the
index.

In our three-parameter family of quasi-K\"ahler $\sut$-structures, we choose
a basis of $8\times8$ $\gamma$-matrices
$\widetilde\gamma\,^{\a}, \widetilde\gamma\,^{\ab}= \big(\,\widetilde\gamma\,^{\a}\,)^\dag$ for the
Clifford algebra on $\F_3$ compatible with the metric $\widetilde \sfg$
in (\ref{2.66}) and
the orthonormal one-forms $\widetilde\Th^\a,\widetilde\Th^\ab $. In
this basis the Clifford relations read
\beq
\big\{\widetilde\gamma\,^\a\,,\, \widetilde\gamma\,^{\bar\b}
\big\}=\delta^{\a\b} \, {\bf 1}_8 \qquad \mbox{and}
\qquad \big\{\widetilde\gamma\,^\alpha\,,\, \widetilde\gamma\,^\beta
\big\} =0= \big\{
\widetilde\gamma\,^{\bar\alpha}\,,\, \widetilde\gamma\,^{\bar\beta} \big\}
\label{F3Cliffrels}\eeq
with $\a,\b=1,2,3$ (complex) orthonormal indices with respect to 
the metric $\widetilde \sfg$. The map identifying differential forms on
$\F_3$ with elements of the
Clifford algebra is given by
\bea
&& \eta= \eta_{\a_1\dots\a_r\bb_1\dots\bb_s}\, \widetilde\Th^{\a_1}\wedge
\cdots \wedge \widetilde\Th^{\a_r}\wedge
\widetilde\Th^{\bb_1}\wedge\cdots\wedge \widetilde\Th^{\bb_s}
\label{diffcliffmap} \\ &&
\qquad \qquad \qquad \qquad \ \longmapsto \ \eta\kern -5pt/ = \eta_{\a_1\dots\a_r\bb_1\dots\bb_s}\,
\widetilde\gamma\,^{[\a_1} \cdots \widetilde\gamma\,^{\a_r}\,
\widetilde\gamma\,^{\bb_1}\cdots\widetilde\gamma\,^{\bb_s]} \ , \nonumber
\eea
where the square brackets denote antisymmetrization over all indices
with weight one; this map defines the $G$-equivariant Clifford module
$\bigwedge^\bullet T^*\F_3$.

$\sut$-equivariant bundles over the coset space $\F_3$ are homogeneous vector bundles
induced by representations of the maximal torus $T=\uo\times\uo$ in
$\sut$; they are parametrized by charges $(q,m)_n\in
W^{k,l}\subset\Z^2$ which lie in the weight
lattices of irreducible representations $\widehat{V}^{k,l}$ of
$\sut$ (see Appendix~A). Every such bundle is thus a sum of line
bundles of the form $\Ocal_{\F_3}(q,m):= (\Lcal_{(1)})^{\otimes
  q}\otimes(\Lcal_{(2)})^{\otimes m}$, where the line
bundles $\Lcal_{(i)}\to \F_3$ for $i=1,2$ correspond to the generators of
$\HQ^2(\F_3;\Z)=\Z\oplus \Z$; the fibre restriction $\Lcal_{(1)}\big|_{\C P^1}$ is the Dirac
monopole line bundle corresponding to the generator of $\HQ^2(\C
P^1;\Z)=\Z$, while $\Lcal_{(2)}$ is the pullback by the twistor
fibration $\pi_2$ from (\ref{2.1}) of the monopole line bundle corresponding to the
generator of $\HQ^2(\C
P^2;\Z)=\Z$.
The Dirac operator acting on eight-component spinor fields on $\F_3$,
in the background bimonopole
field corresponding to weight vector
$(q,m)_n\in W^{k,l}$, is given by
\beq
\Dirac_{q,m} = \pa_{\F_3}^{\,\sigma} + m\; b\kern -5pt/ - q\,
\widehat{a}_+\kern -13pt/ \quad .
\label{Diracbimon}\eeq
Here $\pa_{\F_3}^{\, \sigma}$ is the Dirac operator on $\F_3$ involving
only the (torsional) spin connection; it can be written as
\beq
\pa_{\F_3}^{\, \sigma} = \pa^{\,0}_{\F_3}-\mbox{$\frac32$}\, i\sigma \,
H\!\!\!\!/ \ \ ,
\label{sigmaDirac}\eeq
where $\pa^{\,0}_{\F_3}$ is the Dirac operator associated
to the (torsion-free) Levi--Civita connection for the metric $\widetilde \sfg$, while
$H=\sqrt3 \ {\rm Im}\, \Omega$
is the skew-symmetric torsion
three-form of the canonical connection on $\F_3$ and we have used the
structure constants~(\ref{2.65}); the intrinsic torsion of the coset
space $\F_3$ can be
identified with the three-form $H$-flux of heterotic supergravity. The real parameter
\beq
\sigma=\mbox{$\frac1{24\, \sqrt3}$}\, \big(\s_1\, \s_2\, \s_3 \big)^{-1}
\label{sigmadef}\eeq
formally interpolates between the Dirac operator corresponding to the canonical
connection (\ref{2.37}) at $\sigma=1$ and the Levi--Civita connection at
$\sigma=0$. The former limit defines a surface of
quasi-K\"ahler structures in the moduli space which contains the
nearly K\"ahler point $\s_\a=\frac1{2\,\sqrt3}$, $\a=1,2,3$, whereas
the latter limit cannot be reached by any continuous variation of the
metric moduli. The one-parameter family of Dirac operators
(\ref{sigmaDirac}) is a special subclass of the more general families
studied in~\cite{Agricola} (see also~\cite{HKWY}).
The volume form $\widetilde\omega\,^{\wedge 3}/3!$ determines a
$\Z_2$-grading of the eight-dimensional spinor module
$\Delta_{\F_3}=\Delta_{\F_3}^+\oplus\Delta_{\F_3}^-$ on $\F_3$
such that the chirality operator
\beq
\widetilde{\gamma}:= 
 \big[\,\widetilde{\gamma}\,^1\,,\,\widetilde{\gamma}\,^{\bar1}\,
\big]\, \big[\,\widetilde{\gamma}\,^2\,,\,\widetilde{\gamma}\,^{\bar2}\,
\big]\, \big[\,\widetilde{\gamma}\,^3\,,\,\widetilde{\gamma}\,^{\bar3}\,
\big]
\label{F3chiralityop}\eeq
acts as multiplication by
$\pm\,1$ on $\Delta_{\F_3}^\pm$. In a suitable basis for the Clifford
algebra, the operator (\ref{Diracbimon}) correspondingly has a chiral decomposition
\beq
\Dirac_{q,m}=\begin{pmatrix}
0 & \Dirac^+_{q,m} \\ -\Dirac^-_{q,m} & 0 \end{pmatrix} \ ,
\label{Diracchiraldecomp}\eeq
with the twisted Dolbeault--Dirac operators $\Dirac_{q,m}^\pm$ acting on four-component positive/negative
chirality spinor fields on $\F_3$.

The (reduced)
K-theory of the homogeneous space $\F_3$ is generated by the two monopole line
bundles $\Lcal_{(i)}\to \F_3$ together with
$\Lcal_{(i)}\otimes\Lcal_{(2)}$ and
$\Lcal_{(1)}\otimes\Lcal_{(2)}\otimes\Lcal_{(2)}$ for $i=1,2$. 
The index of the Dirac operator (\ref{Diracbimon}) is computed by the Atiyah--Singer index theorem
\bea
\nu_{q,m} \ := \ \ind(i\Dirac_{q,m}) &=& \int_{\F_3}\,\ch\big((\Lcal_{(1)})^{\otimes
  q}\otimes(\Lcal_{(2)})^{\otimes m}\big)
\wedge\widehat{A}\big(\F_3\big) \nonumber \\[4pt] &=& \int_{\F_3}\,
\Big(\, \frac16\, f_{q,m}\wedge f_{q,m}\wedge
f_{q,m}-\frac1{24} \, f_{q,m}\wedge p_1(T\F_3) \, \Big) \ ,
\label{indexthm}\eea
where
\beq
f_{q,m}= m \, \diff b - q\, \diff\widehat{a}_+
\label{curvbimon}\eeq
is the curvature of the bimonopole line bundle $\Ocal_{\F_3}(q,m)$. For $\F_3$ the first
Pontrjagin class of the tangent bundle vanishes, $p_1(T\F_3)=0$, and
using (\ref{2.39a})--(\ref{2.39b}) we find explicitly~\cite{Dolan,LPS3}
\beq
\nu_{q,m}=\mbox{$\frac18$}\,q\,\big(m^2-q^2\big) \ .
\label{indDiracqm}\eeq
The index $\nu_{q,m}$ is an integer since $q$ and $m$ have the same
parity for $(q,m)_n\in W^{k,l}$. 

\subsection{Harmonic spinors\label{Harmonic}}

For each weight $(q,m)_n\in W^{k,l}$, the vector space
$\ker(\Dirac_{q,m})$ of harmonic spinors on $\F_3$ is also independent of the
choice of connection on the tangent bundle; the irreducible
$\sut$ representation $\widehat{V}^{k',l'}$ isomorphic to
$\ker(\Dirac_{q,m})$, when non-zero, is described in
e.g.~\cite[Thm.~8.4]{Landweber}. However, this is not true in
general of the chiral subspaces
$\ker(\Dirac_{q,m}^\pm)$. There are two members of the family of
Dirac operators (\ref{sigmaDirac}) where the $\sut$-module structure
of these subspaces is also known explicitly.

\bigskip

\noindent
{\bf $\mbf{\sigma=1}$ . \ }
The Dirac spectrum was computed in~\cite{Dolan}
from the canonical spin connection on $\F_3$ with torsion associated
to (\ref{2.37}); in this
case the four operators occurring in the Weitzenb\"ock formula for
$(\im \Dirac_{q,m})^2$ mutually commute and hence simultaneously
diagonalise to quadratic Casimir invariants for the Lie algebras
$\sutL$ and $\uoL\oplus \uoL$ (see also~\cite[Prop.~3.4]{Agricola}). 
For $q<0$ the chiral case $\ker(\Dirac_{q,m}^-)=\{0\}$ corresponds to background
gauge field configurations on $\F_3$ with $q^2\geq m^2$, for which
$\ker(\Dirac_{q,m}^+)$ is isomorphic to the $\sut$-module
$\widehat V^{k',l'}$ having $|q|=k'+l'$, $|m|=k'-l'$. The corresponding
dimension $d^{k',l'}$ from (\ref{dimkl}) coincides with the index
(\ref{indDiracqm}) after shifting $q\to q\pm2$. The antichiral case
$\ker(\Dirac_{q,m}^+)=\{0\}$ corresponds to $q^2\leq m^2$, for which
the Dirac kernel $\ker(\Dirac_{q,m}^-)$ is isomorphic to the $\sut$
representation $\widehat V^{k',l'}$ with $|q|=k'$, $|m|=k'+2l'$. The
corresponding dimension $d^{k',l'}$ agrees with minus the index
(\ref{indDiracqm})
after the shifts $q\to q\pm1$, $m\to m\pm3$. In both cases the charge shifts
account for the contributions of the intrinsic spin of the fermions to the
$\uo$ monopole charges. A similar description applies to the $q>0$
case, with chiral and antichiral interchanged. On the other hand, for
$q=0$ the index (\ref{indDiracqm}) is zero and it is possible for an
equal number of chiral and antichiral harmonic spinors to coexist.
We correspondingly decompose the set of weights $W^{k,l}$ into
disjoint subsets
\bea
W^{k,l}_+&=&\big\{(q,m)_n\in W^{k,l}~\big|~m^2>q^2 \,,\, q>0
\quad \hbox{or}\quad m^2<q^2 \,,\, q<0\big\} \ , \nonumber\\[4pt]
W^{k,l}_-&=&\big\{(q,m)_n\in W^{k,l}~\big|~m^2<q^2 \,,\, q>0
\quad \hbox{or}\quad
 m^2>q^2 \,,\, q<0\big\} \ , \nonumber\\[4pt]
W^{k,l}_0&=&\big\{(q,m)_n\in W^{k,l}~\big|~ m^2= q^2 \quad \hbox{or} \quad
 q=0 \big\}
\ .
\label{nonsymQ0pm}\eea
Note that for some choices of $(k,l)$, the
index vanishes on all weights $(q,m)_n\in W^{k,l}$, i.e. $W^{k,l}=W^{k,l}_0$.
For example, the fundamental representation with $(k,l)=(1,0)$ has weights $(q,m)_n=(1,1)_1$, $(-1,1)_1$
and $(0,-2)_0$, for each of which either $q=0$ or $q^2=m^2$ and
the index (\ref{indDiracqm}) evaluates to $\nu_{q,m}=0$. This
also happens for the six-dimensional representation of $\sut$ with
$(k,l)=(2,0)$. On the other hand, there are zero
eigenspinors of the Dirac operator for the adjoint representation with
$(k,l)=(1,1)$ and for the ten-dimensional decuplet
representation with 
$(k,l)=(3,0)$. 
The canonical connection naturally appears in the
supersymmetry condition for heterotic supergravity.

\bigskip

\noindent
{\bf $\mbf{\sigma=\frac13}$ . \ }
The operator $\pa_{\F_3}^{\, 1/3}$ is a
geometric realization of Kostant's cubic Dirac operator (see
e.g.~\cite{Agricola,Landweber}) for the coset $G/T$, whose square may
be expressed in a simple way in terms of Casimir operators and scalars
alone. When restricted to an irreducible representation
of $T$, the eigenvalue of $\big(\im \pa_{\F_3}^{\, 1/3}\, \big)^2$ is the
difference of quadratic Casimir invariants for the Lie algebras
$\sutL$ and $\uoL\oplus \uoL$. Moreover, $\pa_{\F_3}^{\, 1/3}$ is
the unique Dirac operator which also respects the sign decomposition
in the homogeneous Weyl formula for $G=\sut$, and hence the $\sut$
representation $\widehat V^{k',l'}$ isomorphic to $\ker(\Dirac_{q,m})$
in this case can occur in just one of the chiral subspaces $\ker(\Dirac^\pm_{q,m})$. From
these facts one may identify the vector spaces
$\ker(\Dirac^\pm_{q,m})$ with irreducible $\sut$-modules of dimension
given by the index (\ref{indDiracqm}), analogously to the $\sigma=1$
case above. The appearance of the cubic Dirac operator in the gaugino
Dirac equation of heterotic supergravity was observed
in~\cite{Kimura:2006af}, as was the completing of squares in the
Weitzenb\"ock formula for $\big(\im \pa_{\F_3}^{\, 1/3}\, \big)^2$ (see
also~\cite{Bismut,HKWY,Kimura:2007a}).

\bigskip

\noindent
{\bf $\mbf{\sigma\in\R}$ . \ }
For parameter values (\ref{sigmadef}) which 
represent generic points in the moduli space, a type of Parthasarathy
formula for $\big(\im \pa_{\F_3}^{\, \sigma} \big)^2$ involving invariant
operators on the coset $G/T$ is derived
in~\cite[Thm.~3.2]{Agricola}; its form is rather complicated and does
not seem to provide any useful information about the kernels
$\ker(\Dirac^\pm_{q,m})$ beyond the cases $\sigma=1$ and
$\sigma=\frac13$ above. For simplicity we
will for the most part assume \emph{assume} that the total number of zero eigenspinors of $\Dirac_{q,m}$
for a given weight $(q,m)_n\in W^{k,l}$ with $\nu_{q,m}\neq0$ is the same as the index,
i.e. $|\nu_{q,m}|= \dim \ker(\Dirac_{q,m})$,
but this need not be true. In the generic quasi-K\"ahler case there are two free parameters (apart
from an overall scale), and it could happen that
extra zero modes emerge at specific parameter values.
Although the index of the Dirac operator (\ref{indDiracqm}) 
cannot change, being a topological invariant, it is possible for an equal number of
positive and negative chirality states to appear or disappear at some points $\s_r$
of the moduli space without changing the value $\nu_{q,m}$.
If this happens then some of the
summation formulas of \S\ref{Invspinor} and \S\ref{IndYukawa} below have to be extended to incorporate any extra eigenspinors; while these extra zero modes come in $\sut$
representations, they cannot have weights in $W^{k,l}$. For simplicity we shall ignore this complication here and 
assume that there are no such extra spinor harmonics; amongst other
things this will simplify some of the notation in already cumbersome
formulas. The same convention will be applied for the most part to the
possible states with vanishing index.

\bigskip

We fix a basis of chiral/antichiral harmonic spinors
$\chi_{q,m;\ell}^\pm\in\ker\big(\Dirac_{q,m}^\pm\big)$,
$\ell=1,\dots,|\nu_{q,m}|$ on $\F_3$ for each weight
$(q,m)_n\in W^{k,l}_\pm$. They may be chosen to be orthogonal and
normalised such that
\bea
\int_{\F_3}\,\frac{\widetilde{\omega}\,^{\wedge3}}{3!} \
\chi^\pm_{q,m;\ell'}{}^\dag\; \chi^\pm_{q,m;\ell} =
\delta_{\ell,\ell'} \qquad \mbox{and} \qquad
\int_{\F_3}\,\frac{\widetilde{\omega}\,^{\wedge3}}{3!} \
\chi^\mp_{q,m;\ell'}{}^\dag\; \chi^\pm_{q,m;\ell} = 0 \ ,
\label{0modenorm}\eea
where the second equality follows from $W_+^{k,l}\cap
W_-^{k,l}=\emptyset$. Note the vector space $\ker(\Dirac_{q,m})$ of harmonic spinors is
independent of the multiplicity label $n$ of the weight state $(q,m)_n$.

To explicitly construct the spinor harmonics,
following~\cite{Dolan:2009nz} we associate the
Clifford generators $\widetilde\gamma\,^{\ab}$ and $\widetilde\gamma\,^{\a}$ respectively with creation and
annihilation operators acting on a fermionic Fock space with vacuum
vector $|\Omega\rangle$ such that
$\widetilde\gamma\,^{\a}|\Omega\rangle=0$ for $\a=1,2,3$. A
general Fock space state
\bea
\chi(z,\bar z)&:=& \chi_0(z,\bar z)\otimes
|\Omega\rangle + \chi_{\ab}(z,\bar z) \otimes
\widetilde\gamma\,^{\ab}|\Omega\rangle \nonumber \\ && +\, \mbox{$\frac 12$}\,
\chi_{\ab\,\bb}(z,\bar z) \otimes\widetilde\gamma\,^{\ab}\,
\widetilde\gamma\,^{\bb}|\Omega\rangle +
\mbox{$\frac16$}\,\chi_{\ab\,\bb\, \bar\rho}(z,\bar z) \otimes
\widetilde\gamma\,^{\ab}\, \widetilde\gamma\,^{\bb}\,
\widetilde\gamma\,^{\bar \rho}|\Omega\rangle
\label{FockFermion}
\eea
corresponds to a Dirac spinor field on $\F_3$, where $z^\a,\zb\,^\ab$ denote
local complex coordinates on $\F_3$ and the component of $\chi(z,\zb)$
obtained from $k$ creation operators is a $(0,k)$-form. This
identifies the space of spinors on $\F_3$ with the space of
anti-pseudo-holomorphic differential forms $\bigwedge^{0,\bullet} T^*\F_3$. The chirality
operator on $\F_3$ is given by (\ref{F3chiralityop}), and we see that
$\chi_0$ and $\chi_{\ab\,\bb}$ are the four components of a positive
chirality spinor, while $\chi_\ab$ and $\chi_{\bar1\,\bar2\,\bar3}$
are the four components of a negative chirality spinor.

\bigskip

\section{Equivariant dimensional reduction over
 $\mbf{\sut/\uo\times\uo}$\label{dimredeq}}

\noindent
In this section we describe the reduction of $\sut$-equivariant vector
bundles with connection over the flag manifold $\F_3$, again referring
to~\cite{LPS3} and~\cite{Popov:2010rf} for further details. We also
present the $\sut$-equivariant dimensional
reduction of twisted spinor fields over $\F_3$.

\subsection{Invariant gauge fields}

We are interested in the structure of $G$-equivariant gauge fields
on manifolds of the form
\beq
\man:=M\times \F_3 \ ,
\label{M2ntimesSU3H}\eeq
where $M$ is a closed oriented manifold of dimension $d$; the group $G=\sut$ acts
trivially on $M$ and in the standard way by isometries of the coset
space $\F_3=G/T$. 
By standard induction and reduction, there is an equivalence between smooth
$G$-equivariant vector bundles $\Ecal$ over $\man$ and smooth
$T$-equivariant vector bundles $E$ over $M$, where the maximal torus
$T=\uo \times\uo$ of $\sut$ acts trivially
on~$M$. Let $\Ecal^{k,l}\to \man$ be a rank~$N$ hermitean vector bundle over
the space (\ref{M2ntimesSU3H}), associated to an irreducible
representation $\widehat{V}^{k,l}$ of $\sut$, with structure group
$\urm(N)$. Given a $T$-equivariant bundle
$E^{k,l}\to M$ of rank $N$ associated to the representation
$\widehat{V}^{k,l}\big|_T$ of $T$, the corresponding $G$-equivariant
bundle over $\man$ is defined by induction as
\beq
\Ecal^{k,l}=G\times_T E^{k,l} \ .
\label{Ecalklind}\eeq
The action of the maximal torus $T$ on $E^{k,l}$ is defined by the
isotopical decomposition
\beq
E^{k,l}=\bigoplus_{(q,m)_n\in W^{k,l}}\,E_{(q,m)_n}
\otimes\, V_{(q,m)_n}\qquad \mbox{with} \quad
E_{(q,m)_n}=\Hom_T\big(V_{(q,m)_n}\,,\,E^{k,l}\,
\big) \ , 
\label{Eklisotopical}\eeq
where $V_{(q,m)_n} \cong\C$ are the irreducible $T$-modules occurring in
the decomposition of the restriction $\widehat{V}^{k,l}\big|_T$ and as before
$W^{k,l}\subset\Z^2$ is the set of weights for the irreducible $\sut$
representation $\widehat{V}^{k,l}$ (see Appendix~A). The vector bundles
$E_{(q,m)_n}\to M$ have rank $N_{(q,m)_n}$ and trivial $T$-actions. The
rank $N$ of $E^{k,l}$ is given by
\beq
N=\sum_{(q,m)_n\in W^{k,l}}\,N_{(q,m)_n} \ .
\label{rankdimv}\eeq

The action of the $\sut$ operators $I_\alpha^-,I_{\bar\alpha}^+$ for $\alpha=1,2,3$ is implemented by means of
bifundamental Higgs fields
$$ \phi^{1 \ (\pm)}_{(q,m)_n} \ \in \
\Hom\big(E_{(q,m)_n} \,,\,
E_{(q-1,m+3)_{n\pm1}}\big) \ ,\qquad 
\phi^{2 \ (\pm)}_{(q,m)_n} \ \in \ \Hom\big(E_{(q,m)_n} \,,\,
E_{(q-1,m-3)_{n\pm1}} \big) \ , 
$$ 
\beq
\mbox{and} \qquad
\phi^3_{(q,m)_n} \ \in \
\Hom\big(E_{(q,m)_{n}} \,,\,
E_{(q+2,m)_n}\big) 
\label{bifundkl}\eeq
together with their adjoints.
These
bundle morphisms realize the $G$-action of the coset
generators which twists the naive dimensional reduction by
``off-diagonal'' terms. This construction defines a representation of
the quiver $Q^{k,l}$ with vertex set $W^{k,l}$ and arrow set
$I_{\alpha}^-$, $\alpha=1,2,3$, whose underlying lattice is just the
weight diagram of the representation $\widehat{V}^{k,l}$; it explicitly breaks the gauge
group of the bundle $E^{k,l}$ as
\beq
\urm(N)~\longrightarrow~
\prod_{(q,m)_n \in W^{k,l}}\,\urm(N_{(q,m)_n})  \ .
\label{gaugegroupbreak}\eeq
The structure group of the
principal bundle associated to (\ref{Ecalklind}) is then
\beq
\uo\times\uo\times
\prod_{(q,m)_n \in W^{k,l}}\,\urm(N_{(q,m)_n})
\ .
\eeq

The generic form of a $G$-equivariant connection one-form
$\ca$ on the vector bundle $\Ecal^{k,l}\to \man$ is determined by
specifying a unitary connection $A^{(q,m)_n}$ on each bundle $E_{(q,m)_n}$. Let $\widehat\Pi_{(q,m)_n}$ be the
  hermitean projection of the $T$-restriction of $\widehat V^{k,l}$ onto the
  one-dimensional representation of $T=\uo\times\uo$ with weight vector
  $(q,m)_n\in W^{k,l}$, and let $\Pi_{(q,m)_n}$ be
    the hermitean projection onto the sub-bundle $E_{(q,m)_n}$ of the
    bundle (\ref{Eklisotopical}) over $M$. Then an $\sut$-equivariant gauge connection $\Acal$ on the
corresponding bundle (\ref{Ecalklind}) over $\man$ is given by
\bea
\Acal &=& \sum_{(q,m)_n\in W^{k,l}}\, \bigg[\, A^{(q,m)_n}\otimes
\widehat\Pi_{(q,m)_n} + \Pi_{(q,m)_n}\otimes \big(m \,b-q\, \widehat a_+\big)\,\widehat\Pi_{(q,m)_n} \nonumber \\ && +\,
2\sqrt{3}\,\sum_\pm\, 
\s_1\,\Big(\phi^{1 \ (\pm)}_{(q,m)_n}{}^\dag \otimes
\widehat\Pi_{(q,m)_n} \, I_1^-\, \widehat\Pi_{(q-1,m+3)_{n\pm1}}\,
\widetilde\Theta^1 \nonumber \\ && \qquad\qquad\qquad + \,\phi^{1 \ (\pm)}_{(q,m)_n} \otimes
\widehat\Pi_{(q-1,m+3)_{n\pm1}} \, I_\1^+ \, \widehat\Pi_{(q,m)_n}\,
\widetilde\Theta^\1\Big) \\ && +\, 
2\sqrt{3}\,\sum_\pm\, 
\s_2 \Big( \phi^{2 \ (\pm)}_{(q,m)_n}{}^\dag \otimes
\widehat\Pi_{(q,m)_n} \, I_2^-\, \widehat\Pi_{(q-1,m-3)_{n\pm1}}\,
\widetilde\Theta^2 \nonumber \\ && \qquad\qquad\qquad + \,\phi^{2 \ (\pm)}_{(q,m)_n}\otimes
\widehat\Pi_{(q-1,m-3)_{n\pm1}} \, I_\2^+\, \widehat\Pi_{(q,m)_n}\,
\widetilde\Theta^\2\Big) \nonumber\\ && +\,
2\sqrt{3}\,\s_3\, 
\,\Big(\phi^{3}_{(q,m)_n}{}^\dag \otimes
\widehat\Pi_{(q,m)_n} \, I_3^-\, \widehat\Pi_{(q+2,m)_{n}}\,
\widetilde\Theta^3 + \phi^{3}_{(q,m)_n} \otimes
\widehat\Pi_{(q+2,m)_{n}} \, I_\3^+ \, \widehat\Pi_{(q,m)_n}\,
\widetilde\Theta^\3\Big) \, \bigg] \ .\nonumber
\label{AcalQkl}\eea
The corresponding fields at the K\"ahler locus of the moduli space
can be obtained from (\ref{AcalQkl}) by setting
$\s_1=\s_2=\frac1{\sqrt2\, R}$, $\s_3=\frac1{2R}$, and interchanging the generators
$I_3^-\leftrightarrow I_{\bar 3}^+$ and
$\widetilde\Theta^3\leftrightarrow \widetilde\Theta^{\bar 3}$ (see Appendix~A).

The explicit expressions for the matrix elements of the curvature two-form
\beq
\cf=\diff\ca+\ca\wedge\ca
\label{cfca}\eeq
of the connection (\ref{AcalQkl}) are somewhat complicated; they are
listed in Appendix~B. Here we content ourselves with displaying
the example corresponding to the antifundamental $3\times3$ representation
$\widehat V^{0,1}$ of $\sut$; the underlying quiver $Q^{0,1}$ is
\beq
\xymatrix{
& (0,2)_0  \ar[dr]|{ \ I_1^- \ } & \\
 (-1,-1)_1 \ar[ur]|{ \ I^-_2 \ }  \  & & \ar[ll]|{ \ I_3^- \ }\ (1,-1)_1 
}
\label{Q01quiver}\eeq
Using the generators (\ref{2.58}) we obtain an SU(3)-invariant gauge
connection $\Acal$ as
\begin{equation}\label{4.16}
\Acal = \begin{pmatrix}A^1\otimes 1 + {\bf 1}_{N_1}\otimes 2\, b&
\s_2\,\p^{\+}_2\otimes\widetilde\Th^2&-\s_1\,\p_1\otimes\widetilde\Th^{\1}\\[6pt]
-\s_2\,\p_2\otimes\widetilde\Th^{\2}&A^2\otimes 1 + {\bf 1}_{N_2}\otimes (\widehat a_+-b)&
\s_3\,\p^{\+}_3\otimes\widetilde\Th^3 \\[6pt]
\s_1\,\p^{\+}_1\otimes\widetilde\Th^1&-\s_3\,\p_3\otimes\widetilde\Th^{\3}
&A^3\otimes 1 - {\bf 1}_{N_3}\otimes (\widehat a_+ +b)
\end{pmatrix} \ ,
\end{equation}
where $A^1$, $A^2$ and $A^3$ are $\urmL(N_1)$-, $\urmL(N_2)$- and $\urmL(N_3)$-valued gauge
potentials on hermitean vector bundles $E_1$, $E_2$ and $E_3$ over $M$ with ranks
$N_1$, $N_2$ and $N_3$, respectively, such that
\beq
N_1+N_2+N_3=N={\rm rank}(\Ecal^{0,1}) \ , 
\eeq
while ${\bf 1}_{N_\a}$ denotes the $N_\a\times N_\a$ identity matrix. The bundle morphisms
$\phi_1\in \Hom (E_3, E_1)$, $\phi_2\in \Hom (E_1, E_2)$
and $\phi_3\in \Hom (E_2, E_3)$ are
bifundamental scalar fields on $M$.
The bundles $E_\a$ are sub-bundles of the quiver bundle
\begin{equation}\label{4.17}
E^{0,1}=E_1\otimes\C \ \oplus \ E_2\otimes\C \ \oplus \ E_3\otimes\C
\end{equation}
over $M$, where the factors $\C$ denote trivial
$T$-equivariant line bundles over $M$ arising from the decomposition
of the representation $\widehat V^{0,1}\cong\C^3$ into irreducible representations
of $T=\uo\times \uo$. For the curvature $\Fcal =\diff\Acal+\Acal\wedge\Acal=(\Fcal^{\a\b})$
of the invariant connection (\ref{4.16}) we obtain
\begin{eqnarray}
\Fcal^{11}&=&F^{1}-\s_1^2\,\big({\bf 1}_{N_1}-\p_1\, \p_1^\+ \big)\, 
\widetilde\Th^1\wedge\widetilde\Th^{\1}+\s_2^2\, \big({\bf
  1}_{N_1}-\p_2^\+\, \p_2\big) \,
\widetilde\Th^2\wedge\widetilde\Th^{\2}\ ,\nonumber \\[4pt]
\Fcal^{22}&=&F^{2}-\s_2^2\,\big({\bf 1}_{N_2}-\p_2\, \p_2^\+ \big)\,
\widetilde\Th^2\wedge\widetilde\Th^{\2}+\s_3^2\,\big({\bf
  1}_{N_2}-\p_3^\+\, \p_3 \big) \,
\widetilde\Th^3\wedge\widetilde\Th^{\3}\ ,\nonumber\\[4pt]
\Fcal^{33}&=&F^{3}- \s_3^2\, \big({\bf
  1}_{N_3}-\p_3 \, \p_3^\+\big) \,
\widetilde\Th^3\wedge\widetilde\Th^{\3}+\s_1^2\, \big({\bf 1}_{N_3}-\p_1^\+\, \p_1
\big)\, 
\widetilde\Th^1\wedge\widetilde\Th^{\1} \ ,\nonumber\\[4pt]
\Fcal^{13}&=& -\s_1\, \big(\diff\phi_{1}+A^{1}\, \phi_{1}-
\phi_{1}\, A^{3} \big)\wedge\widetilde\Th^{\1} - \s_2\,
\s_3\, \big(\p_1-\p_2^\+\, \p^\+_3\big)\, 
\widetilde\Th^2\wedge\widetilde\Th^{3}\ ,\nonumber \\[4pt]
\Fcal^{21}&=& -\s_2\,\big(\diff\phi_2 + A^{2}\, \phi_{2}-
\phi_{2}\, A^{1} \big)\wedge\widetilde\Th^{\2}-\s_1\,
\s_3\, \big(\p_2-\p_3^\+\, \p_1^\+\big)\, 
\widetilde\Th^3\wedge\widetilde\Th^{1}\ ,\nonumber \\[4pt]
\Fcal^{32}&=& -\s_3\,\big(\diff\phi_3 + A^{3}\, \phi_{3}-
\phi_{3}\, A^{2}\big)\wedge\widetilde\Th^{\3}-\s_1\,
\s_2\, \big(\p_3-\p_1^\+\, \p_2^\+\big)\, 
\widetilde\Th^1\wedge\widetilde\Th^{2}\ ,\label{4.23}
\end{eqnarray}
plus their hermitean conjugates $\Fcal^{\b\a}=-(\Fcal^{\a\b})^{\+}$ for $\a\ne\b$.
In (\ref{4.23}) the superscripts $\a,\b$ label $N_\a\times N_\b$
blocks in $\Fcal$, and we have suppressed tensor products in order to simplify
notation. Here $F^\a=\diff A^\a + A^\a\wedge A^\a$
is the curvature of the connection $A^\a$ on the complex vector bundle
$E_\a\to M$.

\subsection{Invariant spinor fields\label{Invspinor}}

We assume henceforth
that $M$ is a spin manifold of even dimension $d$ for
simplicity, together with a fixed metric. The equivariant dimensional reduction of Dirac spinors on
$\man= M\times\F_3$ is defined with respect to (twisted) symmetric
fermions on $M$. They act as intertwining operators connecting induced
representations of the maximal torus $T=\uo\times\uo$ in the
$\urm(N)$ gauge group, and also in the twisted spinor module
$\Delta$ over $M$ which admits the isotopical decomposition
\beq
\Delta^{k,l}=\bigoplus_{(q,m)_n\in W^{k,l}}\, \Delta_{q,m}
\otimes V_{(q,m)_n}\qquad \mbox{with} \quad
\Delta_{q,m}=\Hom_T\big(V_{(q,m)_n}\,,\,\Delta^{k,l}\, \big)
\label{spinmoddecomp}\eeq
obtained by restricting $\Delta$ to induced representations
$\widehat{V}^{k,l}\big|_T$ 
of $T\subset\spin(d)$. Using Frobenius
reciprocity, the multiplicity spaces may be identified as
\beq
\Delta_{q,m} = \Hom_G\Big(\Delta^{k,l}\,,\, {\rm
  L}^2\big(\F_3;(\Lcal_{(1)})^{\otimes q}\otimes (\Lcal_{(2)})^{\otimes m}\big) \Big) \ ,
\eeq
and hence the isotopical decomposition (\ref{spinmoddecomp}) can be realized explicitly by constructing symmetric fermions
on $M$ as $\sut$-invariant spinors on $M \times\F_3$. They are
associated with the spinor harmonics of the twisted Dirac
operator on $\F_3$.
Dirac zero modes can 
be used to construct an $\sut$-equivariant spinor field $\Psi$ 
on ${\cal M}=M\times \F_3$ which  
decomposes into invariant subspaces.
The decomposition is simplest when the index reflects the actual 
number of zero-modes, {\it i.e.} $\nu_{q,m}=\dim\ker(\Dirac^+_{q,m})$
for $(q,m)_n\in W^{k,l}_+$ and $\nu_{q,m}=-\dim\ker(\Dirac^-_{q,m})$
for $(q,m)_n\in W^{k,l}_-$.  Then the decomposition is
$\Psi=\Psi^+ \oplus \Psi^-$ with
\bea 
\Psi^+&:=&\bigoplus_{(q,m)_n\in W^{k,l}_+}\, \Psi^+_{(q,m)_n}
\qquad\hbox{with}\qquad \Psi^+_{(q,m)_n}=\sum_{\ell=1}^{\nu_{q,m}}\, \psi_{(q,m)_n;\ell}\otimes \chi^+_{q,m;\ell}\ , \nonumber \\[4pt]
\Psi^-&:=&\bigoplus_{(q,m)_n\in W^{k,l}_-}\, 
\Psi^-_{(q,m)_n}
\qquad\hbox{with}\qquad \Psi^-_{(q,m)_n}=\sum_{\ell=1}^{|\nu_{q,m}|} \,
\psid_{(q,m)_n;\ell}
\otimes \chi^-_{q,m;\ell} \ ,
\label{PsiPM}
\eea
where $\psi_{(q,m)_n;\ell}$ and $\psid_{(q,m)_n;\ell}$ are Dirac spinors on
$M$, and $\chi^\pm_{q,m;\ell}$ the positive/negative 
chirality zero modes on~$\F_3$. 

More generally let $n^\pm_{q,m}=\dim\ker(\Dirac^\pm_{q,m})$,  with $\nu_{q,m}=n^+_{q,m} - n^-_{q,m}$, then
the decomposition is $\Psi=\Psi^+ \oplus \Psi^0 \oplus \Psi^-$ with
\bea 
\Psi^+&:=&\bigoplus_{(q,m)_n\in W^{k,l}_+}\, \Psi^+_{(q,m)_n}
\qquad\hbox{with}\qquad \Psi^+_{(q,m)_n}=\sum_{\ell=1}^{n^+_{q,m}}\, \psi_{(q,m)_n;\ell}\otimes \chi^+_{q,m;\ell}  + \sum_{\ell'=1}^{n^-_{q,m}}\, \eta_{(q,m)_n;\ell'}
\otimes \chi^-_{q,m;\ell'} \ , \nonumber \\[4pt]
\Psi^0&:=&\bigoplus_{(q,m)_n\in W^{k,l}_0}\, \Psi^0_{(q,m)_n}
\qquad\hbox{with}\qquad \Psi^0_{(q,m)_n}=\sum_{\ell=1}^{n^+_{q,m}}\, \psi_{(q,m)_n;\ell}\otimes \chi^+_{q,m;\ell}  + \sum_{\ell'=1}^{n^+_{q,m}}\, \eta_{(q,m)_n;\ell'}
\otimes \chi^-_{q,m;\ell'} \ , \nonumber \\[4pt]
\Psi^-&:=&\bigoplus_{(q,m)_n\in W^{k,l}_-}\, 
\Psi^-_{(q,m)_n}
\qquad\hbox{with}\qquad \Psi^-_{(q,m)_n}=\sum_{\ell=1}^{n^-_{q,m}} \,
\eta_{(q,m)_n;\ell}
\otimes \chi^-_{q,m;\ell}+\sum_{\ell'=1}^{n^+_{q,m}}\, \psi_{(q,m)_n;\ell'}
\otimes \chi^+_{q,m;\ell'} \ . \nonumber \\
\label{PsiPMzero}
\eea
For convenience we shall give formulae below only for the case
(\ref{PsiPM}), the extension to the more general case (\ref{PsiPMzero})
is straightforward but notationally more cumbersome (an exception to
this will be the specific analysis of the adjoint $(k,l)=(1,1)$,
where $(q,m)=(0,0)$ has index $\nu_{0,0}=0$ but $n^+_{0,0}=n^-_{0,0}=1$
and $\Psi^0$ is non-zero).

To describe the Dirac operator acting on the invariant spinor fields (\ref{PsiPM}), we
need to choose bases for the pertinent Clifford algebras. For the
Clifford algebra of $M$ we first choose $2^{d/2}\times 2^{d/2}$ anti-hermitean
gamma-matrices $\gamma^\mu$ obeying the anticommutation relations
\beq
\{\gamma^\mu,\gamma^\nu\}= 2\, \eta^{\mu\nu} \, {\mbf 1}_{2^{d/2}} \ ,
\eeq
with $\mu,\nu=1,\ldots ,d$ (real)
orthonormal indices. The basis of gamma-matrices on $\F_3$ obeys the
Clifford relations (\ref{F3Cliffrels}). Then a suitable basis for the
Clifford algebra on $M\times\F_3$ is given by the gamma-matrices
$\big\{\, \widetilde\Gamma^A\, \big\}=
\big\{\,
\widetilde\Gamma^\mu\,,\,\widetilde\Gamma^\a \,,\,\widetilde\Gamma^\ab\, \big\}$ for
$\mu= 1,\dots,d$, $\a=1,2,3$ 
defined as
\beq 
\widetilde\Gamma^\mu = \gamma^\mu\otimes\mathbf{1}_{8} \ ,
\qquad
\widetilde\Gamma^\alpha = \gamma \otimes \widetilde\gamma\,^\alpha
\qquad\hbox{and}\qquad \widetilde\Gamma^{\bar\alpha} = \gamma \otimes
\widetilde\gamma\,^{\bar\alpha} \ ,
\label{GammacalM}\eeq
where $\gamma=(-1)^{d\,(d-1)/2}\, \gamma^{1}\cdots \gamma^{d}$ is the chirality operator on $M$ which anticommutes with all
generators~$\gamma^\mu$, and satisfies $\gamma=\gamma^\dag$ and
$(\gamma)^2={\bf 1}_{2^{d/2}}$.

The $E^{k,l}$-twisted Dirac operator on $M$ is given by
\beq
\Dirac=\sum_{(q,m)_n\in W^{k,l}}\,\big( \pa_M +A^{(q,m)_n}\kern -32pt/
\kern+24pt \ 
\big) \otimes \Pi_{(q,m)_n}
\label{DiracM}\eeq
where $\pa_M$ is the Dirac operator on $M$ involving only the Levi--Civita
spin connection. Together with the Dirac operators (\ref{Diracbimon}) on
$\F_3$ and the map (\ref{diffcliffmap}), the $\Ecal^{k,l}$-twisted Dirac operator
$\Diraccal$ on $\man=M\times\F_3$,
corresponding to the equivariant gauge potential $\ca$ in
(\ref{AcalQkl}) and acting on the spinor fields
(\ref{PsiPM}), can be written as
\bea
\Diraccal = \Dirac\otimes\Idd_8+\gamma\otimes\Dirac_{\F_3} +\DiracYtilde
\label{Diracgradeddef}\eea
where
\beq
\Dirac_{\F_3}=\sum_{(q,m)_n\in W^{k,l}}\, \Dirac_{q,m} \otimes
\widehat{\Pi}_{(q,m)_n}
\label{DiracF3}\eeq
is the $\sut$-equivariant twisted Dirac
operator acting on eight-component spinor fields on $\F_3$, and
\bea\label{YcalQkltilde}
\DiracYtilde&=&2\sqrt{3}\kern-11pt\sum_{(q,m)_n\in W^{k,l}}\, \bigg[ \, \sum_\pm\,
\s_1\,\Big(\phi^{1 \ (\pm)}_{(q,m)_n}{}^\dag \, \gamma \otimes
\widehat\Pi_{(q,m)_n} \, I_1^-\, \widehat\Pi_{(q-1,m+3)_{n\pm1}}\,
\widetilde\gamma\,^1 \nonumber \\ && \qquad\qquad \qquad\qquad
\qquad\qquad +
\,\phi^{1 \ (\pm)}_{(q,m)_n} \, \gamma \otimes
\widehat\Pi_{(q-1,m+3)_{n\pm1}} \, I_\1^+ \, \widehat\Pi_{(q,m)_n}\,
\widetilde\gamma\,^\1\Big) \nonumber\\ && \qquad\qquad \qquad+\,
\sum_\pm\, \s_2\, \Big( \phi^{2 \ (\pm)}_{(q,m)_n}{}^\dag \, \gamma \otimes
\widehat\Pi_{(q,m)_n} \, I_2^-\, \widehat\Pi_{(q-1,m-3)_{n\pm1}}\,
\widetilde\gamma\,^2 \\ && \qquad\qquad\qquad
\qquad\qquad\qquad + \,\phi^{2 \ (\pm)}_{(q,m)_n}\, \gamma \otimes
\widehat\Pi_{(q-1,m-3)_{n\pm1}} \, I_\2^+\, \widehat\Pi_{(q,m)_n}\,
\widetilde\gamma\,^\2\Big) \, \nonumber \\ &&
+\, \s_3 \,
\Big(\phi^{3}_{(q,m)_n}{}^\dag \, \gamma \otimes
\widehat\Pi_{(q,m)_{n}} \, I_3^-\, \widehat\Pi_{(q+2,m)_n}\,
\widetilde\gamma\,^3 + \phi^{3}_{(q,m)_n} \, \gamma \otimes
\widehat\Pi_{(q+2,m)_n} \, I_\3^+ \, \widehat\Pi_{(q,m)_{n}}\,
\widetilde\gamma\,^\3\Big)  \ \bigg] \ .\nonumber
\eea

This construction also demonstrates that there is a bijective
correspondence between states (\ref{FockFermion}) of the eight-dimensional spinor module
$\Delta_{\F_3}$ and the weight space of the adjoint representation of
$\sut$: From (\ref{YcalQkltilde}) it follows that the action of the
$\gamma$-matrices $\tgamma\,^\alpha$ on the quiver lattice $W^{k,l}$ is given by
\bea
\tgamma\,^{1}\,&:&\,  (q-1,m+3)_{n\pm 1}\longrightarrow  (q,m)_n\
, \nonumber \\[4pt]
\tgamma\,^{2}\,&:&\,  (q-1,m-3)_{n\pm 1} \longrightarrow (q,m)_n\ ,
 \\[4pt]
\tgamma\,^{3}\,&:&\,  (q+2,m)_{n} \longrightarrow (q,m)_n\
,\nonumber
\label{tgammaminus}\eea
and hence
\bea
\tgamma\,^{2\, 3}\,&:&\,  (q,m)_{n}\longrightarrow
 (q-1,m+3)_{n\pm1} \ , \nonumber \\[4pt]
\tgamma\,^{3\, 1}\,&:&\, (q,m)_{n}\longrightarrow 
  (q-1,m-3)_{n\pm1} \ ,  \\[4pt]
\tgamma\,^{1\, 2}\,&:&\, (q,m)_n\longrightarrow  (q+2,m)_n 
\ ,\nonumber
\label{tgammaplus}\eea
while $\tgamma\,^{1\,2\,3}$ and the identity
${\bf 1}_8$ both leave a state
with weight $(q,m)_n$ fixed. These transformation properties reflect
the $\uo\times\uo$ charges of the Clifford algebra generators, which
fill out the weight lattice $W^{1,1}$ of the adjoint representation
$\widehat{V}^{1,1}\big|_T$. In particular, there is a
natural correspondence $\tgamma\,^{\alpha}\leftrightarrow
2\, \sqrt{3}\,I^-_\alpha$ and $\widetilde\gamma^{\ab} \leftrightarrow 2\sqrt{3} I_{\,\ab}^+$ (this correspondence is, of course, not an algebra isomorphism). 
The significance of the adjoint representation here
will
become clear in \S\ref{W1}; recall from \S\ref{Harmonic} that it is the smallest
irreducible $\sut$-module that accommodates harmonic spinors on $\F_3$
with non-zero index.

\bigskip

\section{Vacuum structure of induced Yang--Mills--Higgs
  theory\label{VacYMH}}

\noindent
In this section we study the quiver gauge theory on $M$ obtained by
dimensional reduction of Yang--Mills theory on $\Mcal= M\times \F_3$
over $\F_3$. The reduction of hermitean Yang--Mills equations for such
systems, describing the consistent stable solutions of the quiver
gauge theory, are analysed in detail in~\cite{Popov:2010rf} and
related to quiver vortex equations in certain instances; in the
present paper we are interested in the detailed vacuum structure of
these solutions.

\subsection{Reduction of the Yang--Mills action}

We will begin with the case
of the antifundamental representation $\widehat{V}^{0,1}$. The
dimensionally reduced action $S^{\rm YM}_M$ on $M$ is obtained by 
integrating the Yang--Mills action $S^{\rm YM}_{\cal M}$ on ${\cal M}$
over the coset space $\F_3$.
The $(d+6)$-dimensional lagrangean on ${\cal M}$ is the $(d+6)$-form
\begin{equation}
{\cal L}^{\rm YM}_{\cal M}=
\frac{1}{4 \tilde g^2} \, \Tr_{N}\bigl(\Fcal \wedge \tilde\star \,\Fcal^\dagger \bigr)=
-\frac{1}{4 \tilde g^2}\,\Tr_{N}\Big(\, \bigoplus_{\a,\b=1}^3\, \Fcal^{\a\b}
\wedge \tilde\star \,\Fcal^{\b\a}\,\Big) \ ,
 \end{equation}
where $\Tr_N$ is the trace in the fundamental representation of
$\urm(N)$, $\tilde \star$ is the Hodge duality operator on ${\cal M}$ 
compatible with the orthonormal frame $\widetilde\Theta^\alpha$, and a dimensionful Yang--Mills
coupling constant $\tilde g\in \R$ has been included.
Using (\ref{4.23}) this gives
\bea
{\cal L}^{\rm YM}_{\cal M} &=&
\frac{1}{4 \tilde g^2} \, \sum_{\a=1}^3\, 
\Tr_{N_\a}\bigl(  
-F^\a\wedge \tilde\star\, F^\a + 
\s_\a^2 \, (D\phi_\a)\wedge \tilde\star (D\phi_\a)^\dagger + \s_{\a+1}^2
\, (D\phi_{\a+1})^\dag\wedge \tilde\star (D\phi_{\a+1}) \bigr) \nonumber
\\ &&
+\, \widetilde V\,^{\rm YM} (\phi) \wedge \,\tilde\star \,1 \ ,
\eea
where throughout we read the indices $\a=1,2,3$ modulo~$3$ and
\begin{equation}
D\phi_\a:= \diff\phi_\a +A^\a\, \phi_\a - \phi_\a\, A^{\a+2}
\end{equation}
are bifundamental covariant derivatives of the Higgs fields $\phi_\a \in\Hom(E_{\a+2},E_\a)$. The Higgs
potential is given by
\begin{eqnarray}
\widetilde V \,^{\rm YM}(\phi)&=&  
\frac{1}{4\tilde g^2}\, \sum_{\a=1}^3 \, \Tr_{N_\a}\Big(\s_\a^4
\,\bigl({\bf 1}_{N_\a}-\phi_\a \, \phi_\a^\dagger \bigr)^2 +
\s_{\a+1}^4\,\big({\bf 1}_{N_\a} - \phi_{\a+1}^\dag\, \phi_{\a+1}\big)^2
\nonumber \\
& & \qquad \qquad \qquad \qquad +\, \s_{\a+2}^2\, \s_\a^2\,\bigl|\phi_{\a+1}
-\phi_{\a+2}^\dagger \, \phi_\a^\dagger \bigr|^2 \Big) \ ,
\end{eqnarray}
where we use the matrix notation $|\phi|^2:=\frac12\,\big(\phi^\dag\,
\phi+\phi\,\phi^\dag\big)$.

The action $S^{\rm YM}_{\cal M}=\int_{\cal M} \, {\cal L}^{\rm YM}_{\cal M}$ is made
dimensionless here by taking the coupling $\tilde g$ to have mass dimension
$-\frac{d}{2}-1$ and the metric moduli $\s_\a$ mass dimension 1, so that $\phi_\a$ are
then dimensionless while $A^\a$ have mass dimension~$1$.
This gives the dimensionally reduced lagrangean $d$-form
\begin{eqnarray}
{\cal L}^{\rm YM}_M&:=& \int_{\F_3} \,  {\cal L}^{\rm YM}_{\cal M} \nonumber \\[4pt] &=&
\frac{1}{4 g^2} 
\, \sum_{\a=1}^3\, \Tr_{N_\a}\bigl(
-F^\a \wedge \star F^\a + 
\s_\a^2 \, (D\phi_\a) \wedge \star (D\phi_\a)^\dagger + \s_{\a+1}^2
\, (D\phi_{\a+1})^\dag\wedge \star (D\phi_{\a+1}) \bigr)
\nonumber \\ && +\, V^{\rm YM} (\phi) \wedge \star 1 \label{Lem} 
\end{eqnarray}
with 
\begin{eqnarray}
V ^{\rm YM}(\phi)&=&  
\frac{1}{4 g^2}\, \sum_{\a=1}^3 \, \Tr_{N_\a}\Big(\s_\a^4
\,\bigl({\bf 1}_{N_\a}-\phi_\a \, \phi_\a^\dagger \bigr)^2 +
\s_{\a+1}^4\,\big({\bf 1}_{N_\a} - \phi_{\a+1}^\dag\, \phi_{\a+1}\big)^2
\nonumber \\
& & \qquad \qquad \qquad \qquad +\, \s_{\a+2}^2\, \s_\a^2\,\bigl|\phi_{\a+1}
-\phi_{\a+2}^\dagger \, \phi_\a^\dagger \bigr|^2 \Big) \ ,
\label{VYM}\end{eqnarray}
where 
$g^{-2} := {\rm Vol}({\F_3}) \, \tilde g^{-2}$ with ${\rm Vol}({\F_3})
=\int_{\F_3} \,\widetilde\omega\,^{\wedge 3}/3!$ the volume of $\F_3$, so that the coupling constant
$g$ has dimension $2-\frac {d}{2}$,
and $\star$ is the $d$-dimensional Hodge duality operator on $M$.

\subsection{Reduction of the Chern--Simons torsion coupling}

As a subsector of heterotic supergravity, the Yang--Mills gauge theory
also contains a coupling to a torsional three-form flux $\cal H$ on $\cal
M$. There is a natural candidate for a 3-form on $\F_3$, which
extends trivially to a 3-from on ${\cal M}$, 
\beq
\Hcal=\tilde\star\, \diff \Sigma = \star_{\F_3}\, \diff \widetilde\omega \qquad
\mbox{with} \quad \Sigma:= \widetilde\omega\wedge\star 1 \,
\eeq
which satisfies $\diff\tilde\star\Hcal=0$,
with $\star_{\F_3}$ the Hodge duality operator on $\F_3$ with
respect to the orthonormal frame $\widetilde\Theta^\alpha$. The torsion coupling is
then given via the Chern--Simons three-form of the gauge potential
$\cal A$ through the lagrangean $(d+6)$-form
\beq
{\cal L}^{\rm CS}_{\cal M} = \tilde \kappa \, \Tr_N\big(\Fcal\wedge\Acal
- \mbox{$\frac13$}\, \Acal\wedge\Acal\wedge
\Acal\big)\wedge\tilde\star\, \Hcal  \ ,
\label{calLCScalM}\eeq
where the coupling constant $\tilde\kappa\in\R$ has mass dimension $d$. We may
compute the reduction of (\ref{calLCScalM}) over the coset space $\F_3$ by first observing
that integration by parts of this form yields the equivalent form
${\cal L}^{\rm CS}_{\cal M} = \tilde\kappa\, \Tr_N\big(\Fcal
\wedge \Fcal\big) \wedge \Sigma$ when $M$ is closed, and then substituting (\ref{2.66}) and
(\ref{4.23}). After integration over $\F_3$, this
becomes solely
an additional contribution to the Higgs potential (\ref{VYM}) given by
\bea
V^{\rm CS}(\phi) &=& \frac\kappa2 \, \sum_{\a=1}^3\,
\Tr_{N_\a}\Big( 2\, \s_\a^2\, \s_{\a+1}^2\,\big({\bf 1}_{N_\a}-\phi_\a\,
\phi_\a^\dag\big)\, \big({\bf 1}_{N_\a}-\phi_{\a+1}^\dag\,
\phi_{\a+1}\big) \nonumber \\ && \qquad \qquad \qquad \qquad - \,
\s_{\a+2}^2\, \s_\a^2\,\bigl|\phi_{\a+1}
-\phi_{\a+2}^\dagger \, \phi_\a^\dagger \bigr|^2 \Big)
\label{torsionHiggspot}\eea
where the coupling constant $\kappa:= {\rm Vol}({\F_3}) \,
\tilde\kappa$ has dimension $d-6$. The effects of this torsion
coupling disappear in the classical field theory at both the K\"ahler
locus (where $\diff \widetilde\omega=0$) and the nearly K\"ahler locus (where
$\diff \widetilde\omega$ is a sum of $(3,0)$- and $(0,3)$-forms) of
the moduli space. 

However, in heterotic string backgrounds not all of the intrinsic
torsion appears in the Neveu--Schwarz three-form $\cal H$; for
example, the non-trivial relation between the torsion classes $W_4$
and $W_5$ determines the dilaton, which we have ignored in light of
our non-supersymmetric analysis. Moreover, the non-trivial Bianchi
identity satisfied by the three-form $\cal H$ leads to very stringent
consistency conditions; see~\cite{Popov:2010rf} for an analysis of the
reductions involving more general torsion fluxes, and~\cite{LNP} for
a description of the general torsion flux constraints.

\subsection{Vacuum states}

We will now study the dynamical symmetry breaking patterns in the
Yang--Mills--Higgs theory derived above. For this,
it is convenient to rescale 
\begin{equation}
\s_\a'=\frac{\s_\a}{2 g} \ , \qquad
\phi'_\a =\s_\a'\, \phi_\a \qquad \hbox{and}\qquad A'\,^\a = \frac1g\, A^\a \ , \label{Rescaledphi}
\end{equation}
so that the scalar field $\phi'_\a$ and the
gauge potential $A'\,^\a$ both
have the conventional physical mass dimension $\frac{d}{2} - 1$.
Then (\ref{Lem}) becomes
\begin{equation} \label{Lemreduced}
{\cal L}_M^{\rm YM} =\sum_{\a=1}^3\, \Tr_{N_\a}\Bigl(- \frac{1}{4} 
\, 
F'\,^\a \wedge \star F'\,^\a + 
(D'\phi'_\a) \wedge \star (D'\phi'_\a)^\dagger +
(D'\phi'_{\a+1})^\dag\wedge \star (D' \phi'_{\a+1}) \Bigr)
+ V^{\rm YM} (\phi'\,) \wedge \star 1 \ ,
\end{equation}
with
\begin{eqnarray}
V ^{\rm YM}(\phi'\,)&=&  
4 g^2 \, \sum_{\a=1}^3 \, \Tr_{N_\a}\bigg(
\bigl(\s_\a'\,^2\, {\bf 1}_{N_\a}-\phi'_\a \, \phi'_\a{}^\dagger \bigr)^2 +
\big(\s'_{\a+1}{}^2\, {\bf 1}_{N_\a} - \phi'_{\a+1}{}^\dag\, \phi'_{\a+1}\big)^2
\nonumber \\
& & \qquad \qquad \qquad \qquad +\, \Bigl|\frac{\s_{\a+2}'\,
  \s_\a'}{\s_{\a+1}'}\, \phi'_{\a+1}
-\phi'_{\a+2}{}^\dagger \, \phi'_\a{}^\dagger \Bigr|^2 \bigg) \ . \label{HiggsPotential}
\end{eqnarray}
Every term in (\ref{HiggsPotential}) is non-negative, hence
the absolute minimum is achieved by the field configurations satisfying
\begin{eqnarray}
\phi'_\a \, \phi'_\a{}^\dagger=\s'_\a\,^2\, {\bf 1}_{N_\a} \ , \qquad
\phi'_{\a+1}{}^\dag\,\phi'_{\a+1}{}=\s'_{\a+1}{}^2\, {\bf 1}_{N_\a}
\qquad \mbox{and} \qquad \s'_{\a+1}\, \s_{\a+2}'\, \phi'_\a = \s'_\a\,
\phi'_{\a+1}{}^\dagger\, \phi'_{\a+2}{}^\dagger
\label{vaceqs}\end{eqnarray}
for $\a=1,2,3$.

Rather than exhibiting the full complication of
the most general case, which rapidly becomes notationally very cumbersome,
for illustrative purposes we shall only consider the case $N\equiv 0$ mod~$3$
with $N_\a=p:= \frac {N}{3}$ for $\a=1,2,3$.
The gauge group is then $\urm(p)^3$, with $3p^2$ gauge bosons.
The vacuum solution which minimises the Higgs potential and kinetic energies is then given by
\begin{equation}
\phi'_\a=\s'_\a\,  V_\a \ ,
\end{equation}
with $V_\a\in\urm(p)$ constant unitary matrices satisfying the condition
\begin{equation}
V_3 \, V_2 \, V_1={\bf 1}_{p} \ .
\end{equation}
As in~\cite{Dolan:2009nz}, the Higgs vacuum thus provides a representation of the relations of the
\emph{double quiver} $\Qbar\,^{0,1}$ associated to the antifundamental
representation $\widehat{V}^{0,1}$ of $\sut$, obtained by adding an
arrow in the opposite direction to each arrow of the quiver
$Q^{0,1}$, which ensure that there
are no non-trivial oriented cycles in $\Qbar\,^{0,1}$ (and hence that
finite-dimensional quiver representations are possible); it defines a flat
connection of $\urm(p)$ lattice gauge theory on the finite quiver lattice.

We obtain the mass matrix for the gauge bosons by substituting this
vacuum solution into the bicovariant derivative terms $D'\phi'_\a$ in (\ref{Lemreduced}) and extracting the
part quadratic in the gauge potentials $A'\,^\a$ from the Higgs field kinetic terms.  The mass
matrix $\mbf M$ is then given by
\begin{equation}
\mbf M^2= 4g^2\, 
\left(\begin{matrix}
\s_1'{}^2 + \s_2'{}^2 & -\s_2'{}^2  & -\s'_1{}^2 \\
-\s'_2{}^2 & \s_2'{}^2 + \s_3'{}^2 & -\s'_3{}^2 \\
-\s'_1{}^2 & -\s'_3{}^2 & \s_1'{}^2 + \s_3'{}^2 \\
\end{matrix}\right) \ .
\end{equation}
There is a single zero eigenvalue, corresponding to the
massless diagonal combination 
$\frac{1}{\sqrt{3}}\,\big(A^1 + A^2 + A^3\big)$,  
and two non-zero eigenvalues
\begin{equation}
M^2_\pm = 4g^2\,\Big(\, \s'_1{}^2 + \s'_2{}^2 + \s'_3{}^2 
\pm\sqrt{\s'_1{}^4 + \s'_2{}^4 + \s'_3{}^4 -
\bigl( \s'_1{}^2\, \s'_2{}^2 + \s'_2{}^2\, \s'_3{}^2 + \s'_3{}^2\,
\s'_1{}^2 \bigr)} \ \Big) \ .
\label{GaugeMasses}\end{equation}
Thus the gauge group $\urm(p)^3$ 
is broken to its diagonal  subgroup $\urm(p)_{\rm diag}$ 
and the remaining $2p^2$ gauge bosons acquire the masses
$M_\pm$. In the nearly K\"ahler and K\"ahler reductions of the general
$\sut$-structure on $\F_3$ these masses become
\bea
\mbox{Nearly K\"ahler:} && \quad 
M_+= M_-= \frac{\sqrt3}{2R}  \ , \nonumber \\[4pt]
\mbox{K\"ahler:} && \quad M_+=\frac{\sqrt6}{2R} \ , \qquad
M_-=\frac1{R} \ .
\label{KahlerMpm}\eea
Note that the nearly K\"ahler locus in the moduli space yields degenerate mass
eigenvalues.

At a generic point in the moduli space one should also incorporate the
potential (\ref{torsionHiggspot}) arising from the torsion coupling in
the supergravity equations. The main effect of this addition
is that it generally introduces non-positive terms in the potential and
leads to a larger vacuum moduli space of solutions. In particular, at the special
coupling value
\beq
\kappa=-\frac1{2g^2}
\label{kappagrel}\eeq
the first lines of the potentials (\ref{VYM}) and
(\ref{torsionHiggspot}) complete to a perfect square and the total
potential is a sum of non-negative terms; the vacuum
equations (\ref{vaceqs}) are then modified to 
\beq
\phi'_\a \,
\phi'_\a{}^\dagger -
\phi'_{\a+1}{}^\dag\,\phi'_{\a+1} =\big( \s'_\a\,^2 - \s'_{\a+1}{}^2\big) \, {\bf 1}_{N_\a}
\qquad \mbox{and} \qquad \s'_{\a+1}\, \s_{\a+2}'\, \phi'_\a = \s'_\a \,
\phi'_{\a+1}{}^\dagger\, \phi'_{\a+2}{}^\dagger \ .
\label{vaceqstorsion}\eeq
These equations represent respectively the moment map equations and
relations of the double quiver $\Qbar\,^{0,1}$~\cite{Popov:2010rf}.
Any solution of (\ref{vaceqs}) is also a solution of
(\ref{vaceqstorsion}), but not conversely. The relationship
(\ref{kappagrel}) between the Yang--Mills and Chern--Simons coupling
constants also appears in heterotic supergravity, wherein
both couplings are proportional to $g^2_s\,\alpha'$ with $g_s$ the
string coupling constant and $\alpha'$ the Regge slope.

\bigskip

\noindent
{\bf $\mbf{N=3}$ . \ }  To understand the structure of the Higgs
potential, we first consider the
case $p=1$ which gives rise to an abelian gauge theory: 
$\urm(1)^3$ is broken to $\urm(1)_{\rm diag}$. There
are three complex Higgs fields, two gauge bosons acquire a mass and,
of the six real scalar
fields in the Higgs sector, four are physical.
The Higgs vacuum is of the form 
\beq \label{YewOneHiggsVacuum}
\phi_\a'=\s_\a'\, \e^{\im\zeta_\a} \ ,
\eeq
where $\zeta_\a$, $\a=1,2,3$ are three phases constrained
by the condition $\zeta_1+\zeta_2+\zeta_3=2\pi\, k$ for some $k\in\Z$. 
A general configuration can be parameterized as
\beq
\phi_\a'=\e^{\im\zeta_\a}\, \big(\s_\a'+h_\a\big)\ ,
\eeq
with $h_\a$, $\a=1,2,3$ real scalar fields.  The fourth physical degree of
freedom is encoded into the $\urm(1)_{\rm diag}$ invariant combination
$\theta=\zeta_1+\zeta_2+\zeta_3$.  The Higgs potential is then
\bea
V^{\rm YM}(h_1,h_2,h_3,\theta)&=& 4 g^2 \, \sum_{\a=1}^3\, \Big(
2h_\a^2\, \big(2\s_\a'+ h_\a\big)^2
+\big(\s_1'\, \s_2'\, \s_3'\big)^2\, \frac{\big(\s_\a'+h_\a\big)^2}{\s_\a'{}^4}
\nonumber \\ & & \qquad \qquad \qquad +\, 
\big(\s_\a' +h_\a\big)^2\, \big(\s_{\a+1}' +h_{\a+1}\big)^2 \nonumber \\
& & \qquad \qquad \qquad -\, \frac{2\,\s_1'\, \s_2'\, \s_3'\, \big(\s_1'+h_1\big)\, \big(\s_2'+h_2\big)\,
  \big(\s_3'+h_3\big)}{\s_\a'{}^2} \ \cos\theta  \Big) \ .
\label{Higgscostheta}\eea

Expanding (\ref{Higgscostheta}) to second order in $h_\a$ and setting $\theta$ to zero, the Higgs mass matrix $\mbf m$ is read off as
\beq \label{HiggsMassMatrix}
\mbf m^2=4 g^2 \, \left( {\begin{matrix}
8\s_1'{}^2 + \frac{\s_1 '{}^2\, \s_2'{}^2 + \s_2'{}^2\, \s_3'{}^2 + \s_3'{}^2\, \s_1'{}^2 } {\s_1'{}^2}  & 
\frac{\s_1'{}^2\, \s_2'{}^2 - \s_2'{}^2\, \s_3'{}^2 -
  \s_1'{}^2\, \s_3'{}^2}{\s_1'\, \s_2'}
& \frac{\s_1'{}^2\, \s_3'{}^2 - \s_1'{}^2\, \s_2'{}^2 -
  \s_2'{}^2\, \s_3'{}^2}{\s_1'\, \s_3'} \\
\frac{\s_1'{}^2 \, \s_2'{}^2 - \s_2'{}^2\, \s_3'{}^2 -
  \s_1'{}^2\, \s_3'{}^2}{\s_1'\, \s_2'}
& 8 \s_2'{}^2  + \frac{\s_1 '{}^2\, \s_2'{}^2 + \s_2'{}^2\, \s_3'{}^2 + \s_3'{}^2\, \s_1'{}^2 } {\s_2'{}^2}
&
\frac{\s_2'{}^2\, \s_3'{}^2 - \s_1'{}^2 \, \s_3'{}^2 -
  \s_1'{}^2 \, \s_2'{}^2}{\s_2'\, \s_3'}
\\
 \frac{\s_1'{}^2\, \s_3'{}^2 - \s_1'{}^2 \, \s_2'{}^2 -
   \s_2'{}^2\, \s_3'{}^2}{\s_1'\, \s_3'} &
\frac{\s_2'{}^2\, \s_3'{}^2 - \s_1'{}^2\, \s_3'{}^2 -
  \s_1'{}^2 \, \s_2'{}^2}{\s_2'\, \s_3'}
& 8 \s_3'{}^2 +\frac{\s_1 '{}^2\, \s_2'{}^2 + \s_2'{}^2\, \s_3'{}^2 +
  \s_3'{}^2 \, \s_1'{}^2 } {\s_3'{}^2}
\\
\end{matrix} } \right) \ .
\eeq
The eigenvalues of (\ref{HiggsMassMatrix}) at generic points $\s_\a'$
of the moduli space are symmetric under
permutations of the indices $\a=1,2,3$, consistently with the centre
$\Z_3$-symmetry of the weight lattices $W^{k,l}$,
but the explicit expressions are not very instructive.
At the nearly K\"ahler and K\"ahler loci they reduce to
\bea 
\mbox{Nearly K\"ahler:} && \quad m_1=m_2=\frac{\sqrt3}{R} \ , \qquad
m_3= \frac{3}{2R} \ , \nonumber\\[4pt]
\mbox{K\"ahler:} && \quad m_\pm=\frac{\sqrt{\frac{9\pm\sqrt5}2}}R \ ,
\qquad m_3=\frac{\sqrt5}R \ .
\eea
Incorporating the torsional Chern--Simons coupling with
(\ref{kappagrel}) is easily seen to yield a qualitatively similar mass matrix.

The fourth physical Higgs degree of freedom associated with the scalar
field $\theta$ introduces degenerate
vacua: The true vacuum state is a linear superposition of
(\ref{YewOneHiggsVacuum}) with different integers $k\in\Z$ in the
flatness condition $\theta=\theta_k:=2\pi
\, k$ for the $\uo$ quiver lattice gauge theory. Let us look more
closely at the effective field theory for this scalar. We turn off the
gauge fields $A^\a$ and the Higgs fields $h_\a$, and use a $\uo\times \uo$
gauge transformation to choose a symmetric gauge in which all three phases
$\zeta_\a$ are equal; this gauge choice preserves the $\Z_3$-symmetry under
permutations of $\s_1'$, $\s_2'$ and $\s_3'$. Then we obtain from (\ref{Lemreduced}) and
(\ref{Higgscostheta}) the lagrangean
\beq
{\cal L}_M(\theta)= \mbox{$\frac12$}\, \diff\vartheta  \wedge \star \diff\vartheta +
\lambda\,\big(1-\cos\beta\, \vartheta \big) \wedge \star 1 \ ,
\label{SGmodel}\eeq
where
$\beta\, \vartheta= \theta$ with
\beq
\beta=\frac3 {\sqrt{2\big(\s_1'{}^2+\s_2'{}^2+\s_3'{}^2 \big)}}
=\frac{6 g}{\sqrt{M_+^2 + M_-^2}}
\eeq
and
\beq
\lambda=8 g^2 \, \big(\s_1'{}^2\, \s_2'{}^2+\s_2'{}^2\,
\s_3'{}^2+\s_3'{}^2\, \s_1'{}^2 \big)
=\frac{M_+^2 \, M_-^2}{6 g^2} \ .
\eeq
On
$M=\R^{1,1}$, this is just the lagrangean of the sine-Gordon model~\cite{Rajaraman}; in
this case the gauge coupling $g$ has dimensions of mass and the metric moduli
$\s_\a'$ are dimensionless. Expanding the potential in (\ref{SGmodel})
for $\beta\to0$ shows that the perturbative spectrum consists of
scalar particles of mass
\beq
M_{\rm pert}=\sqrt{\lambda}\, \beta = \frac{\sqrt6\, M_+\,
  M_-}{\sqrt{M_+^2+M_-^2}}
\eeq
which, like the gauge and Higgs boson masses (\ref{GaugeMasses}) and
(\ref{HiggsMassMatrix}), is independent of the gauge coupling
$g^2$. On the other hand, this field theory admits well-known
nonperturbative soliton solutions of mass
\beq
M_{\rm sol}=\frac{\sqrt{\lambda}}\beta=\frac{\sqrt6}{36g^2} \, M_+\,
  M_-\, \sqrt{M_+^2+M_-^2} \label{SolitonMass}
\eeq
which is dynamically determined by the induced gauge theory.
Since all
soliton solutions $\theta=\theta(t,x)$ must approach a vacuum solution $\theta_k$ at
spatial infinity $|x|\to \infty$, one can associate to each of them a topological
charge
\beq
Q=\frac1{2\pi}\,\int_\R\,\diff x \ \frac{\partial\theta}{\partial x} =
\frac{\theta(t,+\infty)-\theta(t,-\infty)}{2\pi} = k_+-k_- \qquad
\mbox{with} \quad k_\pm\in\Z \ ,
\eeq
corresponding to the conserved topological current $J=
\frac1{2\pi}\, \star\diff\theta$ with $\diff\star J=\frac1{2\pi}\,
\diff^2\theta= 0$.
Such a field configuration tunnels between the degenerate vacua
$\theta_{k_+}$ and $\theta_{k_-}$. Note that the critical coupling of
the sine-Gordon model $\beta^2=8\pi$ (where the renormalized coupling
has a simple pole) corresponds to gauge boson masses
$M_\pm$ with $M_+^2+M_-^2=\frac{9g^2}{2\pi}$; at the nearly K\"ahler
and K\"ahler loci (\ref{KahlerMpm}) of the moduli space, the base $\C P^1$
radius $R$ may then be regarded as being dynamically induced and is
inversely proportional to the gauge coupling $g$, i.e. large radius
corresponds to the perturbative regime of the induced
Yang--Mills--Higgs theory.

\bigskip

\noindent
{\bf $\mbf{N>3}$ . \ } Now we consider the case $p>1$. We start with $3p^2$ complex Higgs fields,
giving $6p^2$ real scalar fields,
$2p^2$ of which are absorbed by the Higgs mechanism thus
leaving $4p^2$ physical Higgs fields.

The physical Higgs masses can be obtained by using a polar decomposition
of the Higgs fields~$\phi'_\a$: Any square complex matrix $\phi$ can be
uniquely decomposed into
a unitary matrix $U$ and a hermitean matrix $H$ as $\phi=U\, H$. 
To compute the masses
it is sufficient to use constant matrices, thus minimising the Higgs field
kinetic energy, so we 
can expand about the vacuum solution and write
\begin{equation}
\phi'_\a= V_\a\, \big(\s_\a'\, {\bf 1}_{p} + h_\a \big) \ , 
\end{equation}
with $h_\a$ hermitean.
We are free to use a $\urm(p)  \times \urm(p)$
gauge transformation to choose a gauge in which all three fields $V_\a$ are equal,
$V_\a={\mbf U}$ for $\a=1,2,3$, while at the same time leaving 
the diagonal subgroup $\urm(p)_{\rm diag}$
intact.  The physical degrees of freedom are now the 
$3p^2$ 
fields in the hermitean matrices $h_\a$ together with the $p^2$
angular variables in the hermitean matrix $\mbf\theta$ 
defined by ${\mbf U}^3=: \exp(\im\bm\theta)$.
The 
Higgs potential is now much more involved than in the abelian case, since 
the fields $h_\a$ do not commute with $\mbf\theta$ in general. Nevertheless, one can easily
expand it to second order in $h_\a$ about the vacuum solution with ${\mbf U}^3={\bf 1}_{p}$
to find the mass matrix $\mbf m^2\otimes {\bf 1}_{p}$, with $\mbf
m^2$ the mass matrix of the abelian case (\ref{HiggsMassMatrix}). 

The $p^2$ angular variables in the hermitean matrix field $\bm\theta$ 
lead to a very interesting vacuum structure, in analogy
to that of the sine-Gordon solitons for the abelian case above. In
particular, with all other fields turned off there is a matrix Higgs
potential given by $\lambda\,\Tr_{p}\big({\bf 1}_{p}
-\cos\mbf\theta\big)$ for the symmetric gauge choice $V_\a={\mbf
  U}=\exp(\im{\mbf\theta}/3)$, $\a=1,2,3$; gauge equivalence classes of vacuum states are obtained by setting
the eigenvalues of $\mbf\theta$ to $\mbf\theta_i=\mbf\theta_{i,k_i}:= 2\pi\, k_i$ for
$k_i\in\Z$, $i=1, \dots,p$. On $M=\R^{1,1}$, soliton field
configurations are parametrized by a sine-Gordon soliton $\mbf\theta_i$ for
each $i=1,\dots,p$, together with an element $P\in S_p$ of
the Weyl group of $\urm(p)_{\rm diag}$ which
permutes the eigenvalues of the matrix $\mbf\theta$; such a soliton
carries a topological charge vector $\vec Q\in\Z^{p}$ with entries
\beq
\vec Q_i=\frac{\mbf\theta_{i,k_+}-\mbf \theta_{P(i),k_-}}{2\pi}
\eeq
interpolating among the $p$ families of infinitely degenerate vacuum sectors.

\subsection{Flat connections in quiver lattice gauge theory\label{Flatconns}}

For any pair of non-negative integers $(k,l)$, the dimensional
reduction of Yang--Mills gauge theory on $\cal M$ to a
Yang--Mills--Higgs theory on $M$ associated to the irreducible $\sut$
representation $\widehat{V}^{k,l}$ can be described using the formulas
of Appendix~B; the general expression for the induced quiver gauge
theory action on $M$ is rather lengthy and not very informative. We
will therefore satisfy ourselves with studying the
solutions of the vacuum equations with constant Higgs fields and
vanishing gauge potentials. The Higgs potential is a sum of squares,
one for each term multiplying the
$\widetilde\Theta\wedge\widetilde\Theta$ forms in
(\ref{Fcalkldiag})--(\ref{Fcalkloffdiag23}). The vacuum equations are
thus obtained by setting each of these terms to zero, and after
rescaling the Higgs fields analogously to (\ref{Rescaledphi}) they read as
\bea
&& \sum_\pm\, \bigg[\, \mbox{$\frac{n\mp
    q+1\pm1}{n+1}$}\, \lambda_{k,l}^\pm(n,m)^2\,
\Big(\s_1'{}^2\, {\bf
  1}_{N_{(q,m)_n}}- \phi^{1 \ (\pm)\, \prime}_{(q,m)_n}{}^\dag\, \phi^{1 \
  (\pm)\, \prime}_{(q,m)_n}\Big) \nonumber\\ && \qquad \qquad
 -\, \mbox{$\frac{n\mp
    q+1\mp1}{n+1\mp1}$}\, \lambda_{k,l}^\pm(n\mp 1,m-3)^2\, \Big(\s_1'{}^2\, {\bf
  1}_{N_{(q,m)_n}}- \phi^{1 \ (\pm)\, \prime}_{(q+1,m-3)_{n\mp1}}\, \phi^{1 \
  (\pm)\, \prime}_{(q+1,m-3)_{n\mp1}}{}^\dag \Big)\, \bigg] \= 0 \ , \nonumber
\\[4pt] && \sum_\pm\, \bigg[\, \mbox{$\frac{n\mp
    q+1\pm1}{n+1\pm 1}$}\, \lambda_{k,l}^\mp(n\pm 1,m-3)^2\,\Big(\s_2'{}^2\, {\bf
  1}_{N_{(q,m)_n}}- \phi^{2 \ (\pm)\, \prime}_{(q,m)_n}{}^\dag \, \phi^{2 \
  (\pm)\, \prime}_{(q,m)_n}\Big) \nonumber\\ && \qquad\qquad
-\, \mbox{$\frac{n\mp
    q+1\mp1}{n}$}\, \lambda_{k,l}^\mp(n,m)^2\,
\Big(\s_2'{}^2\, {\bf
  1}_{N_{(q,m)_n}}- \phi^{2 \ (\pm)\, \prime}_{(q+1,m+3)_{n\mp1}}\, \phi^{2 \
  (\pm)\, \prime}_{(q+1,m+3)_{n\mp1}}{}^\dag  \Big)\, \bigg] \= 0 \ , 
\nonumber \\[4pt]  &&
 (n-q)\, (n+q+2)\, \Big(\s_3'{}^2\, {\bf
  1}_{N_{(q,m)_n}}- \phi^{3\, \prime}_{(q,m)_n}{}^\dag\, \phi^{3\,
  \prime}_{(q,m)_n} \Big) \nonumber\\ && \qquad \qquad -\,
(n+q)\, (n-q+2)\, \Big(\s_3'{}^2\, {\bf
  1}_{N_{(q,m)_n}}- \phi^{3\, \prime}_{(q-2,m)_n}\,
\phi^{3\, \prime}_{(q-2,m)_n}{}^\dag \Big) \= 0 \ , 
\label{phi1kl}\eea
together with
\bea
&& \sqrt{\mbox{$\frac{((n+1\pm 1)^2-q^2) \, (n\pm
        q+1\pm 1)}{2(n+1)}$}}\, \Big( \s_2'\, \s_3'\, \phi^{1 \
    (\pm)\,\prime}_{(q,m)_n}- \s_1'\, \phi^{3\, \prime}_{(q-1,m+3)_{n\pm 1}}{}^\dag \,
\phi^{2\, (\mp)\,\prime}_{(q+1,m+3)_{n\pm1}}{}^\dag \Big) \nonumber\\ &&
\qquad\qquad -\, \sqrt{\mbox{$\frac{(n\pm q+1\mp 1)\, (n-q+2)\,
        (n+q)}{2(n+1)}$}}\, \Big( \s_2'\,\s_3'\, \phi^{1 \
    (\pm)\, \prime}_{(q,m)_n}- \s_1'\, \phi^{2\, (\mp)\,\prime}_{(q-1,m+3)_{n\pm1}}{}^\dag\, \phi^{3 \, \prime}_{(q-2,m)_n}{}^\dag
\Big) \= 0 \ , \nonumber \\
[4pt] && \sqrt{\mbox{$\frac{((n+1\pm1)^2-q^2)\, (n\pm
        q+1\pm1)}{2(n+1\pm1)}$}}
\, \Big(\s_1'\,\s_3'\, \phi^{2\, (\pm)\,\prime}_{(q,m)_n}- \s_2'\, \phi^{1 \
  (\mp)\, \prime}_{(q-1,m-3)_{n\pm 1}}{}^\dag \,\phi^{3 \
  \prime}_{(q-2,m)_{n}}{}^\dag  \Big)\nonumber \\ && \qquad \qquad -\,
\sqrt{\mbox{$\frac{(n\pm q +1 \mp 1) \, (n-q+2)\, (n+q)}{2(n+1\pm 1)}$}}\,
 \Big( \s_1'\,\s_3'\, \phi^{2\,(\pm)\,
  \prime}_{(q,m)_n}-  \s_2'\, \phi^{3\, \prime}_{(q-1,m-3)_{n\pm 1}}{}^\dag \, \phi^{1 \
  (\mp)\, \prime}_{(q+1,m-3)_{n\pm1}}{}^\dag \Big) \= 0 \ , \nonumber \\
[4pt] && \sum_\pm\, \bigg[\,
\lambda_{k,l}^\pm(n\mp1,m-3)^2 \, \Big( \s_1'\, \s_2'\, \phi^{3 \, \prime}_{(q,m)_n}- \s_3'\, \phi^{2 (\mp) \,\prime}_{(q+2,m)_{n}}{}^\dag \, \phi^{1 \
  (\pm)\, \prime}_{(q+1,m-3)_{n\mp 1}}{}^\dag\Big) \nonumber\\ && \qquad \qquad -\,
\lambda_{k,l}^\pm(n,m)^2 \, \Big( \s_1'\, \s_2'\, \phi^{3 \,
     \prime}_{(q,m)_n}- \s_3'\, \phi^{1 \
  (\pm)\, \prime}_{(q+2,m)_{n}}{}^\dag \, \phi^{2\,(\mp)\,\prime}_{(q+1,m+3)_{n\pm1}}{}^\dag
\Big)\, \bigg] \= 0 \ ,  \label{phi3rel}
\eea
and
\bea
\phi^{3\, \prime}_{(q,m)_{n}} \, \phi^{1 \
  (\pm)\,\prime}_{(q,m)_{n}}{}^\dag &=& \phi^{1 \
  (\pm)\,\prime}_{(q+2,m)_{n}}{}^\dag \, \phi^{3\,
  \prime}_{(q-1,m+3)_{n\pm1}} \ , \nonumber \\[4pt]
\phi^{3 \,\prime}_{(q,m)_{n}} \, \phi^{2 \
  (\pm)\,\prime}_{(q,m)_{n}}{}^\dag &=& \phi^{2 \
  (\pm)\,\prime}_{(q+2,m)_{n}}{}^\dag \, \phi^{3 \,\prime}_{(q-1,m-3)_{n\pm1}} \ , \nonumber \\[4pt]
\phi^{2 \
  (\mp)\,\prime}_{(q+2,m)_{n}} \, \phi^{1\ (\pm)\, \prime}_{(q+2,m)_{n}}{}^\dag &=&
\phi^{1\ (\pm)\, \prime}_{(q+1,m-3)_{n\mp1}}{}^\dag \, \phi^{2 \
  (\mp)\,\prime}_{(q+1,m+3)_{n\pm1}}
\label{phi4rel}\eea
for each weight $(q,m)_n\in W^{k,l}$.

When $N_{(q,m)_n}=p$ for all weights $(q,m)_n\in W^{k,l}$, the gauge
symmetry reduction is given by
\beq
\urm(N) \ \longrightarrow \ \urm(p)^{d^{k,l}}
\qquad \mbox{with} \quad N=p\, d^{k,l} \ ,
\label{gensymred}\eeq
where $d^{k,l}=\big|W^{k,l}\big|$ is the dimension (\ref{dimkl}) of the irreducible
$\sut$ representation
$\widehat{V}^{k,l}$. In this instance an explicit solution of (\ref{phi1kl}) is given by
\beq
\phi^{1\ (\pm)\,\prime}_{(q,m)_n}=\s_1'\, V^{1\ (\pm)}_{(q,m)_n} \ , \qquad
\phi^{2\ (\pm)\, \prime}_{(q,m)_n}=\s_2'\, V^{2\ (\pm)}_{(q,m)_n}  \qquad
\mbox{and} \qquad \phi^{3\,\prime}_{(q,m)_n}=\s_3'\, V^{3}_{(q,m)_n}
\label{phiQklgensol}\eeq
with unitary matrices $V^{1\ (\pm)}_{(q,m)_n}, V^{2\  (\pm)}_{(q,m)_n}, V^3_{(q,m)_n}\in \urm(p)$ for each $(q,m)_n\in
W^{k,l}$. Substituting into (\ref{phi4rel}) 
yields the commutation relations
\bea
V^{1\ (\pm)}_{(q+2,m)_{n}}{} \, V^3_{(q,m)_{n}} &=&
V^{3}_{(q-1,m+3)_{n\pm1}}\, V^{1\ (\pm)}_{(q,m)_{n}}{} \ ,\nonumber \\[4pt]
V^{2 \
  (\pm)}_{(q+2,m)_{n}}{} \, V^{3}_{(q,m)_{n}} &=& V^{3}_{(q-1,m-3)_{n\pm1}} \, V^{2 \
  (\pm)}_{(q,m)_{n}}{} \ ,  \nonumber \\[4pt]
V^{1 \
  (\pm)}_{(q+1,m-3)_{n\mp1}}{} \, V^{2 \ (\mp)}_{(q+2,m)_{n}} &=& V^{2 \
  (\mp)}_{(q+1,m+3)_{n\pm1}} \, V^{1 \
  (\pm)}_{(q+2,m)_{n}}{} \ ,
\label{Vcommrel}\eea
and then (\ref{phi3rel}) leads to the conditions
\bea
V^{1 \
    (\mp)}_{(q-1,m-3)_{n\pm1}}\, V^{2\ (\pm)}_{(q,m)_{n}}\, V^{3}_{(q-2,m)_{n\pm1}} = {\bf 1}_p
\label{r3qmn}\eea
for each $(q,m)_n\in W^{k,l}$. In contrast to the BPS equations
derived in~\cite{Popov:2010rf} (which do not involve the curvature
matrix elements (\ref{Fcalkloffdiag23})), the requirements (\ref{r3qmn}) are much
stronger than the set of relations (\ref{phi3rel}) of the double
quiver $\Qbar\,^{k,l}$ associated to $\widehat{V}^{k,l}$; as before
they specify a flat connection of $\urm(p)$ lattice gauge theory on the finite quiver lattice
$Q^{k,l}$, which is a tessellation of the plane $\R^2$ by equilateral triangles associated
to the vector representations $\widehat{V}^{0,1}$ and
$\widehat{V}^{1,0}$ of $\sut$. Including a Chern--Simons coupling at
the special value (\ref{kappagrel})  removes the conditions
(\ref{phi4rel}) and modifies (\ref{phi1kl}) into a single set of
equations 
representing the moment map equations
of~$\Qbar\,^{k,l}$~\cite{Popov:2010rf}.

See~\cite[\S2.2]{LPS3} for the explicit construction of the nodes and
arrows for a generic quiver $Q^{k,l}$; below we use these results in the
combinatorics of physical fields.

\bigskip

\noindent
{\bf Triangular quivers. \ } To enumerate the physical degrees
of freedom remaining after the dynamical symmetry breaking mechanism, we first
consider the representations $\widehat{V}^{k,0}$ (and their complex
conjugates $\widehat{V}^{0,k}$) of dimension $d^{k,0} = \frac12\,
  (k+1)\, (k+2)$. In this case the boundary of the weight diagram
  $W^{k,0}$ is an inverted equilateral triangle, the weight states are all unique, and there are $k+1$ hypercharge
  levels (see Appendix~A and Fig.~\ref{weight50}). 
\begin{figure}[htb]
 \centering
  \includegraphics[width=4cm]{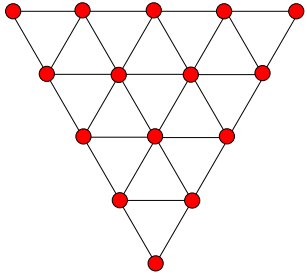}
 \caption{Weight diagram $W^{4,0}$.}
 \label{weight50}
\end{figure}
Starting at the lowest hypercharge
  eigenvalue $m=-2k$, we tessellate the interior of the boundary
  triangle with equilateral triangles of the form (\ref{Q01quiver})
  (and its inversion representing the quiver $Q^{1,0}$);
  at level $i$ there are $2i-1$ triangles with $3i$ edges in total for $i=1,\dots,
  k$. Hence the solution (\ref{phiQklgensol}) involves
  $\frac32\, k\, (k+1)$ unitary degrees of freedom, one for each
  link of the quiver lattice $Q^{k,0}$. We can use a gauge transformation in
the $\urm(p)^{d^{k,0}-1}$ subgroup of (\ref{gensymred}) for $l=0$ to
set $d^{k,0}-1$ of these
lattice gauge fields all equal to a constant unitary matrix $\mbf U$ on the lattice with $\mbf
U^3={\bf 1}_p$, and then solve for the remaining fields using the $k^2$
plaquette relations (\ref{r3qmn}). Thus the
solution (\ref{phiQklgensol}) breaks the gauge symmetry of the
$d$-dimensional field theory on $M$ to the diagonal subgroup
$\urm(p)_{\rm diag}$, leaving in this case $\frac12\, k\, (k+3) \,p^2$ massive gauge
bosons (with physical masses proportional to $\frac1R$ at the K\"ahler
and nearly K\"ahler loci of the moduli space) and $\frac12\, k\,
(5k+3)\, p^2$ real physical Higgs fields. Of these scalar fields,
$p^2$ of them reside in the $\urm(p)_{\rm diag}$
invariant hermitean field $\mbf\theta$ defined by $\mbf U^3=:\exp(\im
\mbf\theta)$, whose vacuum structure is qualitatively analogous to that associated with the
antifundamental representation $\widehat{V}^{0,1}$ from before.

For
example, the quiver $Q^{2,0}$ associated to the six-dimensional representation $\widehat V^{2,0}$ is
\beq
\xymatrix{
(-2,2)_2 \  \ar[dr]|{ \ I^-_1 \ } & & \ar[ll]|{ \ I^-_3 \ }\ (0,2)_2 \ 
 \ar[dr]|{ \ I^-_1 \ }  & & \ar[ll]|{ \ I^-_3 \ } \ (2,2)_2  \\
 & (-1,-1)_1 \ \ar[ur]|{ \ I^-_2 \ } \ar[dr]|{ \ I^-_1 \ } \  & &\ar[ll]|{ \ I^-_3 \ } \ (1,-1)_1
\ar[ur]|{ \ I^-_2 \ }  & \\ 
 & & (0,-4)_0 \ \ar[ur]|{ \ I^-_2 \ }  & & 
}
\label{Q20quiver}\eeq
and the vacuum field content consists of $5p^2$ massive gauge bosons
plus $13p^2$ real physical scalar fields.

\bigskip

\begin{figure}[htb]
\centering
  \includegraphics[width=8cm]{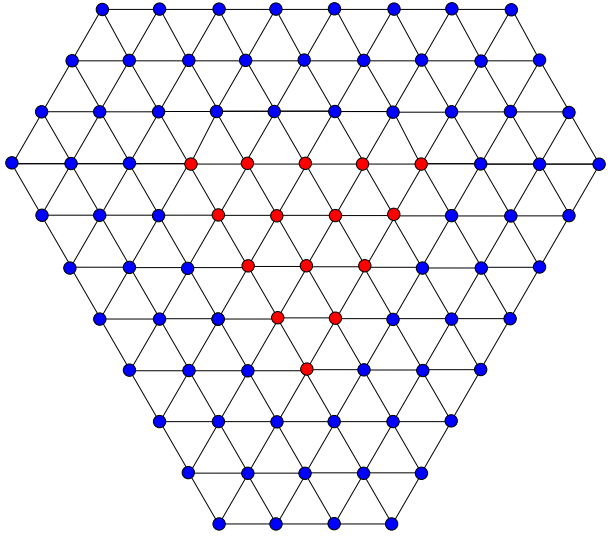}
\caption{Weight diagram $W^{7,3}$ (multiplicities are not shown).}
 \label{weight73}
\end{figure}

\noindent
{\bf Hexagonal quivers. \ } For a generic representation
$\widehat{V}^{k,l}$ with $k,l\neq0$, $k\geq l$ (or its complex
conjugate $\widehat{V}^{l,k}$), the boundary of the weight
diagram $W^{k,l}$ is a hexagon, symmetric about $(0,0)\in\Z^2$, with $k+1$ weights on the upper edge, $l+1$
weights on the lower edge, and $k+l+1$ hypercharge levels from the
lower to the upper edge (see Fig.~\ref{weight73}).
The outer $l$ layers are all hexagonal, while inner
layer $l+1$ is triangular (and hence so are all of its interior
layers). The counting of physical degrees of freedom
is further complicated now by the existence of degenerate weight states:
States in the $i$-th hexagonal layer
have multiplicity $i$ for $i=1,\dots,l$, while states in the
inverted triangular layers all have multiplicity $l+1$. In the
corresponding quiver diagram, diagonal links
around layers of fixed multiplicity $i$ are mapped to $i$ arrows between the
nodes, and diagonal links from a layer with multiplicity $i$ to a
layer with multiplicity $i+1$ are mapped to $2i$ arrows; horizontal links always
map to single arrows. 
Starting from the lowest hypercharge eigenvalue
$m=-2k-l$, the tip of the first inverted triangular layer starts $l$
levels up at $m=-2k+2l$. The enumeration within the interior triangle is
thus that of the triangular representation $\widehat{V}^{k-l,0}$ as
derived above,
adjusted by the multiplicity factor $l+1$; in particular, it contains
$d^{k-l,0}=\frac12\, (k-l+1)\, (k-l+2)$ nodes
and
\beq
\mbox{$\frac32$}\, (k-l)\, (k-l+1)\, (l+1)-\mbox{$\frac12$}\, l\,
(k-l)\, (k-l+1) =\mbox{$\frac12$} \,
(k-l)\, (k-l+1)\, (2l+3)
\label{triangleno}\eeq
complex Higgs fields, where the subtraction compensates the
overcounting of horizontal arrows in the quiver diagram. The boundary of the weight
diagram is generated by $k$ applications of the $\sut$ operators
$I^-_\alpha$ and $l$ applications of $I^+_{\bar\alpha}$ for
$\alpha=1,2,3$; hence there are $3(k+l)$ boundary nodes (each of
multiplicity one). The $i$-th hexagonal
layer contains $3(k+l-2i+2)$ weights each of multiplicity $i$ for
$i=1,\dots,l$. By suitably compensating the enumeration of horizontal
arrows and the two ``corner'' nodes, it is straightforward to see that there
are $3i\, (k+l-2i+2)-(k+l-2i+2)\, (i-1)$ arrows around layer $i$,
$3(k+l-2i+2)-(k+l-2i+4)$ additional horizontal arrows, and $8i\,
(k+l-2i+1)$ diagonal arrows to layer $i+1$. Adding
everything, there are altogether
\bea
\sum_{i=1}^l\, \big((10i+3)\, (k+l-2i+2) -2(4i+1) \big) 
\label{hexagonno}\eea
complex Higgs fields on the outer $l$ hexagonal layers. The total
number of unitary matrices parametrizing the vacuum solution
(\ref{phiQklgensol}) is then the sum of (\ref{triangleno}) and
(\ref{hexagonno}) which is
\beq
\mbox{$\frac12$}\, k^2\, (2l+3) +\mbox{$\frac32$}\, k\, \big(2l^2+4l+1
\big) -\mbox{$\frac16$}\, l\, (l+1)\, (4l-13) \ .
\eeq
Again, we eliminate these lattice gauge fields in
favour of a unitary matrix $\mbf U$ with $\mbf U^3={\bf 1}_p$ by using a $\urm(p)^{d^{k,l}-1}$ gauge transformation and the
plaquette relations (\ref{r3qmn}). The gauge symmetry is broken to $\urm(p)_{\rm diag}$, leaving $\big(d^{k,l}-1\big)\,p^2$ massive gauge
bosons and a total of
\bea
\mbox{$\frac12$}\, k^2\, (3l+5) + \mbox{$\frac12$}\, k\,
\big(11l^2+20l+3\big) -\mbox{$\frac16$}\,l\, \big(8l^2-15l-17\big)
\eea
real physical Higgs fields.

For
example, the quiver $Q^{1,1}$ associated to the eight-dimensional
adjoint representation $\widehat V^{1,1}$ is
\beq
\xymatrix{
 & (-1,3)_1 \  \ar@/^/[dr]|{ \ I^-_1 \ } \ar@/_/[dr]|{ \ I^-_1 \ }  
& & \ (1,3)_1\  \ar[dr]|{ \ I^-_1  \ }\ar[ll]|{ \ I^-_3 \ }
 &  \\
(-2,0)_2 \  \ar[dr]|{ \ I^-_1 \ } \ar[ur]|{ \ I^-_2 \ } & & \ (0,0)_{0,2} \ \ar[ll]|{ \ I^-_3 \ }  \ar@/_/[ur]|{ \ I^-_2 \ } \ar@/^/[ur]|{ \ I^-_2 \ }
 \ar@/^/[dr]|{ \ I^-_1 \ } \ar@/_/[dr]|{ \ I^-_1 \ }
& & \ (2,0)_2 \ \ar[ll]|{ \ I^-_3 \ }  \\
 & (-1,-3)_1 \   \ar@/^/[ur]|{ \ I^-_2 \ }
\ar@/_/[ur]|{ \ I^-_2 \ } & & \ (1,-3)_1\ \ar[ur]|{ \ I^-_2 \ } \ar[ll]|{ \ I^-_3 \ }
}
\label{Q11quiver}\eeq
leading to $7p^2$ massive gauge fields and $25p^2$ real physical
scalar fields.

\bigskip

\section{Induced Yukawa interactions of symmetric fermions\label{IndYukawa}}

\noindent
In this section we will describe the $\sut$-equivariant dimensional
reduction of Yang--Mills--Dirac theory on $\man=M\times\F_3$ for invariant spinor fields over the flag manifold $\F_3$, and the
induced Yang--Mills--Higgs--Dirac theory on $M$. We will focus
particular attention to the possible emergence of Yukawa couplings between the Higgs and
fermion fields. 

\subsection{Reduction of the Dirac action\label{DiracRed}}

Consider the minimally-coupled Dirac lagrangean $(d+6)$-form on
$\man$ given by
\beq
\Lcal_\Mcal^{\rm D}=i\Psi^\dag \wedge \,\tilde\star\, \Diraccal\Psi
\eeq
on the space of massless ${\rm
  L}^2$-spinors (\ref{PsiPM}), where the fermion field $\Psi$ has
canonical dimension $\frac12\, (d+5)$. For simplicity, we take a euclidean signature
metric on $M$; for lorentzian signature the adjoint spinor $\Psi^\dagger$ should be replaced
with the appropriate lorentzian adjoint $\overline\Psi$. We also
assume for definiteness that the spinor field $\Psi$ transforms under the fundamental
representation of the $\urm(N)$ gauge group, but other fermion representations can
be similarly treated. For
representations $\widehat{V}^{k,l}$ that give rise to a non-zero
index for the Dirac operators on $\F_3$, and hence to harmonic spinors, dimensional reduction of this lagrangean yields a
non-trivial fermionic field theory on $M$ coupled to the
Yang--Mills--Higgs theory of~\S\ref{VacYMH}.

By integrating this lagrangean over the coset space $\F_3$, we
arrive at a dimensionally reduced lagrangean $d$-form 
\beq
\int_{\F_3}\, \Lcal_\Mcal^{\rm D} = \Lcal_M^{\rm
  D} + \Lcal_M^{\rm Y}+ \Lcal_M^{\rm Y}\,^\dag
\label{totalfermaction}\eeq
on $M$. The
second term from (\ref{Diracgradeddef}) vanishes on harmonic spinor fields
on $\F_3$, while the first term yields a series of massless twisted Dirac
kinetic terms for the various fermion fields on $M$; with the same
rescalings of the bosonic fields and the metric moduli as in \S\ref{VacYMH},
using the orthogonality relations (\ref{0modenorm}) this
gives
\bea
\Lcal_M^{\rm D}&=& i \kern -17pt \sum_{(q,m)_n\in W_+^{k,l}} \
\sum_{\ell=1}^{\nu_{q,m}}\, \psi_{(q,m)_n;\ell}{}^\dag\wedge
\star \big( \pa_M +g\, A^{\prime\, (q,m)_n}\kern -36pt/
\kern+24pt \ \ 
\big) \psi_{(q,m)_n;\ell} \nonumber \\ && 
+\, i \kern -17pt\sum_{(q,m)_n\in W_-^{k,l}} \
\sum_{\ell=1}^{|\nu_{q,m}|}\, \eta_{(q,m)_n;\ell}{}^\dag\wedge
\star \big( \pa_M +g\, A^{\prime\, (q,m)_n}\kern -36pt/
\kern+24pt \ \ 
\big) \eta_{(q,m)_n;\ell} \ .
\eea
The fermion fields $\psi_{(q,m)_n;\ell}$ and $\eta_{(q,m)_n;\ell'}$
transform in the fundamental representation of ${\rm U}(N_{(q,m)_n})$
for each $\ell=1,\dots, \nu_{q,m}$ and $\ell'=1,\dots, |\nu_{q,m}|$ respectively.

\begin{figure}
  \centering
  \includegraphics[width=2.5cm]{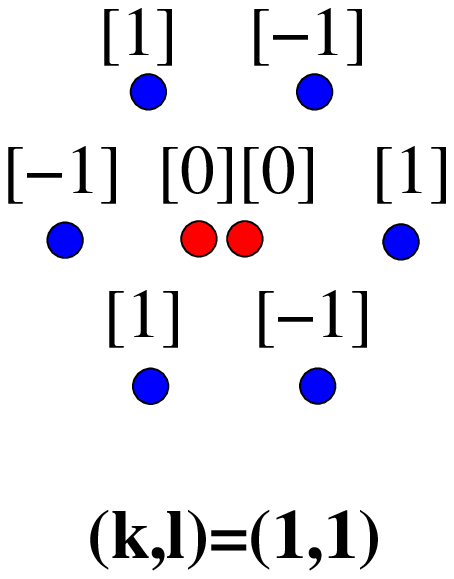} \qquad \qquad 
  \includegraphics[width=3cm]{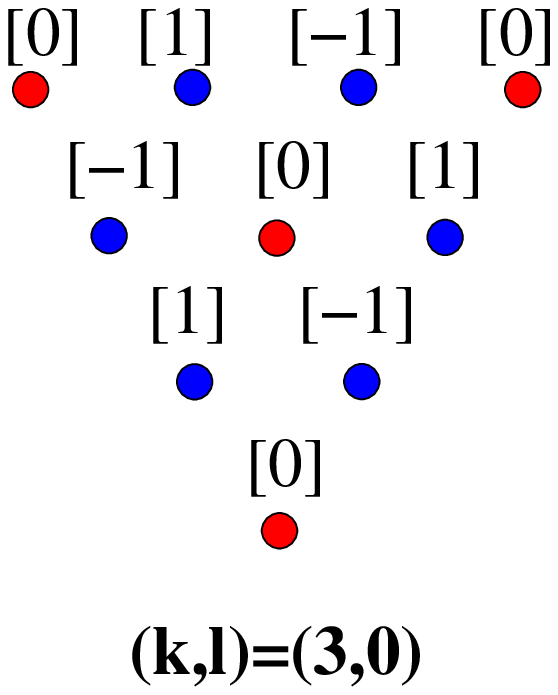} \\[20pt]
  \includegraphics[width=4cm]{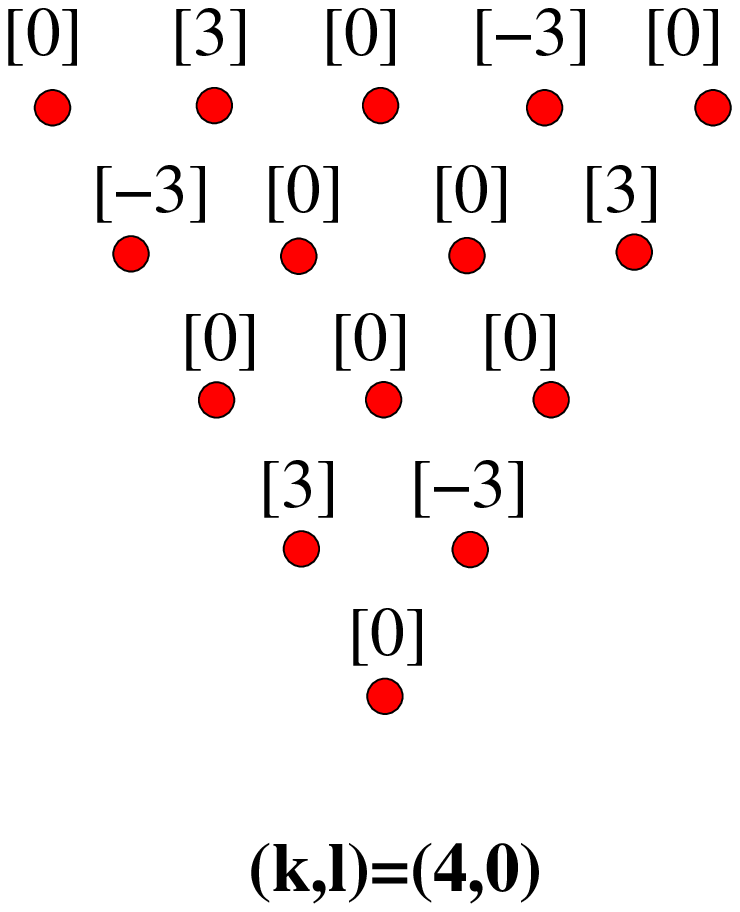} \qquad \qquad \includegraphics[width=5cm]{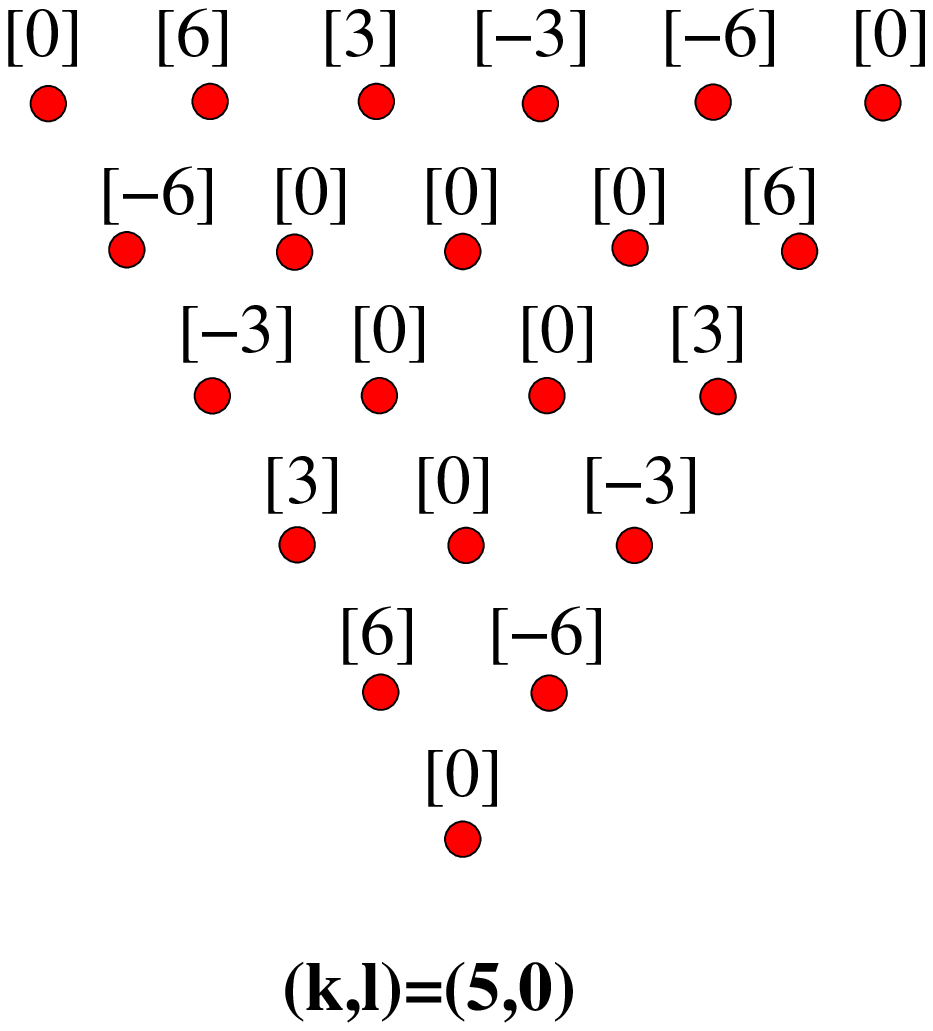} \\[20pt]
  \includegraphics[width=5.5cm]{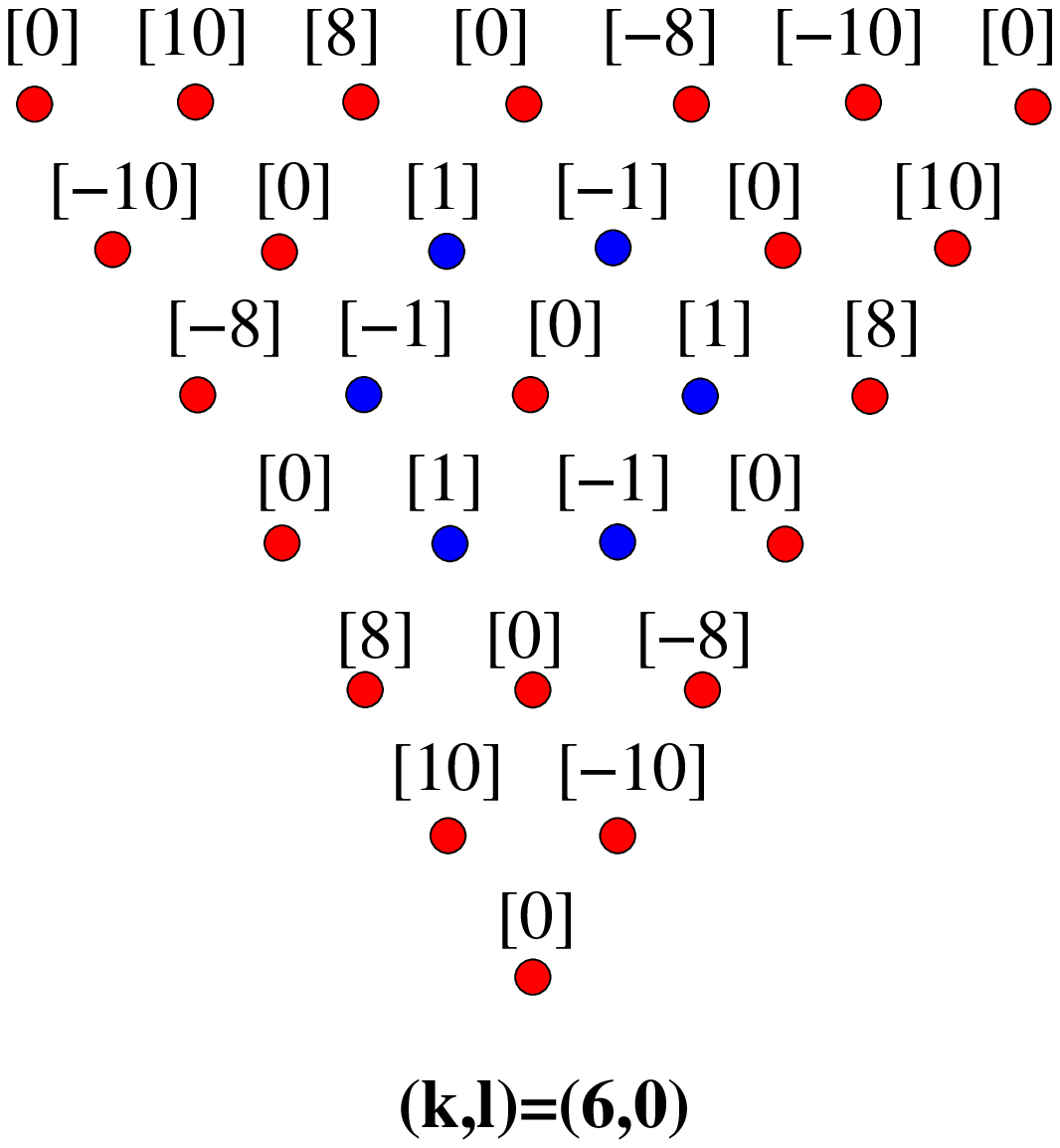}
 \caption{Nodes of lowest $\sut$ weight diagrams accommodating harmonic
   spinors, with the index $\nu_{q,m}$ in square
   brackets attached to each weight $(q,m)_n\in W^{k,l}$.}
 \label{indexweights}
\end{figure}

Zero modes of the Dirac
operator on $\F_3$ can also give rise to Yukawa couplings when the theory
is reduced to $M$. Upon reduction to $M$, Yukawa couplings between $\psi_{(q,m)_n;\ell}$
and $\eta_{(q,m)_n;\ell}$ can arise from integrating the off-diagonal terms from the operator (\ref{YcalQkltilde}), involving
the Higgs fields, in 
\bea
\Psi^\dagger\wedge\, \tilde\star\, \DiracYtilde\, \Psi=
{\Psi^-}\,^\dagger \wedge
\,\tilde\star\,
\DiracYtilde \,\Psi^+ + {\Psi^+}\,^\dagger \wedge \, \tilde\star\, \DiracYtilde \,\Psi^-
\label{YukawaPsiAPsi}\eea 
over $\F_3$, and they depend crucially on the
zero mode structure. The projectors $\widehat\Pi_{(q,m)_n}$
pick out specific zero modes in the expansions (\ref{PsiPM}). Using the explicit matrix elements from
(\ref{offdiagBied}), upon integration over $\F_3$ only the singlet parts of
the fermion bilinears $\chi^{\pm}_{q',m';\,\ell'}{}^\dag \,
\widetilde\gamma\,^\a \chi_{q,m;\,\ell}^\mp$ can survive and
generate Yukawa coupling coefficients, $\widetilde
Y_{+\, \a;\,\ell,\ell'}^{\,(q,m\,)_{n}, \, (q',m')_{n'}}\in\C\,$, given by
\bea
\widetilde Y_{+\, 1;\,\ell,\ell'}^{\,(q,m)_n,\,(q-1,m+3)_{n\pm1}} &=& \sqrt{\mbox{$\frac{n\mp
      q+1\pm1}{2(n+1)}$}} ~ \lambda^\pm_{k,l}(n,m) \, \int_{\F_3}\, \frac{\widetilde{\omega}\,^{\wedge3}}{3!} \
\chi_{q,m;\,\ell}^{+}{}^\dagger \, 
\widetilde\gamma\,^1\chi^-_{q-1,m+3;\,\ell'} \nonumber \\
[4pt]
\widetilde Y_{+\, 2;\,\ell,\ell'}^{\,(q,m)_n,\,(q-1,m-3)_{n\pm1}} 
&=& \sqrt{\mbox{$\frac{n\mp
      q+1\pm1}{2(n+1\pm1)}$}} ~ \lambda^\mp_{k,l}(n\pm1,m-3)
\, 
\int_{\F_3}\, \frac{\widetilde{\omega}\,^{\wedge3}}{3!} \
\chi_{q,m;\,\ell}^{+}{}^\dagger \, 
\widetilde\gamma\,^2\chi^-_{q-1,m-3;\,\ell'} \ , \nonumber \\
[4pt]
\widetilde Y_{+\, 3;\,\ell,\ell'}^{\,(q,m)_n,\,(q+2,m)_n} &=& \frac 1 2 \sqrt{\mbox{${(n-q)\,
        (n+q+2)}$}} \, \int_{\F_3}\,
  \frac{\widetilde{\omega}\,^{\wedge3}}{3!} \
  \chi_{q,m;\,\ell}^{+}{}^\dagger \, 
\widetilde\gamma\,^3\chi^-_{q+2,m;\,\ell'} \ , 
 \label{quasiKahlerYukawa}
\eea
on $M$, plus a completely analogous set of coupling coefficients $\widetilde
Y_{+\, \a;\,\ell,\ell'}^{\,(q,m\,)_n, \, (q',m')_{n'}}\in\C$ which are
obtained from (\ref{quasiKahlerYukawa}) by interchanging chiral and
antichiral spinor labels $\chi^+\leftrightarrow \chi^-$, keeping the same co-efficients,
\bea
\widetilde Y_{-\, 1;\,\ell,\ell'}^{\,(q,m)_n,\,(q-1,m+3)_{n\pm1}} &=& \sqrt{\mbox{$\frac{n\mp
      q+1\pm1}{2(n+1)}$}} ~ \lambda^\pm_{k,l}(n,m) \, \int_{\F_3}\, \frac{\widetilde{\omega}\,^{\wedge3}}{3!} \
\chi_{q,m;\,\ell}^{-}{}^\dagger \, 
\widetilde\gamma\,^1\chi^+_{q-1,m+3;\,\ell'} \nonumber \\
[4pt]
\widetilde Y_{-\, 2;\,\ell,\ell'}^{\,(q,m)_n,\,(q-1,m-3)_{n\pm1}} 
&=& \sqrt{\mbox{$\frac{n\mp
      q+1\pm1}{2(n+1\pm1)}$}} ~ \lambda^\mp_{k,l}(n\pm1,m-3)
\, 
\int_{\F_3}\, \frac{\widetilde{\omega}\,^{\wedge3}}{3!} \
\chi_{q,m;\,\ell}^{-}{}^\dagger \, 
\widetilde\gamma\,^2\chi^+_{q-1,m-3;\,\ell'} \ , \nonumber \\
[4pt]
\widetilde Y_{-\, 3;\,\ell,\ell'}^{\,(q,m)_n,\,(q+2,m)_n} &=& \frac 1 2 \sqrt{\mbox{${(n-q)\,
        (n+q+2)}$}} \, \int_{\F_3}\,
  \frac{\widetilde{\omega}\,^{\wedge3}}{3!} \
\chi_{q,m;\,\ell}^{-}{}^\dagger \, 
\widetilde\gamma\,^3 \chi^+_ {q+2,m;\,\ell'} \ . 
 \label{quasiKahlerYukawaminus}
\eea  
Note that a Yukawa coupling between two weights connected by a quiver arrow
$(q',m')_{n'}\longrightarrow (q,m\,)_n$ can only arise if the
corresponding indices are of opposite sign, due to the change in
spinor chirality induced by multiplication with the $\gamma$-matrices,
$\widetilde\gamma\,^\a$. In
Fig.~\ref{indexweights} we depict some examples of the Dirac
index associated with some low values of $k$ and $l$.
Using $W_+^{k,l}\cap W_-^{k,l}=\emptyset$, the $d$-dimensional Yukawa interactions on $M$
are then given generically by
\bea
\Lcal_M^{\rm Y} &=& 2ig\kern -10pt  \sum_{(q,m)_n\in W_+^{k,l}} \
\sum_{\ell,\ell'=1}^{\nu_{q,m}} \, \Big( \,
\sum_\pm\, \widetilde
Y_{+\, 1;\,\ell,\ell'}^{\,(q,m)_n,\,(q-1,m+3)_{n\pm1}} \
\psi_{(q,m)_n;\ell}{}^\dagger \wedge \star \phi^{1 \
  (\pm)\,\prime}_{(q,m)_n}{}^\dag\,\psid_{(q-1,m+3)_{n\pm1};\,\ell'} \,
\,
\nonumber \\
&& \qquad \qquad \qquad \qquad \quad +\,  \sum_\pm \, \widetilde
Y_{+\,2;\,\ell,\ell'}^{\,(q,m)_n,\,(q-1,m-3)_{n\pm1}} \
\psi_{(q,m)_n;\,\ell}{}^\dagger \wedge\star
\phi^{2 \ (\pm)\,\prime}_{(q,m)_n}{}^\dag\,\psid_{(q-1,m-3)_{n\pm1};\ell'} \nonumber \\
&& \qquad \qquad \qquad \qquad \quad +\, 
\widetilde Y_{+\,3;\,\ell,\ell'}^{\,(q,m)_n,\,(q+2,m)_n} \
\psi_{(q,m)_n;\ell}{}^\dagger \wedge \star
\phi^{3 \,\prime}_{(q,m)_n}{}^\dag \,\psid_{(q+2,m)_n;\,\ell'} \Big)
\label{LMYukawagen} \\ 
&& +\, 2ig\kern-10pt
 \sum_{(q,m)_n\in W_-^{k,l}} \
\sum_{\ell,\ell'=1}^{|\nu_{q,m}|} \, \Big( \, \sum_\pm \,
\widetilde 
Y_{-\, 1;\,\ell,\ell'}^{\,(q,m)_n,\,(q-1,m+3)_{n\pm1}} \
\psid_{(q,m)_n;\ell}{}^\dagger \wedge \star \phi^{1 \
  (\pm)\,\prime}_{(q,m)_n}{}^\dag\, \psi_{(q-1,m+3)_{n\pm1};\,\ell'}
\nonumber \\
&& \qquad \qquad \qquad \qquad \qquad +\, 
\sum_\pm\, \widetilde
Y_{-\, 2;\,\ell,\ell'}^{\,(q,m)_n,\,(q-1,m-3)_{n\pm1}} \
\psid_{(q,m)_n;\ell}{}^\dagger \wedge\star 
\phi^{2 \ (\pm)\,\prime}_{(q,m)_n}{}^\dag\,\psi_{(q-1,m-3)_{n\pm1};\,\ell'}
 \nonumber  \\
&& \qquad \qquad \qquad \qquad \qquad +\,
\widetilde Y_{-\, 3;\,\ell,\ell'}^{\,(q,m)_n,\,(q+2,m)_n} \
\psid_{(q,m)_n;\ell}{}^\dagger \wedge \star \phi^{3
  \,\prime}_{(q,m)_n}{}^\dag\, \psi_{(q+2,m)_n;\,\ell'}
 \, \Big), \nonumber
\eea 
where $\psid_{(q,m)_n;\ell} := \gamma\, \eta_{(q,m)_n;\ell}$,
together with its hermitean conjugate in (\ref{totalfermaction}).

Without an explicit construction of the harmonic spinors we cannot
evaluate the Yukawa coupling coefficients (\ref{quasiKahlerYukawa})
and (\ref{quasiKahlerYukawaminus}), 
nor indeed say which of them will be generally non-zero for
a given choice of $\sut$ representation $\widehat{V}^{k,l}$.
In the case of equivariant dimensional reduction of Yang--Mills--Dirac
theory over the projective spaces $\C
P^n$ with $n=1,2$, which are K\"ahler manifolds, it was observed
in~\cite{Dolan:2009ie,Dolan:2009nz} that Yukawa couplings only arise
in the contributions from harmonic spinors which are \emph{constant} on the coset space; for $\C
P^1=\su/\uo$ such spinors exist in every even-dimensional irreducible
representation of $\su$, while for $\C P^2$ there is a unique choice
of spin$^c$ structure for each irreducible $\sut$ representation
$\widehat{V}{}^{k,l}$ which accommodates constant spinor
harmonics. Moreover, when they exist, the constant spinor harmonics
are unique and hence lie in states of index $\pm\,1$. In the following
we will demonstrate that an analogous construction applies to the
nearly K\"ahler coset space $\F_3$, except that we shall also find
constant spinor harmonics for a class of nodes of index zero. We will furthermore compare the Yukawa interactions (\ref{LMYukawagen}) at different points of the moduli space.

\subsection{Symmetric spinors of torsion class $W_1$\label{W1}}

We will begin by classifying the $\sut$ representations
$\widehat{V}^{k,l}$ which permit non-vanishing Yukawa couplings
(\ref{LMYukawagen}), and lead to dynamical mass generation for the
fermion fields via spontaneous symmetry breaking, at the locus
$\sigma=1$ of the moduli space; recall that this surface contains the
nearly K\"ahler point (\ref{2.53}). In this case the Dirac operator
$\pa_{\F_3}^{\, 1}$ from (\ref{sigmaDirac}) is associated to the
canonical connection (\ref{2.37}). We will explicitly construct
harmonic spinors of the corresponding Dirac operators (\ref{Diracbimon}).

For this, we decompose the complex $(1,0)$-forms $\widetilde\Theta^\a$ into an
invariant, local real orthonormal basis $e^a$, $a=1,\dots,6$ of the
cotangent bundle $T^*\F_3$ as
\beq
\widetilde\Theta^\a=\mbox{$\frac12$}\bigl(e^{2\a-1}+ \im e^{2\a}\bigr)
\eeq
for $\a=1,2,3$. Similarly, we decompose the corresponding complex
gamma-matrices $\widetilde\gamma\,^\a$ into hermitean gamma-matrices
$\gamma^a$, $a=1,\dots,6$ as
\beq
\widetilde\gamma\,^\a=\mbox{$\frac12$}\,\big(\gamma^{2\a-1}+\im
\gamma^{2\a} \big)
\label{complexrealgamma}\eeq
for $\a=1,2,3$; they obey the Clifford relations
\beq
\big\{\gamma^a\,,\,\gamma^b\big\}=2\,\delta^{ab}\, {\bf 1}_8 \ .
\label{realCliffrels}\eeq

Then the canonical torsion three-form can be expressed as
\beq
H=-\mbox{$\frac1{4\sqrt3}$}\, f_{abc}\,e^{abc} = \mbox{$\frac14$}\,
\bigl(e^{135} + e^{425} + e^{416} + e^{326}
\bigr) \ ,
\eeq
where generally $e^{a_1\dots a_r}:=e^{a_1} \wedge \cdots \wedge e^{a_r}$
with $a_i=1,\ldots,6$ and we have used the structure constants (\ref{fabc}). Using the map (\ref{diffcliffmap}) we define
the corresponding hermitean matrix
\beq
\H=\mbox{$\frac\im4$}\, 
\bigl( \gamma^{135}+ \gamma^{425}+ \gamma^{416}+\gamma^{326}\bigr) \ ,
\eeq
where generally $\gamma^{a_1\dots
  a_r}:=\gamma^{[a_1}\dots\gamma^{a_r]}$.
Using the Clifford algebra
(\ref{realCliffrels}) it is straightforward to check that
\beq
\H{}^2=\mbox{$\frac14$}\, \bigl
({\bf 1}_8 - \gamma^{1234} - \gamma^{3456} - \gamma^{1256} 
\bigr) \qquad \mbox{and} \qquad \H{}^4 = \H{}^2 \ .
\eeq
Hence $\H{}^2$ is a projector, and $\Tr_8({\H{}^2})=2$ so it is of rank
two.  The matrix 
$\H$ itself is traceless with real eigenvalues, so it has six zero
eigenvalues and two non-vanishing eigenvalues $\pm\, 1$ each of
multiplicity one.
The eigenspinors of $\H$ are easily determined. Going back to the
complex basis (\ref{complexrealgamma}) we can write
\beq
\H=\im\bigl(\, \tgamma\,^{\bar 1 \, \bar 2 \, \bar
  3}+\tgamma\,^{123}\bigr) \ .
\eeq
Expanding spinors on $\F_3$ as in (\ref{FockFermion}), one
then has
\beq\label{chinoughtwo}
\H \bigl( |\Omega\rangle \pm \im
\tgamma\,^{\bar 1\, \bar 2\, \bar 3}|\Omega\rangle\bigr) =
\pm\, \bigl( |\Omega\rangle \pm\im
\tgamma\,^{\bar 1 \, \bar 2\, \bar 3}|\Omega\rangle\bigr)
\eeq
and
\beq
\H\bigl(\, \tgamma\,^{\bar\alpha}|\Omega\rangle\bigr)=0 =
\H\bigl(\, \tgamma\,^{\bar\alpha\, \bar\beta}|\Omega\rangle\bigr) \ .
\label{Hkernel}\eeq

The Dirac operator (\ref{Diracbimon}) for
the canonical connection on the non-symmetric coset space $\F_3$
squares to~\cite{Dolan}
\beq
\bigl(\im \Dirac_{q,m} \bigr)^2 = \nabla_{q,m}^2
+\mbox{$\frac14$}\,\Rscal \ {\bf 1}_8 - \mbox{$\frac18$}\, \sfR_{abcd}\,
\gamma^{abcd}- \mbox{$\frac\im2$}\, \big(f_{q,m}\big)_{ab}\, \gamma^{ab} \ ,
\label{canDiracsquare}\eeq
where $\nabla_{q,m}^2$ is the spinor laplacean including both the
canonical spin connection and the bimonopole gauge connection, and
$f_{q,m}$ is the
$\uo\times\uo$ field strength (\ref{curvbimon}).
The structure constants (\ref{fabc}) and the explicit expression
for the Riemann curvature tensor of the
canonical connection~\cite{Dolan}
\beq
\sfR_{abcd} = f_{ab7}\, f_{cd7} + f_{ab8}\, f_{cd8}
\eeq
yield curvature two-forms
\bea
\sfR_{12}&=& \mbox{$\frac16$}\, \bigl(2 \, e^{12} -e^{34} - e^{56} \bigr)
\ , \nonumber\\[4pt]
\sfR_{34}&=& \mbox{$\frac16$}\, \bigl(- e^{12} +2 \, e^{34} - e^{56}
\bigr) \ , \nonumber \\[4pt]
\sfR_{56}&=& \mbox{$\frac16$}\, \bigl(- e^{12} -e^{34} +2\, e^{56} \bigr)
\ .
\eea
The Ricci tensor is then ${\sf Ric}_{ab}= \frac13\, \delta_{ab}$,
the scalar curvature is $\Rscal =2$, and
\beq
\sfR_{abcd}\,\gamma^{abcd}=-\mbox{$\frac43$}\, \bigl(\gamma^{1234} + \gamma^{3456}
+\gamma^{1256} \bigr) \ .
\eeq
Combining these expressions we find that the square of the Dirac
operator (\ref{canDiracsquare}) can be written as
\beq
\bigl(\im \Dirac_{q,m}\bigr)^2 = \nabla_{q,m}^2 +
\mbox{$\frac23$}\, \bigl({\bf 1}_8 -\H{}^2\bigr)- \mbox{$\frac\im2$}\,
\big(f_{q,m}\big)_{ab}\, \gamma^{ab} \ .
\label{canDiracsquareH}\eeq
The rank-6 projector ${\bf 1}_8 -\H{}^2$ is related to the $U(1)\times U(1)$
holonomy group in the following way: the generators of the holonomy group
in the spinor representation
can be constructed, using the Clifford algebra (\ref{realCliffrels}), as
\bea\label{TeeSeven}
T_7&=&-\frac14 f_{ab7}\gamma^{ab}=\frac{i}{2\sqrt{3}}\bigl(\tgamma^\1\gamma^1
+\tgamma^\2\tgamma^2-2\tgamma^\3\tgamma^3\bigr),\\
\label{TeeEight}T_8&=&-\frac14 f_{ab8}\gamma^{ab}=-\frac{i
}{2}\bigl(\tgamma^\1\gamma^1
-\tgamma^\2\tgamma^2\bigr),
\eea
with
\beq
[T_7, T_8]=0.
\eeq
Then 
\beq
\mbox{$\frac13$}\, \bigl({\bf 1}_8 -\H{}^2\bigr)=-
\bigl(T_7^2 +T_8^2\bigr)
\eeq
is the second order Casimir.  It is straightforward to check,
using (\ref{I78Bied}),
that $-3\bigl(I_7^2 +I_8^2\bigr)$ is indeed a rank-6 projector
in the adjoint representation: the zero eigenvectors are of course the
weights $(0,0)_2$ and $(0,0)_0$.  In fact $[T_a,\H]=0$, for $a=7,8$.

Let us now look at the zero modes of (\ref{canDiracsquareH}),
beginning with the case where there are no monopole backgrounds,
i.e. $(q,m)=(0,0)$ and $\Dirac_{0,0}=\pa_{\F_3}^{\, 1}$ is the
untwisted canonical Dirac operator from (\ref{sigmaDirac}). By~\cite[Thm.~10.8]{FI} every harmonic spinor is also parallel for the canonical
connection. Now the matrix ${\bf 1}_8 - \H{}^2$ is a projector of rank six, with a two-dimensional
kernel spanned by the Fock space states $\big({\bf 1}_8 \pm\im
\tgamma\,^{\bar 1 \, \bar 2\, \bar 3}\,\big) |\Omega\rangle$. Since
the background flux vanishes, $f_{0,0}=0$, these states generate
harmonic modes which comprise
two covariantly constant spinors $\chi_{0,0}^\pm$,
i.e. $\nabla_{0,0}\chi_{0,0}^\pm=0$. By~\cite[Thm.~4.2]{Agricola} a spinor
field is
parallel with respect to the canonical connection if and only if it is
constant, whence there are two constant zero modes with opposite
chirality given by
\beq
\chi_{0,0}^+= |\Omega\rangle \qquad \mbox{and} \qquad \chi_{0,0}^-=
-\tgamma\,^{\bar 1 \, \bar 2 \, \bar 3}|\Omega\rangle \ .
\label{chipmconst}\eeq
These states are unique, up to a normalisation and a phase,
consistent with the index theorem (\ref{indDiracqm}) which in this case
gives $\nu_{0,0}=0$, as then $\dim\ker(
\Dirac_{0,0}^+)=\dim\ker(\Dirac_{0,0}^-) =1$. The states (\ref{chipmconst}) correspond to the doubly
degenerate weight $(0,0)_{0,2}$ of the adjoint representation, as discussed
in~\S\ref{Invspinor}; these states are represented
by the red nodes in the $(k,l)=(1,1)$ diagram of
Fig.~\ref{indexweights}.
They form the components of the canonical
covariantly constant spinor $\chi_{0,0}=\chi_{0,0}^+\oplus \chi_{0,0}^-$ associated to the
family of $\sut$-structures on $\F_3$ at the nearly K\"ahler point (\ref{2.53})
of the moduli space.  

For $(q,m)\neq(0,0)$, the constant spinors (\ref{chipmconst}) are no
longer zero modes, because of both the inhomogeneous field
strength term in (\ref{canDiracsquareH}) and the bimonopole
connection featuring in the Dirac laplacean $\nabla^2_{q,m}$. However,
by the above analysis
there are also six constant eigenspinors of $\im \pa_{\F_3}^{\, 1}$ with eigenvalues $\pm\,
\sqrt{2/3}$ each of multiplicity three from (\ref{Hkernel}). Moreover,
from (\ref{curvbimon}) and (\ref{2.39a})--(\ref{2.39b}) we find
\beq
\big(f_{q,m}\big)_{ab}\, \gamma^{ab} = -
\mbox{$\frac\im{24}$}\,\big(\, (m-q)\, \tgamma\,^{1\,\bar
  1}-(m+q)\, \tgamma\,^{2\,\bar 2} + 2\, q\, \tgamma\,^{3\,\bar 3} \, \big) \ .
\eeq
Hence by choosing the background bimonopole charges $(q,m)$
appropriately we can cancel both the flux $\frac23\,\big({\bf 1}_8 -
\H{}^2\big)$ in (\ref{canDiracsquareH}) and the corresponding
contribution to
(\ref{Diracbimon}) of the canonical connection; note that this
cancellation relies crucially on the fact that the canonical
connection (\ref{2.37}) is valued in the Cartan subalgebra $\uoL\oplus\uoL$,
like the background bimonopole gauge potentials. From
(\ref{2.37}) it follows that these charges are precisely the non-zero
weights of the adjoint representation, and for each such charge there
is a unique constant harmonic spinor $\chi_{q,m}^\pm$; these states are represented
by the blue nodes in the $(k,l)=(1,1)$ diagram of
Fig.~\ref{indexweights}. Each of these charges will turn one of the
six constant but non-zero modes of the untwisted Dirac operator $\im
\pa_{\F_3}^{\, 1}$ into a constant zero mode, while at the same time
turning the two constant zero modes of $\im
\pa_{\F_3}^{\, 1}$ into constant non-zero modes. From the $\uo\times\uo$ charge
assignments given by (\ref{tgammaminus})--(\ref{tgammaplus}), the three
positive chirality zero modes are
\beq
\chi^+_{-2,0} = \tgamma\,^{\bar1\, \bar2}|\Omega\rangle \ ,
\qquad \chi^+_{1,3} = \tgamma\,^{\bar3\, \bar1}|\Omega\rangle \qquad
\mbox{and} \qquad \chi^+_{1,-3} = \tgamma\,^{\bar2\,
  \bar3}|\Omega\rangle \ ,
\label{chiplusconst}\eeq
while the three negative chirality zero modes are
\beq
\chi^-_{-1,3} = \tgamma\,^{\bar1}|\Omega\rangle \ , \qquad
\chi^-_{-1,-3} = \tgamma\,^{\bar2}|\Omega\rangle \qquad \mbox{and}
\qquad \chi^-_{2,0} = \tgamma\,^{\bar3}|\Omega\rangle \ .
\label{chiminusconst}\eeq
In fact, these are the \emph{only}
bimonopole fields which give rise to twisted spinor bundles that admit
unique, constant chiral harmonic modes: For a given representation
$\widehat{V}^{k,l}$, it is
straightforward to show from the index formula (\ref{indDiracqm}) that
there are only six possible weights $(q,m)_n\in W^{k,l}$ yielding index
$\nu_{q,m}=\pm\,1$, given as
\beq
(q,m)=(1,3) \ , \quad (-2,0) \ , \quad (1,-3) \qquad \mbox{with} \quad
\nu_{q,m}=+1
\label{qmnuplus}\eeq
and
\beq
(q,m)=(-1,-3) \ , \quad (2,0) \ , \quad (-1,3) \qquad \mbox{with} \quad
\nu_{q,m}=-1 \ ,
\label{qmnuminus}\eeq
consistently with the $\uo\times\uo$ charge assignments in (\ref{chiplusconst})--(\ref{chiminusconst}).

Returning to the zero modes (\ref{chipmconst}) we shall now show
that they are actually linear 
combinations of $SU(3)$ weights.
The $U(1)\times U(1)$ action (\ref{TeeSeven}) and (\ref{TeeEight}) on spinors
can be extended to $SU(2)\times U(1)$ by constructing two more generators,
\beq 
T_5=- \frac 1  4 f_{ab5}\gamma^{ab},\qquad
T_6=- \frac 1  4 f_{ab6}\gamma^{ab}\,,
\label{TeeFivenSix}
\eeq
where $\gamma^7:=i\gamma^{123456}=\tgamma$ is the chirality operator in (\ref{F3chiralityop}).\footnote{There is no $\gamma^8$ in this construction but it
works because, with the structure constants in (\ref{fabc}),
no $\gamma^8$ ever appears on the right hand sides of (\ref{TeeFivenSix}).}
Raising and lowering operators can now be built,
\beq\label{TeePlus}
T_+= \frac 1 2 (T_5+iT_6)= \frac{1}{2\sqrt{3}}(\tgamma^{\bar 1\bar 2} - i \tgamma^3\tgamma),\eeq
\beq\label{TeeMinus}
T_-= \frac 1 2 (T_5-iT_6)= \frac{1}{2\sqrt{3}}(\tgamma^{1 2} + i \tgamma^{\bar 3}\tgamma)\, 
\eeq
which, together with $T_7$ in equation (\ref{TeeSeven}), generate $SU(2)$, 
\beq
[T_+,T_-]= \frac{1}{2\sqrt{3}} T_7, \qquad [T_7,T_\pm]=\pm\frac{i}{2\sqrt{3}} T_\pm\,.
\eeq
Equations (\ref{TeePlus}) and (\ref{TeeMinus}) give, 
with (\ref{chiplusconst}) and (\ref{chiminusconst}), 
\beq
T_+\chi^-_{2,0}=\frac{i}{2\sqrt{3}}\bigl(|\Omega> -i \tgamma^{\bar 1 \bar 2 \bar 3}|\Omega>\bigr)=\frac{i}{2\sqrt{3}}\bigl(\chi^+_{0,0} -i \chi^-_{0,0}\bigr)\,,\eeq
\beq
T_-\chi^+_{-2,0}=-\frac{1} {2\sqrt{3}}\bigl(|\Omega> -i \tgamma^{\bar 1 \bar 2 \bar 3}|\Omega>\bigr)=-\frac{1} {2\sqrt{3}}\bigl(\chi^+_{0,0} -i \chi^-_{0,0}\bigr)\,.
\eeq
Thus $\chi^-_{2,0}$, $\frac 1 {\sqrt{2}} (\chi^+_{0,0} - i \chi^-_{0,0})$ and $\chi^-_{2,0}$ form a natural $SU(2)$ triplet and we can identify, up to inconsequential phases,
\beq
\chi_{(0,0)_2}=\frac 1 {\sqrt{2}} (\chi^+_{0,0} - i \chi^-_{0,0}).
\label{chi002}\eeq
The orthogonal combination
\beq
\chi_{(0,0)_0}=\frac{1} {\sqrt{2}} (-i\chi^+_{0,0} +\chi^-_{0,0})\,,
\label{chi000}\eeq
is an $SU(2)$ singlet annihilated by $T_\pm$.
$\chi_{(0,0)_2}$ and $\chi_{(0,0)_0}$ are in fact the eigenspinors of $\H$ in 
(\ref{chinoughtwo}).
Thus the two zero modes associated with the weights $(0,0)_2$ and
$(0,0)_0$ are not themselves Weyl spinors, they are Dirac spinors 
which are linear combinations of two Weyl spinors.

It is possible to classify completely all representations $\widehat{V}^{k,l}$
which give rise to such spinor harmonics.
For a given irreducible representation of $\sut$, one can work out the
index for each weight using (\ref{indDiracqm}) (see Fig.~\ref{indexweights}); note that for triangular
quivers $Q^{k,0}$ the index (\ref{indDiracqm}) can
be parametrized using (\ref{mnqjpm}) as
\beq
\nu_{k;q,n}= \mbox{$\frac18$}\, q\, \big(q^2-(2k-3n)^2\big) \ ,
\label{indDirack0}\eeq
where $n=0,1,\dots,k$ labels the hypercharge levels and
$q\in\{-n,-(n-2),\dots,n-2, n\}$.
The smallest irreducible representation in which the weights
(\ref{qmnuplus})--(\ref{qmnuminus}) appear is the adjoint representation of
$\sut$ with $(k,l)=(1,1)$, while the next one is the decuplet
representation with $(k,l)=(3,0)$; in both cases these six
possibilities, together with the nodes $(0,0)$ at the centre, represent the only zero modes which arise. They can also
occur in higher irreducible representations $\widehat{V}^{k,l}$, but
only in those with weight diagrams $W^{k,l}$ which have the $\Z_3$-symmetry
that allows them to contain the outer hexagonal layer of the adjoint
diagram $W^{1,1}$ (Fig.~\ref{indexweights}). From the explicit construction of
the weights $(q,m)_n$ given in (\ref{mnqjpm}), it is straightforward to see that
the only weight diagrams $W^{k,l}$ which contain the weights
(\ref{qmnuplus})--(\ref{qmnuminus}) correspond to representations
\beq
\widehat{V}^{k,l} \qquad \mbox{with} \quad k-l\equiv 0 \ \ {\rm mod}
\ 3 \ .
\label{Yukawareps}\eeq
By (\ref{qmnuplus})--(\ref{qmnuminus}), each of the blue nodes in Fig.~\ref{indexweights} corresponds to a
twisted spinor bundle over $\F_3$ which admits a unique constant zero
mode of either positive or negative chirality.

For a generic representation in the class
(\ref{Yukawareps}) with $k\geq l$, the bottom edge of the adjoint hexagon has nodes
with hypercharge $m=-3$. Recall from our analysis of the combinatorics of weight diagrams from
\S\ref{Flatconns} that the inverted triangular
layer of a weight diagram $W^{k,l}$ starts at level $m=-2(k-l)$ (see
Fig.~\ref{weight73}). Hence for $k>l$ the adjoint hexagon lies inside
the triangle and each of its weight states therefore has multiplicity $l+1$, while for
$k=l$ there are no triangular layers and the states have multiplicity
$l$; in both cases the node $(0,0)_n$ in the centre of the hexagon has
degeneracy $l+1$. 
For a
fixed weight $(q,m)_n$ in the adjoint hexagon, from (\ref{mnqjpm}) it
follows that the isospin labels can be parametrized as $n=n_j$
where
\beq
n_j=2j+\mbox{$\frac23$}\, (k-l)+\mbox{$\frac m3$} \qquad \mbox{with}
\quad j:=2j_-=0,1,\dots, l
\eeq
for $k>l$, while for $k=l$ we have $j\neq0$ when $m=-3$ or
$m=0,q=\pm\,2$ and $j\neq l$ for $m=3$.

If we normalise the fermionic Fock vacuum
$|\Omega\rangle$ such that
\beq
\langle\Omega|\Omega\rangle= {\rm Vol}({\F_3})^{-1} \ ,
\eeq
then the non-zero Yukawa couplings around the adjoint hexagon are denoted by
\bea
\widetilde x_j^{\,+} &:=& -\widetilde Y_{-1;\,1,1}^{\,  (2,0)_{n_j} ,\,
(1,3)_{n_{j}-1}} \=  \sqrt{\mbox{$\frac{n_{j}+2}{2(n_{j}+1)}$}} \
\lambda_{k,l}^-(n_{j},0)
\, , \nonumber\\
[4pt] \widetilde x_j^{\,-} &:=& \widetilde Y_{-1;\,1,1}^{\,  (-1,-3)_{n_j} ,\,
(-2,0)_{n_{j}+1}} \= \sqrt{\mbox{$\frac{n_{j}+3}{2(n_{j}+1)}$}} \
\lambda_{k,l}^+(n_j,-3) \ , \nonumber\\
[4pt]
 \widetilde y_j^{\,+} &:=& \widetilde Y_{-2;\,1,1}^{\,  (2,0)_{n_j} ,\,
(1,-3)_{n_{j}-1}} \= \sqrt{\mbox{$\frac{n_{j}+2}{2n_{j}}$}} \ 
\lambda_{k,l}^+(n_{j}-1,-3) \, , \nonumber\\
[4pt] 
\widetilde y_j^{\,-}&:=& -\widetilde Y_{-2;\,1,1}^{\,(-1,3)_{n_j},\,
   (-2,0)_{n_{j}+1}} \= \sqrt{\mbox{$\frac{n_{j}+3}{2(n_{j}+2)}$}} \
\lambda_{k,l}^-(n_{j}+1,0)\, , 
\label{xyYuk}\eea
for the diagonal links, and
\beq
\widetilde z_j^0:=  
\widetilde Y_{-3;\,1,1}^{\,  (-1,3)_{n_j},\,
(1,3)_{n_j}}= -\widetilde Y_{-3;\,1,1}^{\,  (-1,-3)_{n_j},\,
(1,-3)_{n_j}}  =\mbox{$\frac{n_{j}+1}{2}$}\label{zYuk}
\eeq
for the horizontal links (all the relevant fermion bilinears in
(\ref{quasiKahlerYukawa}) evaluate to~$\pm1$ for the constant spinor
harmonics (\ref{chiplusconst})--(\ref{chiminusconst})). 

To evaluate the non-vanishing Yukawa couplings corresponding to arrows 
associated with the centre nodes $(0,0)_{n_j}$ we must allow for possible mixing between the $SU(3)$ weights and  the chiral zero modes.
Let  the central $SU(3)$ weights be $(0,0)_{n_>}$ and $(0,0)_{n_<}$, 
with $n_>=n_<+2$, then the corresponding zero modes can be taken to be 
$\chi_{(0,0)_{n_>}}$ and $\chi_{(0,0)_{n_<}}$, where  
\bea
\chi_{(0,0)_{n_<}}&=&\bigl(u_{n_<}\chi_{0,0}^+ + v_{n_<}\chi_{0,0}^-\bigr)\\
\chi_{(0,0)_{n_>}}&=&\bigl(u_{n_>}\chi_{0,0}^+ + v_{n_>}\chi_{0,0}^-\bigr)
\label{Chi12}
\eea
is a unitary transformation.
We adopt a slight modification in notation for the central weights $(0,0)_{n_>}$ and $(0,0)_{n_<}$
of the adjoint under consideration here: because $\chi_{(0,0)_{n_>}}$ and 
$\chi_{(0,0)_{n_<}}$ are not in general chiral the two associated spinor fields on $M$ will 
be denoted $\rho_{(0,0)_{n_>}}$ and $\rho_{(0,0)_{n_<}}$
and the expansion of $\Psi_0$ in (\ref{PsiPMzero}) becomes
\beq
\Psi_0=\rho_{(0,0)_{n_>}} \otimes \chi_{(0,0)_{n_>}} +
\rho_{(0,0)_{n_<}} \otimes \chi_{(0,0)_{n_<}}\,.
\eeq
The relevant Fermion bilinears then evaluate to give Yukawa couplings
\bea
\widetilde\xi_j^{\,\pm} &:=& \widetilde Y_{+1;\,1,1}^{\,
(0,0)_{n_{j}}, \,   (-1,3)_{n_{j}\mp1}} \=
 \sqrt{\mbox{$\frac{n_{j}\mp1+1}{2(n_{j}+1)}$}} \ \lambda_{k,l}^\mp(n_{j},0) \,u_{n_j}^*
\ , \nonumber \\[4pt]
\widetilde\xi_j^{\,\prime}\,^{\pm} &:=& \widetilde Y_{+1;\,1,1}^{\,
(1,-3)_{n_{j}}, \,   (0,0)_{n_{j}\mp1}} \=
\mbox{$\frac1{\sqrt{2}}$} \ \lambda_{k,l}^\mp(n_{j},-3)\,v_{n_j\mp1} \ , \nonumber\\[4pt]
\widetilde\zeta_j^{\,\pm} &:=& \widetilde Y_{+2;\,1,1}^{\,
 (0,0)_{n_{j}}  , \,(-1,-3)_{n_{j}\mp1}} \= \mbox{$\frac1{\sqrt2}$} \ \lambda_{k,l}^{\pm}(n_{j}\mp1,-3)\, u^*_{n_j} \ ,
\nonumber\\[4pt] 
\widetilde\zeta_j^{\,\prime}\,^{\pm} &:=& \widetilde Y_{+2;\,1,1}^{\,
(1,3)_{n_{j}}, \,   (0,0)_{n_{j}\mp1}} \=
\sqrt{\mbox{$\frac{n_{j}+1}{2(n_{j}\mp 1+1)}$}} \
\lambda_{k,l}^{\pm}(n_{j}\mp1,0)\,v_{n_j\mp1}\,
\label{ycentrediag}
\eea
for the diagonal links, while
\bea\label{zeta0Yuk}
\widetilde\zeta_j^{\,0}&:=& 
\widetilde Y_{+3;\,1,1}^{\,(0,0)_{n_j},\,(2,0)_{n_j}}
=\frac12\sqrt{\mbox{$ {n_j\, (n_j+2) }$}}\, u_{n_j}^*\\
\widetilde\zeta_j^{\,\prime}\,^{\,0}&:=& 
\widetilde Y_{+3;\,1,1}^{\,  (-2,0)_{n_j}\,,(0,0)_{n_j}}
=\frac12\sqrt{\mbox{$ {n_j\, (n_j+2) }$}}\,v_{n_j}\,,
\label{zetaprime0Yuk}\eea
for the horizontal links, where the $\pm$ superscripts on the left
hand side of (\ref{ycentrediag}) are
both included in the $k=l$ cases to account for the doubling of
arrows between layers of different weight multiplicities, and we have
utilized the combinatorial counting of physical fields from
\S\ref{Flatconns}. 

In the following we abbreviate $\psi_{(q,m)_n}:= \psi_{(q,m)_n;1}$,
$\psid_{(q,m)_n}:= \psid_{(q,m)_n;1}=\gamma\,\eta_{(q,m);1}$ and
$\breve\rho_{(q,m)_n}:=\gamma\,\rho_{(q,m)_n}$. 
Then 
the constant spinor contributions to the Yukawa interaction
lagrangean (\ref{LMYukawagen}), including the contribution from $\Psi^0$
in (\ref{PsiPM}), read as
\bea
\Lcal_M^{\rm Y}\,^0 &=& 2ig\, \sum_{j} \,\Big( 
 -\widetilde x_j^{\,+}\, \psid_{(2,0)_{n_j}}{}^\dagger \wedge\star 
\phi^{1 \
  (-)\,\prime}_{(2,0)_{n_j}}{}^\dag\,\psi_{(1,3)_{n_{j}-1}}
 +\,\widetilde
x_j^{\,-}\, \psid_{(-1,-3)_{n_j}}{}^\dagger \wedge\star 
\phi^{1 \
  (+)\,\prime}_{(-1,-3)_{n_j}}{}^\dag\,\psi_{(-2,0)_{n_{j}+1}}
\nonumber \\ & & \qquad \qquad
+\widetilde y_j^{\,+}\, \psid_{(2,0)_{n_j}}{}^\dagger \wedge\star 
\phi^{2 \
  (-)\,\prime}_{(2,0)_{n_j}}{}^\dag\,\psi_{(1,-3)_{n_{j}-1}}
-\,\widetilde
y_j^{\,-}\, \psid_{(-1,3)_{n_j}}{}^\dagger \wedge\star 
\phi^{2 \
  (+)\,\prime}_{(-1,3)_{n_j}}{}^\dag\,\psi_{(-2,0)_{n_{j}+1}} 
\nonumber \\ & & \qquad \qquad
 +\, \widetilde
z_j^{\,0}\,\big(- \psid_{(-1,-3)_{n_j}}{}^\dagger \wedge\star 
\phi^{3 \,\prime}_{(-1,-3)_{n_j}}{}^\dag\,\psi_{(1,-3)_{n_j}} +
\psid_{(-1,3)_{n_j}} {}^\dagger \wedge\star 
\phi^{3 \,\prime}_{(-1,3)_{n_j}}{}^\dag\, \psi_{(1,3)_{n_j}}\big)
\nonumber \\ & & \qquad \qquad 
 +\, \widetilde
\xi_j^{\,\mp}\, \rho_{(0,0)_{n_{j}}}{}^\dagger \wedge\star 
\phi^{1 \
  (\pm)\,\prime}_{(0,0)_{n_{j}}}{}^\dag\,\psid_{(-1,3)_{n_{j}\pm1}} +
\widetilde \xi_j^{\,\prime}\,^{\mp}\, 
\psi_{(1,-3)_{n_{j}}}{}^\dagger \wedge\star \phi^{1 \
  (\pm)\,\prime}_{(1,-3)_{n_{j}}}{}^\dag\,\breve\rho_{(0,0)_{n_{j}\pm1}}
\nonumber \\ & & \qquad \qquad
+\, \widetilde
\zeta_j^{\,\mp}\, \rho_{(0,0)_{n_{j}}}{}^\dagger \wedge\star 
\phi^{2 \
  (\pm)\,\prime}_{(0,0)_{n_{j}}}{}^\dag\,\psid_{(-1,-3)_{n_{j}\pm1}}
 +\widetilde \zeta_j^{\,\prime}\,^{\mp}\, \psi_{(1,3)_{n_{j}}}{}^\dagger \wedge\star 
\phi^{2 \
  (\pm)\,\prime}_{(1,3)_{n_{j}}}{}^\dag\,\breve\rho_{(0,0)_{n_{j}\pm1}} \nonumber \\ & & \qquad \qquad +\, \widetilde
\zeta_j^{\,0}\, \psi_{(0,0)_{n_j}}{}^\dagger \wedge\star 
\phi^{3 \,\prime}_{(0,0)_{n_j}}{}^\dag\,\psid_{(2,0)_{n_j}} 
+\zeta^{\,\prime}\,^{\,0}\psi_{(-2,0)_{n_j}}{}^\dagger \wedge\star 
\phi^{3\,\prime}_{(-2,0)_{n_j}}{}^\dag\,\breve\rho_{(0,0)_{n_j}} \Big)\ , \nonumber \\
\label{LMYukawa0} 
\eea
together with its hermitian conjugate,
with an implicit sum over $\pm$ for the real representations with
$k=l$. Thus the quiver gauge theory contains Yukawa interactions for
every quiver derived from an $\sut$ representation of the form
(\ref{Yukawareps}). If the Higgs fields appearing in (\ref{LMYukawa0})
acquire a non-zero vacuum expectation value through dynamical symmetry
breaking, then the $8(l+1)-6\,\delta_{kl}$ fermion fields occurring in
(\ref{LMYukawa0}) acquire a mass matrix. In the special case
(\ref{phiQklgensol}), the eigenvalues of this mass matrix, like the perturbative
induced gauge and Higgs boson masses, are independent of the gauge
coupling $g$ and determined entirely by the metric moduli $\s_\a$ of
the coset space $\F_3$.

As an explicit example, let us consider the simplest non-trivial case
of the adjoint representation
with $(k,l)=(1,1)$. The Yukawa couplings in (\ref{LMYukawa0}) can be
associated with the arrows in the quiver diagram
\bea
\xymatrix{
 & (-1,3)_1 \ \ar@/^/[dr]|{ \ \widetilde \xi\,^+\,\s'_1 \ } \ar@/_/[dr]|{ \ \widetilde \xi\,^-\,\s'_1 \ } & & \ (1,3)_1
 \ar[ll]|{ \ \widetilde z\,^0\,\s'_3 \ }\ar[dr]|{ \ -\widetilde x\,^+\,\s'_1
  \ }
 &  \\
(-2,0)_2 \  \ar[ur]|{ \ -\widetilde y\,^-\,\s'_2 \ } \ar[dr]|{ \ \widetilde x\,^-\,\s'_1 \ } & &  
\ (0,0)_{0,2} \ \ar[ll]|{ \ \widetilde \zeta'\,^{0}\,\s'_3 \ }
 \ar@/_/[ur]|{ \ \widetilde \zeta\,^{\prime -}\,\s'_2 }
 \ar@/^/[ur]|{ \ \widetilde \zeta\,^{\prime +}\,\s'_2 \ }
 \ar@/^/[dr]|{ \ \widetilde\xi\,^{\prime -}\,\s'_1 \ } 
\ar@/_/[dr]|{ \ \widetilde\xi\,^{\prime +}\,\s'_1 \ }
 & & 
\ (2,0)_2 \ .\ar[ll]|{ \ \widetilde\zeta\,^0\,\s'_3 \ }  \\
 & (-1,-3)_1 \   \ar@/^/[ur]|{ \ \widetilde\zeta\,^-\,\s'_2 \ }
\ar@/_/[ur]|{ \ \widetilde\zeta\,^+\,\s'_2 \ }
& & \ (1,-3)_1 \ar[ur]|{ \ \widetilde y\,^+\,\s'_2 \ } 
\ar[ll]|{ \ -\widetilde z\,^0\,\s'_3 \ } 
}
\eea

Using (\ref{lambdaklnm}), together with the values of $u_2$, $u_0$, $v_2$
and $v_0$ from (\ref{chi002}) and (\ref{chi000}), the Yukawa coupling coefficients
(\ref{xyYuk}), (\ref{zYuk}), (\ref{ycentrediag}), (\ref{zeta0Yuk}) and 
(\ref{zetaprime0Yuk}) can be computed explicitly, and after
substituting the Higgs vacuum (\ref{phiQklgensol}) with $V^{1\ (\pm)}_{(q,m)_n}= V^{2}_{(q,m)_n}= V^{3\
  (\pm)}_{(q,m)_n}={\mbf 1}_p$ the induced
$8\times 8$ fermion
mass matrix $\mbf\mu$ can be read off from (\ref{LMYukawa0}).
Then, with the relevant symmetric spinors then organised
into an eight-dimensional vector,
\bea
\deepsigh=\big(\eta_{(2,0)_2}\,,\, \rho_{(0,0)_2}\,,\, \psi_{(-2,0)_2}\,,\,
 \psi_{(1,3)_1} \,,\, \eta_{(-1,3)_1}\,,\,\eta_{(-1,-3)_1}\,,\, \psi_{(1,-3)_1}\,,\,\rho_{(0,0)_0}\, \big)^T\ ,
\eea
the mass matrix in $\deepsigh^\dagger \gamma\, {\mbf\mu}\,\deepsigh$ is
\beq
 \mbf\mu=2ig\left( \begin {array}{cccccccc} 0&-{\s_3}&0&-{\s_1}&0&0&
{\s_2}&0\\\noalign{\medskip} {\s_3}&0&-i{\s_3}&-i\frac{\s_2}{2}
&\frac{\s_1}{2}&\frac{\s_2}{2}&-i\frac{\s_1}{2}&0\\
\noalign{\medskip}0
&-i{\s_3}&0&0&{\s_2}&-{\s_1}&0&0\\
\noalign{\medskip}
{\s_1}&-i\frac{\s_2}{2}&0&0&-{\s_3}&0&0&\frac{\sqrt{3}}{2}\, {\s_2}\\
\noalign{\medskip}0&-\frac{\s_1}{2}&-{\s_2}&{\s_3}&0&0&0&
i\frac{\sqrt{3}}{2}\,{\s_1}\\
\noalign{\medskip}0&-\frac{\s_2}{2} &{\s_1}&0&0&0&-{\s_3}&
i\frac{\sqrt{3}}{2}\,{\s_2}
\\
\noalign{\medskip}-{\s_2}&-i\frac{\s_1}{2} &0&0&0&{\s_3}&0&\frac{\sqrt{3}}{2}\, {\s_1} \\
\noalign{\medskip}0&0&0&-\frac{\sqrt{3}}{2}\,{\s_2} &
i\frac{\sqrt{3}}{2}\,{\s_1}&i\frac{\sqrt{3}}{2}\,{\s_2} &
-\frac{\sqrt{3}}{2}\, {\s_1}&0\end {array} \right)
\label{massmatrixYuk}\eeq
together with the constraint $\s_1\,\s_2\, \s_3=\frac{\sqrt3}{72}$.
The matrix
(\ref{massmatrixYuk}) is hermitian, it can be diagonalised and its eigenvalues
determined as functions of the metric moduli $\s_\a$, $\a=1,2,3$; the
mass eigenvalues come in charge conjugate pairs $\pm\,\mu_i$,
$i=1,\dots,4$.  The explicit expression for the eigenvalues
are not illuminating for generic $\s_\alpha$ 
and here we just note that the determinant of
the mass matrix is given by
\beq
192\,g^8\, \left( {{\s_2}}^{2}+{{\s_1}}^{2} \right) ^{2} \left( {{
\s_2}}^{2}+{{\s_1}}^{2}+2\,{{\s_3}}^{2} \right) ^{2}\,,
\eeq
so there are never any massless fermions in this sector of the spectrum on $M$.

We should stress that this analysis does not necessarily classify all
possible Yukawa interactions, as we have not precluded the possibility
that the coupling coefficients (\ref{quasiKahlerYukawa}) and (\ref{quasiKahlerYukawaminus}) may be
non-vanishing for fermion bilinears associated to pairings between
weight states $(q,m)_n\in
W^{k,l}$ associated to indices of equal magnitude $|\nu_{q,m}|>1$ and
opposite sign (see Fig.~\ref{indexweights}); however, we have shown that such 
harmonic spinors are
necessarily non-constant, and the evaluation of the
integrals in (\ref{quasiKahlerYukawa}), (\ref{quasiKahlerYukawaminus}) 
requires their explicit
construction, which we will not attempt here. Nevertheless, we have
classified all couplings associated to constant harmonic spinors,
equivalently symmetric fermions corresponding to unique Dirac zero modes,
and found a large class of quasi-K\"ahler $\sut$-structures including
the nearly K\"ahler point which admit
chirally paired fermion mass generation.

\subsection{Symmetric spinors of torsion class $W_1\oplus W_2$\label{SpinorW1W2}}

It is interesting to study how the Yukawa couplings vary as we move
around the moduli space. Unfortunately, the situation is far less
under control for generic values of the metric moduli $\s_\a$, as the
constant spinor fields $\chi$ on $\F_3$ are no longer zero modes of
the untwisted Dirac operator $\pa_{\F_3}^{\,
  \sigma}$. By~\cite[Thm.~4.2]{Agricola}, constant spinors are
eigenspinors of the square $\big(\im \pa_{\F_3}^{\, \sigma}
\big)^2$ with the eigenvalue equation
\beq
\big(\im \pa_{\F_3}^{\, \sigma} \big)^2\chi=\mbox{$\frac94$}\,
(1-\sigma)^2\, \chi \qquad \mbox{for} \quad \chi={\rm constant} \ .
\eeq
Hence in the generic case the adjoint octet of constant spinors (\ref{chipmconst}),
(\ref{chiplusconst})--(\ref{chiminusconst}) play no role in the
construction of symmetric fermions, and an explicit determination of the Yukawa coupling
coefficients (\ref{quasiKahlerYukawa}) requires a more detailed
understanding of the non-constant spinor harmonics. The Yukawa couplings
in this case come from weight states $(q,m)_n\in W^{k,l}$
corresponding to higher-dimensional Dirac kernels $\ker(\Dirac_{q,m}^\pm)$, and because they
can only arise on links between nodes of the same index
$\pm\,\nu_{q,m}$, we expect that for a given $\sut$-module
$\widehat{V}{}^{k,l}$ there will be far fewer Yukawa interactions
between symmetric fermions, if any (see Fig.~\ref{indexweights}). One might regard this feature as a
further physical vindication for specifically constraining the
heterotic flux compactification to the nearly K\"ahler locus of the
moduli space, a requirement that usually follows from supersymmetry
considerations~\cite{Klaput:2011mz}.

A representative class of quasi-K\"ahler $\sut$-structures which does
not include the nearly K\"ahler point $\s_\a=\frac1{2\, \sqrt3}$,
$\a=1,2,3$ occurs on the surface $\sigma=\frac13$ in the moduli
space. Recall that this is the only other known case where the index
$|\nu_{q,m}|$ corresponds to the dimension of the vector space
$\ker(\Dirac_{q,m})$ of
harmonic spinors. Moreover, in this case any
constant spinor is a zero mode of the operator $\big(\im
\pa_{\F_3}^{\, 1/3}\, \big)^2- \mbf1_8$. The complexity of the change in structure of
the harmonic spinors in this case is further exemplified by noting
that the analog of the relation (\ref{canDiracsquareH}) for
$\sigma=\frac13$ is given by~\cite{Kimura:2006af,HKWY}
\beq
\bigl(\im \Dirac_{q,m}\bigr)^2 = \nabla_{q,m}^{2} -
\mbox{$\frac23$}\, \H{}^2 + \mbox{$\frac12$}\, \big( \mbf 1_8+
K_{abcd}\, \gamma^{abcd} \big) - \mbox{$\frac\im2$}\,
\big(f_{q,m}\big)_{ab}\, \gamma^{ab} \ .
\label{KostantDiracsquareH}\eeq
The torsional curvature $K:=\diff H=\sqrt3 \ {\rm Im}\, \diff\Omega$ can
be computed by using the Cartan structure equations for the frame
$\widetilde\Th^\a$ from \S\ref{SU3struct} to get
\beq
\diff\Omega = W_1\, \widetilde\omega\wedge \widetilde\omega+ W_2\wedge
\widetilde\omega \ ,
\eeq
where
\beq
W_1=\mbox{$\frac1{36}\, \big(\frac1{\s_1^2} + \frac1{\s_2^2} +
  \frac1{\s_3^2} \big)$}
\eeq
and
\beq
W_2=\mbox{$\frac\im{432}\,\Big( \big(\frac2{\s_1^2} - \frac1{\s_2^2} -
  \frac1{\s_3^2}\big) \, \widetilde\Th^1\wedge\widetilde\Th^{\1}-\big(\frac2{\s_2^2} - \frac1{\s_1^2} -
  \frac1{\s_3^2}\big) \, \widetilde\Th^2\wedge\widetilde\Th^{\2} + \big(\frac2{\s_3^2} - \frac1{\s_1^2} -
  \frac1{\s_2^2} \big) \, \widetilde\Th^3\wedge\widetilde\Th^{\3} \Big)$} \ ,
\eeq
together with the constraint
$\s_1\,\s_2\,\s_3=\frac{\sqrt3}{72}$. While the canonical connection
$\Gamma$ at $\sigma=1$
on $\F_3$ appears in the supersymmetry condition which demands that
the supersymmetry parameter be a covariantly constant spinor with
respect to it, and hence equal to $\chi_{0,0}$ from
(\ref{chipmconst}), the two-parameter family of connections
$\widetilde{\Gamma}$ at $\sigma=\frac13$ is the one
that appears in the Dirac zero mode equation for the $d$-dimensional
gaugino field on $M$. 

\subsection{Symmetric spinors of torsion class $W_5$}

Finally, let us consider the Yukawa couplings for the standard
K\"ahler geometry of the homogeneous space $\F_3$. The relevant
connection in this case is the Levi--Civita connection
$\widehat\Gamma$ from (\ref{2.25}) at the locus (\ref{2.27}), and
hence it formally
corresponds to the $\sigma=0$ member of the family of Dirac operators
(\ref{sigmaDirac}) on $\F_3$. Hence the same remarks concerning the
$\sigma=\frac13$ case of \S\ref{SpinorW1W2} apply here as well, but
with two further complications. Firstly, there is no nice
simplification for the square of the Dirac operator $\bigl(\im
\Dirac_{q,m}\bigr)^2$ in this case, such as that in
(\ref{canDiracsquareH}) for $\sigma=1$ and that of
(\ref{KostantDiracsquareH}) for $\sigma=\frac13$; the rather
cumbersome formula can be found in~\cite[Thm.~3.2]{Agricola}. The
issue here is that $\F_3$ is a non-symmetric coset space, and moreover the
Levi--Civita connection (\ref{2.25}) is valued in the Lie algebra $\sutL$ so there
is no way to cancel its off-diagonal components using solely the
background bimonopole fields which take values in the Cartan subalgebra
$\uoL\oplus\uoL$; as we saw explicitly in \S\ref{W1}, the presence of
torsion gets around the Lichnerowicz theorem which would otherwise
forbid the construction of harmonic spinors as parallel spinors. Secondly, strictly speaking, the K\"ahler structure
on $\F_3$ does not really live in the family of quasi-K\"ahler
$\sut$-structures parametrized by the metric moduli $\s_\a$,
$\a=1,2,3$; passing to the K\"ahler locus corresponds to a discontinuous change of
complex structure $\J_+\leftrightarrow\J_-$ on $T^*\F_3$ which cannot be
implemented by smoothly varying any continuous parameters like
$\sigma$ or~$\s_\a$.

Let us briefly
spell out how discrete changes in the complex structure on $\F_3$
induce discrete changes in the Yukawa couplings.
For the almost K\"ahler structure on $\F_3$ given by (\ref{2.23}), suitable gamma-matrices on
$M\times\F_3$ are constructed as in (\ref{GammacalM}) but now using
gamma-matrices $\widehat\gamma\,^\alpha,\widehat\gamma\,^{\ab}$ with
complex orthonormal indices $\a=1,2,3$ with respect to 
the metric $\widehat\sfg$. 
We then use (\ref{AcalQkl}) with $\widetilde\Th^\a=\frac\Lambda{\sqrt3}\, \Th^\a=\frac\Lambda{\sqrt3}\,
\widehat\t\,^\a$ for
$\a=1,2$ and $\widetilde\Th^3=\frac R{\sqrt3}\, \Th^3=\frac
R{\sqrt3}\, \widehat\t\,^{\3}$, together with the obvious modification of the
Clifford map (\ref{diffcliffmap}), to replace the orthonormal
one-forms $\widehat\theta\,^\a$ by $\widehat\Gamma^\a=\gamma\otimes\widehat\gamma\,^\a$. 
Note that the discrete change of
complex structure $\J_+\leftrightarrow\J_-$ sends
$\widetilde{\gamma}\,^3\leftrightarrow \widehat{\gamma}\,^{\bar 3}$
and hence changes the Yukawa couplings, and also $I_3^-\leftrightarrow
I_{\bar 3}^+=-\big(I_3^-\big)^\dag$ and so it further changes the group theory coefficients
determining the dynamical fermion masses. The $d$-dimensional Yukawa
interactions on $M$ are again of the form (\ref{LMYukawagen}) but now
with the set of Yukawa coupling
coefficients (\ref{quasiKahlerYukawa}) replaced by
\bea
\widehat Y_{+\, 1;\,\ell,\ell'}^{\,(q,m)_n,\,(q-1,m+3)_{n\pm1}} &=& \Lambda\, \sqrt{\mbox{$\frac{n\mp
      q+1\pm1}{6(n+1)}$}} ~ \lambda^\pm_{k,l}(n,m) \, \int_{\F_3}\, \frac{\widehat{\omega}\,^{\wedge3}}{3!} \
\chi^+_{q,m;\,\ell}{}^\dagger \, 
\widehat\gamma\,^1\chi_{q-1,m+3;\,\ell'}^{-} \ , \nonumber \\[4pt]
\widehat Y_{+\, 2;\,\ell,\ell'}^{\,(q,m)_n,\,(q-1,m-3)_{n\pm1}} &=&
\Lambda \, \sqrt{\mbox{$\frac{n\mp
      q+1\pm1}{6(n+1\pm1)}$}} ~ \lambda^\mp_{k,l}(n,m)
\, 
\int_{\F_3}\, \frac{\widehat{\omega}\,^{\wedge3}}{3!} \
\chi^+_{q,m;\,\ell}{}^\dagger \, 
\widehat\gamma\,^2\chi_{q-1,m-3;\,\ell'}^{-} \ , \nonumber \\[4pt]
\widehat Y_{+\, 3;\,\ell,\ell'}^{\,(q,m)_n,\,(q+2,m)_n} &=& R\, 
\sqrt{\mbox{$\frac{(n-q)\,
        (n+q+2)}{12}$}} \, \int_{\F_3}\,
  \frac{\widehat{\omega}\,^{\wedge3}}{3!} \
\chi^+_{q,m;\, \ell}{}^\dagger \, 
\widehat\gamma\,^\3 \chi_{q+2,m;\,\ell'}^ {-} \ , \label{KahlerYukawa}
\eea
and similarly for (\ref{quasiKahlerYukawaminus}).

It is probable that some
Yukawa couplings which are zero in the quasi-K\"ahler case will become non-zero
in this case, and vice-versa. Again, without an explicit construction
of the non-constant
spinor zero-modes on $\F_3$ it is not possible to be more specific, but the important point here
is that the choice of almost K\"ahler structure on the internal coset
space influences the Yukawa interactions in
the dimensionally reduced field theory. Whether or not the choice of
nearly K\"ahler structure described in \S\ref{W1} leads to
phenomenologically more viable heterotic string vacua will require
further detailed investigation.

\bigskip

\section{Conclusions \label{Conclusions}}

\noindent
A detailed study has been carried out of the equivariant dimensional reduction of 
Yang-Mills-Dirac theory over the space $M\times \F_3$, with particular attention paid to the
Higgs and Yukawa sectors of the resulting field theory on $M$.  The study is motivated
by heterotic string theory, and although our model lacks two features of that theory 
(we use the gauge group $\urm(N)$ rather than ${\rm E}_8$ and we do not insist on supersymmetry) we believe that the model retains enough of the features of 
the heterotic model for
the analysis to be instructive.  Indeed the model exhibits enough interesting features to
merit study in its own right. 

The most general family of quasi-K\"ahler SU(3)-structures on $\F_3$, 
including the standard nearly K\"ahler structure was considered. 
We have further shown how equivariant dimensional reduction over $\F_3$ can be extended
to the non-K\"ahler case and can still yield 
a physical particle spectrum that has many features similar to that of the Standard Model. 
We have included fermions in the analysis, albeit only in certain limited cases.

The model has yielded a remarkable vacuum structure with gauge boson, Higgs boson 
and fermion masses induced by the scheme, with the masses expressed 
as functions of the moduli of the ${\rm SU}(3)$ structures.
The Higgs potential in particular has the exciting new feature of having vacua corresponding to solitonic solutions, opening up the
possibility of Higgs masses that are inversely proportional to the gauge coupling,
(\ref{SolitonMass}) --- a new aspect of the Higgs mechanism that has not been noticed
before.

We have also analysed the Yukawa couplings as functions  of the moduli
and computed induced fermion mass matrices arising from dynamical symmetry breaking,
although only in some specific cases, since the general case is technically formidable
and beyond our present techniques. The particular cases, when the spinors are constant
on $\F_3$, the calculation was  tractable and the fermion mass matrix is given
explicitly in terms of ${\rm SU}(3)$ moduli in (\ref{massmatrixYuk}).

It would be very interesting to use what we have learned from this analysis to tackle
the gauge group ${\rm E}_8$ and/or a supersymmetric lagrangean as the starting point.

\section*{Acknowledgments}

\noindent
We thank F.~Pf\"affle, A.~Popov, C.~S\"amann and C.~Stephan for helpful discussions
and correspondence. Part of this work was carried out while RJS was
visiting the Isaac Newton Institute for Mathematical Sciences in
Cambridge, UK in
February/March 2012 under the auspices of
the Programme ``Mathematics and Applications of Branes in String and
M-Theory''; he would like to thank David Berman, Neil Lambert and
Sunil Mukhi for the invitation to participate and hospitality during
the programme.
The work of R.J.S. was supported in part by the
Consolidated Grant ST/J000310/1 from the UK Science and Technology Facilities Council.

\bigskip

\appendix

\section{Representations of SU(3)\label{SU3Reps}}

\noindent
{\bf Generators and relations. \ }  Choose a basis set $\{I_A\}$ for the Lie algebra $\sutL$ with
$A=1,\dots,8$ in such a way that $I_7,I_8$ yield a basis for
the Cartan subalgebra $\uoL\oplus\uoL$. The structure constants $f^C_{AB}$ are defined by the
Lie brackets
\begin{equation}\label{2.47}
[I_A, I_B]=f^C_{AB}\, I_C \ \with g_{AB}:= f^D_{AC}\, f^C_{DB}=\de_{AB}\ ,
\end{equation}
where we have further chosen the basis so that it is orthonormal with respect to the
Cartan--Killing form on $\sutL$. Then $f_{ABC}:=f^D_{AB}\, \de_{DC}$
is totally antisymmetric in $A,B,C$. The structure constants
completely determine the geometry of the homogeneous space $\F_3$.

The non-vanishing structure constants which conform with the nearly K\"ahler structure
of $\F_3$ and the structure equations (\ref{2.38}) are given by~\cite{Popov:2010rf}
\beq
f_{135}=f_{425}=f_{416}=f_{326}=-\mbox{$\frac{1}{2\sqrt{3}}$} \ ,
\nonumber\eeq
\begin{equation}\label{fabc}
f_{127}=f_{347}=\mbox{$\frac{1}{2\sqrt{3}}$} \ ,\qquad
\end{equation}
Correspondingly, we choose the basis for
$3\times3$ matrices of the antifundamental representation of $\sutL$ given by
\begin{equation}\nonumber
I_1=\frac{1}{2\sqrt{3}}\,\begin{pmatrix}0&0&-1\\0&0&0\\1&0&0\end{pmatrix}
\ ,\qquad
I_2=\frac{1}{2\sqrt{3}}\,\begin{pmatrix}0&0&\im\\0&0&0\\\im&0&0\end{pmatrix}
\ ,\qquad
I_3=\frac{1}{2\sqrt{3}}\,\begin{pmatrix}0&1&0\\-1&0&0\\0&0&0\end{pmatrix}\ ,
\end{equation}
\begin{equation}\nonumber
I_4=\frac{1}{2\sqrt{3}}\,\begin{pmatrix}0&\im&0\\\im&0&0\\0&0&0\end{pmatrix}
\ ,\qquad
I_5=\frac{1}{2\sqrt{3}}\,\begin{pmatrix}0&0&0\\0&0&1\\0&-1&0\end{pmatrix}
\ ,\qquad
I_6=\frac{1}{2\sqrt{3}}\,\begin{pmatrix}0&0&0\\0&0&\im\\0&\im&0\end{pmatrix}\ ,
\end{equation}
\begin{equation}\label{2.46}
I_7=\frac{\im}{2\sqrt{3}}\,\begin{pmatrix}0&0&0\\0&-1&0\\0&0&1\end{pmatrix}
\und
I_8=\frac{\im}{6}\,\begin{pmatrix}2&0&0\\0&-1&0\\0&0&-1\end{pmatrix}\ .
\end{equation}
The matrices
\bea\nonumber
I_1^-:=\mbox{$\frac{1}{2}$}\,(I_1-\im  I_2)=\frac{1}{2\sqrt{3}}\,
\begin{pmatrix}0&0&0\\0&0&0\\1&0&0\end{pmatrix} &,& 
I_{\1}^+:=\mbox{$\frac{1}{2}$}\,(I_1+\im  I_2)=\frac{1}{2\sqrt{3}}\,
\begin{pmatrix}0&0&-1\\0&0&0\\0&0&0\end{pmatrix}\ , \\[4pt]
\nonumber
I_2^-:=\mbox{$\frac{1}{2}$}\,(I_3-\im  I_4)=\frac{1}{2\sqrt{3}}\,
\begin{pmatrix}0&1&0\\0&0&0\\0&0&0\end{pmatrix}
&,& 
I_{\2}^+:=\mbox{$\frac{1}{2}$}\,(I_3+\im  I_4)=\frac{1}{2\sqrt{3}}\,
\begin{pmatrix}0&0&0\\-1&0&0\\0&0&0\end{pmatrix}\ , \\[4pt]
\nonumber
I_3^-:=\mbox{$\frac{1}{2}$}\,(I_5-\im \, I_6)=\frac{1}{2\sqrt{3}}
\begin{pmatrix}0&0&0\\0&0&1\\0&0&0\end{pmatrix}  &,&
I_{\3}^+:=\mbox{$\frac{1}{2}$} \,(I_5+\im  I_6)=\frac{1}{2\sqrt{3}}\,
\begin{pmatrix}0&0&0\\0&0&0\\0&-1&0\end{pmatrix}\ , \\[4pt]
-\im I_{7}=\frac{1}{2\sqrt{3}}\,
\begin{pmatrix}0&0&0\\0&-1&0\\0&0&1\end{pmatrix} &\and& 
-\im I_{8}=\frac{1}{6}\,
\begin{pmatrix}2&0&0\\0&-1&0\\0&0&-1\end{pmatrix}
\label{2.58}\eea
form a basis for the complexified Lie algebra $\sltcL$ in the
antifundamental representation. Here complex conjugation acts by
interchanging barred and unbarred indices. 

The
non-vanishing structure
constants $C_{AB}^C$ of $\sltcL$
in the basis (\ref{2.58}) are given by
\begin{equation}\label{2.61}
\begin{matrix}
C^{\1}_{23}=C^{\2}_{31}=C^{\3}_{12}=C^{1}_{\2\3}=C^{2}_{\3\1}
=C^{3}_{\1\2}=-\frac{1}{2\sqrt{3}}\ ,\\[4mm]
C^{1}_{71}=C^{2}_{72}=-C^{\1}_{7\1}=-C^{\2}_{7\2}=\frac{1}{2\sqrt{3}}\ ,
\qquad C^{3}_{73}=-C^{\3}_{7\3}=-\frac{1}{\sqrt{3}}\ ,\\[4mm]
C^{1}_{81}=-C^{\1}_{8\1}=-\frac{1}{2}\ ,
\qquad C^{2}_{82}=-C^{\2}_{8\2}=\frac{1}{2}\ ,\\[4mm]
C^{7}_{1\1}=C^{7}_{2\2}=-\frac{1}{4\sqrt{3}}\ ,\qquad
C^{7}_{3\3}=\frac{1}{2\sqrt{3}}\ ,
\qquad C^{8}_{1\1}=\frac{1}{4} \und C^{8}_{2\2}=-\frac{1}{4}\ .
\end{matrix}
\end{equation}
After the rescaling (\ref{2.64}), the structure constants (\ref{2.61}) are
rescaled as
\begin{equation*}
\widetilde C^{\ab}_{\b\g}=2\sqrt3\ \mbox{$\frac{\s_\b\,\s_\g}{\s_\a}$}\,C^{\ab}_{\b\g}= \mbox{$
-\frac{\s_\b\,\s_\g}{\s_\a}$} \,\ve^{\ab}_{\b\g}\ ,
\end{equation*}
\begin{equation*}
\widetilde C^1_{71}=C^1_{71}=\mbox{$\frac{1}{2\sqrt{3}}$}\ ,\qquad
\widetilde C^2_{72}=C^2_{72}=\mbox{$\frac{1}{2\sqrt{3}}$}\ ,\qquad
\widetilde C^3_{73}=C^3_{73}=-\mbox{$\frac{1}{\sqrt{3}}$} \ ,
\end{equation*}
\begin{equation}
\widetilde C^1_{81}=C^1_{81}=-\mbox{$\frac{1}{2}$} \ ,\qquad
\widetilde C^2_{82}=C^2_{82}=\mbox{$\frac{1}{2}$} \ ,
\label{2.65}\end{equation}
\begin{equation*}
\widetilde C^7_{1\1}={12\s_1^2}\,C^7_{1\1}=-\sqrt3\,\s_1^2 \ ,
\qquad
\widetilde C^7_{2\2}={12\s_2^2}\,C^7_{2\2}=-\sqrt3\, \s_2^2 \ ,
\qquad
\widetilde C^7_{3\3}={12\s_3^2}\,C^7_{3\3}=2\sqrt3\,\s_3^2 \ ,
\end{equation*}
\begin{equation*}
\widetilde C^8_{1\1}={12\s_1^2}\,C^8_{1\1}=3\s_1^2 \ ,\qquad
\widetilde C^8_{2\2}={12\s_2^2}\,C^8_{2\2}=-3\s_2^2 \ ,
\end{equation*}
plus their complex conjugates. 
The non-vanishing structure constants $\widehat C_{AB}^C$ of the Lie algebra $\sutL$
for the complex basis of one-forms $\widehat\t\,^{\a}$ adapted to the K\"ahler structure on $\F_3$ and
the structure equations (\ref{2.26}) are given by
\begin{equation}\label{2.61a}
\begin{matrix}
\widehat C^{\1}_{2\3}=\widehat C^{\2}_{1\3}=-\frac{1}{2\sqrt{6}}\ ,\qquad
\widehat C^{3}_{12}=-\frac{1}{\sqrt{6}}\ ,\\[4mm]
\widehat C^{7}_{1\1}=\widehat C^{7}_{2\2}=\widehat C^{7}_{3\3}=-\frac{1}{4\sqrt{3}}\ ,\qquad
\widehat C^{8}_{1\1}=\frac{1}{4}\und \widehat C^{8}_{2\2}=-\frac{1}{4} \ ,
\end{matrix}
\end{equation}
and their complex conjugates, plus
\beq
\begin{matrix}
\widehat C^{1}_{71}=\widehat C^{2}_{72}=\frac{1}{2\sqrt{3}}\ ,
\qquad \widehat C^{3}_{73}=-\frac{1}{\sqrt{3}}\ ,\qquad
\widehat C^{1}_{81}=-\frac{1}{2}\ ,
\qquad \widehat C^{2}_{82}=\frac{1}{2}\ ,\\[4mm]
\widehat C^\ab_{7\ab}=-\widehat C^\a_{7\a} \und \widehat C^\ab_{8\ab}=-\widehat C^\a_{8\a}
\end{matrix}
\label{2.61b}\eeq
for $\a=1,2,3$. Here we have chosen $R^2=2\La^2=6$.

In the basis
(\ref{2.58}), the Chevalley generators are given by
\beq
E_{\alpha_1}=-2\,\sqrt3\, I_{\3}^+ \ , \qquad E_{\alpha_2}=-2\,\sqrt3\,
I_{\1}^+ \qquad \mbox{and} \qquad E_{\alpha_1+\alpha_2}=2\,\sqrt3 \,
I_{2}^- \ ,
\eeq
where $\alpha_1,\alpha_2$ are the simple roots of $\sut$. Compared to the
representations pertinent to the holomorphic
K\"ahler loci of the moduli space~\cite{LPS3}, the change in Chevalley generator $I_3^-\to I_{\bar 3}^+$ corresponds to the change in sign of the almost complex
structure along the $\C P^1$-fibre direction of~$\F_3$. 

\bigskip

\noindent
{\bf Irreducible modules. \ }  For each fixed pair of
non-negative integers $(k,l)$ there is an irreducible representation
$\widehat V^{k,l}$ of $\sut$ of dimension 
\beq
d^{k,l}=\mbox{$\frac12$}\,(k+1)\,(l+1)\,(k+l+2) \ .
\label{dimkl}\eeq
The integer $k$ is the
number of fundamental representations $\widehat V^{1,0}$ and $l$
the number of antifundamental representations $\widehat V^{0,1}$
appearing in the usual tensor product construction of
$\widehat V^{k,l}$.
All irreducible $T$-modules are
one-dimensional, and the collection of weight vectors of the maximal
torus $T=\uo\times\uo$ in $\sut$ label
points in the weight diagram $W^{k,l}$ for $\widehat V^{k,l}$. We denote
them by $(q,m)_n$, where $q=2I_z$ and $m=3Y$ are respectively isospin and
hypercharge eigenvalues, and the label by the total isospin integer
$n=2I$ is used to keep track of multiplicities of states in the weight
diagram. They may be conveniently
parameterized by a pair of independent $\su$ spins $j_\pm$,
with $2j_+=0,1,\dots,k$ and $2j_-=0,1,\dots,l$, and the corresponding
component spins $m_\pm\in\{-j_\pm,-j_\pm+1,\dots,j_\pm-1,j_\pm\}$,
which are defined in terms of Young tableaux as follows. 
Represent the irreducible $T$-module $V_{(q,m)_n}$ with weight vector
$(q,m)_n=(1,1)_1$ by $\vcenter{\hbox{\tiny$\young(\times)$}}\,$, that
with $(q,m)_n=(-1,1)_1$ by
$\vcenter{\hbox{\tiny$\young(\bullet)$}}\,$, and that with $(q,m)_n =(0,-2)_0$
by $\vcenter{\hbox{\tiny$\young(\circ)$}}\,$. Then the
$\sut\rightarrow\uo\times\uo$ decomposition of the fundamental
representation
\beq
\widehat{V}^{1,0}\big|_T=V_{(1,1)_1}\oplus V_{(-1,1)_1}\oplus
V_{(0,-2)_0}
\eeq
is depicted by
\beq
\young(\ ) \qquad \longrightarrow \qquad \young(\times)\quad \oplus
\quad \young(\bullet)\quad \oplus
\quad \young(\circ) \ . 
\label{3to2tableau}
\eeq
In terms of $\SU(3)$ Young tableaux, the irreducible representation
$\widehat{V}^{k,l}$ corresponds to the diagram
\begin{equation}
\underbrace{\young(\ \Dots\ ,\  \Dots \ )}_l\kern -2.1pt
\raise 6.2pt\hbox{$\underbrace{\young(\ \Dots\ )}_k$}
\end{equation}
and this contains all $\uo \times \uo$ representations
\beq
\underbrace{\young(\times \Dots\times ,\bullet \Dots \bullet
  )}_{l-2j_-}\kern -2.1pt\underbrace{\young(\circ \Dots\circ ,\times \Dots \times
  )}_{j_-+m_-}\kern -2.1pt\underbrace{\young(\circ \Dots\circ ,\bullet
  \Dots \bullet )}_{j_--m_-}\kern -2.1pt 
\raise 6.1pt\hbox{$\underbrace{\young(\circ
    \Dots\circ)}_{k-2j_+}$}\kern -0.5pt 
\raise 6.1pt\hbox{$\underbrace{\young(\times
    \Dots \times)}_{j_++m_+}$}\kern -0.5pt\raise
6.1pt\hbox{$\underbrace{\young(\bullet \Dots\bullet )}_{j_+-m_+}$} 
\eeq
of dimension $2j_++2j_-+1$, isospin charge $2m_++2m_-$, and hypercharge $2(l-k)+6(j_+-j_-)$, with
multiplicity one. This gives
\beq
q=2(m_++m_-) \ , \qquad
m=6(j_+-j_-)-2(k-l) \qquad \mbox{and} \qquad n=2(j_++j_-) \ .
\label{mnqjpm}\eeq
The $\su$ spin $j_+$ (resp.~$j_-$) is the value of the isospin contributed by the upper (resp.~lower) indices of the $\sut$
tensor corresponding to the irreducible module $\widehat V^{k,l}$. The
integers $(q,m)_n$ all have the same even/odd parity. 

\bigskip

\noindent
{\bf Biedenharn basis. \ }  To explicitly represent the coset generators of $\F_3$, we
use the Biedenharn basis for the irreducible representation $\widehat
V^{k,l}$ of $\sut$~\cite{LPS3}. The generators of the complex torus
$T^\C=\C^*\times \C^*$ for the irreducible
module corresponding to the weight vector $(q,m)_n$ in this basis are given by
\beq
-\im I_{7}^{(q,m)_n}=\mbox{$\frac{q}{2\,\sqrt3}$} \und
-\im I_{8}^{(q,m)_n} =
\mbox{$\frac m{6}$} \ ,
\label{I78Bied}\eeq
while the non-vanishing off-diagonal matrix elements of the remaining
generators of $\sltc$ are
\bea
I_1^{- \ (q,m)_n,\, (q-1,m+3)_{n\pm1}}&=& \sqrt{\mbox{$\frac{n\mp
      q+1\pm1}{24(n+1)}$}} ~ \lambda^\pm_{k,l}(n,m) \ ,
\nonumber\\[4pt]
I_2^{- \ (q,m)_n,\,(q-1,m-3)_{n\pm1}}&=& \sqrt{\mbox{$\frac{n\mp
      q+1\pm1}{24(n+1\pm1)}$}} ~ \lambda^\mp_{k,l}(n\pm1,m-3) \ ,
\nonumber \\[4pt] 
I_3^{- \ (q,m)_n,\, (q+2,m)_n} &=& \sqrt{\mbox{$\frac{(n-q)\,
        (n+q+2)}{48}$}} \ , 
\label{offdiagBied}\eea
where
\bea
\lambda_{k,l}^+(n,m)&=&\mbox{$\frac1{\sqrt{n+2}}~
\sqrt{\big(\frac{k+2l}3+\frac n2+\frac
    m6+2\big)\,\big(\frac{k-l}3+\frac n2+\frac m6+1\big)\,\big(
\frac{2k+l}3-\frac n2-\frac m6\big)}$} \ , \nonumber\\[4pt]
\lambda_{k,l}^-(n,m)&=&\mbox{$\frac1{\sqrt n}~
\sqrt{\big(\frac{k+2l}3-\frac n2+\frac
    m6+1\big)\,\big(\frac{l-k}3+\frac n2-\frac m6\big)\,\big(
\frac{2k+l}3+\frac n2-\frac m6+1\big)}$} \ .
\label{lambdaklnm}\eea
The latter constants are defined for $n>0$ and we set
$\lambda_{k,l}^-(0,m):=0$. The analogous relations for $I_{\bar\alpha}^+$ can be derived by hermitean conjugation
of (\ref{offdiagBied}) using the property
$\big(I_{\bar\alpha}^+\big)^\dag = -I_{\alpha}^-$.

\bigskip

\section{Matrix elements of invariant curvatures}

\noindent
The diagonal matrix elements of the curvature (\ref{cfca}) of the gauge potential (\ref{AcalQkl}) at each vertex
$(q,m)_n\in W^{k,l}$ of the weight diagram for $\widehat V^{k,l}$ can
be computed by substituting (\ref{offdiagBied}) and are
given by
\bea
\Fcal^{(q,m)_n\, (q,m)_n} &=& F^{(q,m)_n} \label{Fcalkldiag} \\ && +\,
\frac{\s_1^2}{2}\,
\widetilde\Th^1\wedge \widetilde\Th^\1 \, \sum_\pm\, \bigg[\, 
\mbox{$\frac{n\mp
    q+1\pm 1}{n+1}$}\, \lambda_{k,l}^\pm(n,m)^2\, \Big({\bf
  1}_{N_{(q,m)_n}}- \phi^{1 \ (\pm)}_{(q,m)_n}{}^\dag\, \phi^{1 \
  (\pm)}_{(q,m)_n} \Big)  \nonumber\\ && \qquad  -\, 
\mbox{$\frac{n\mp q+1\mp 1}{n+1 \mp 1}$}\, \lambda_{k,l}^\pm(n\mp
1,m-3)^2\,\Big({\bf
  1}_{N_{(q,m)_n}}- \phi^{1 \ (\pm)}_{(q+1,m-3)_{n\mp 1}}\, 
\phi^{1 \ (\pm)}_{(q+1,m-3)_{n\mp 1}}{}^\dag\Big)\, \bigg]\nonumber \\ && 
+\,\frac{\s_2^2}{2}\, \widetilde\Th^2\wedge
\widetilde\Th^\2 \,   \sum_\pm\, \bigg[\, 
\mbox{$\frac{n\mp
    q+1\pm 1}{n+1\pm1}$}\, \lambda_{k,l}^\mp(n\pm1,m-3)^2\, \Big({\bf
  1}_{N_{(q,m)_n}}- \phi^{2 \ (\pm)}_{(q,m)_n}{}^\dag \, \phi^{2 \ (\pm)}_{(q,m)_n} \Big) \nonumber\\ && \qquad -\, 
\mbox{$\frac{n\mp q+1\mp 1}n$}\, 
\lambda_{k,l}^\mp(n,m)^2\,\Big({\bf
  1}_{N_{(q,m)_n}}- \phi^{2 \ (\pm)}_{(q+1,m+3)_n\mp 1}\, \phi^{2 \ (\pm)}_{(q+1,m+3)_{n\mp 1}}{}^\dag \Big)
\, \bigg]\nonumber \\ && 
+\, \frac{\s_3^2}{4}\,
\widetilde\Th^3\wedge \widetilde\Th^\3 \,
\bigg[ (n-q)\, (n+q+2)\, \Big({\bf
  1}_{N_{(q,m)_n}}- \phi^{3}_{(q,m)_n}{}^\dag\, \phi^{3}_{(q,m)_n} \Big) \nonumber\\ && \qquad \qquad \qquad \qquad -\, (n+q)\, (n-q+2)\, \Big({\bf
  1}_{N_{(q,m)_n}}- \phi^{3}_{(q-2,m)_n} \,
\phi^{3}_{(q-2,m)_n}{}^\dag\Big) \bigg]
 \nonumber
\eea
where $F^{(q,m)_n}= \diff A^{(q,m)_n}+ A^{(q,m)_n}\wedge
A^{(q,m)_n}$ is the curvature of the vector bundle $E_{(q,m)_n}\to M$,
and we suppress tensor products to simplify the notation. 
The remaining non-vanishing off-diagonal matrix elements of the
curvature two-form $\Fcal$ are given by
\bea
\Fcal^{ (q-1,m+3)_{n\pm1}\, (q,m)_n} &=&
\mbox{$\lambda_{k,l}^\pm(n,m)$}\,
  \bigg\{\s_1\,\sqrt{\mbox{$\frac{(n\mp q+1\pm 1)}{2(n+1)}$}} 
~ \widetilde\Th^\1 \wedge D\phi^{1 \
    (\pm)}_{(q,m)_n}  \nonumber
  \\ 
&& + \, \frac{\s_2\, \s_3}{2}\, \widetilde\Th^2\wedge \widetilde\Th^3 \,\bigg[\, 
  \sqrt{\mbox{$\frac{((n+1\pm 1)^2-q^2) \, (n\pm
        q+1\pm 1)}{2(n+1)}$}}\nonumber \\ 
&& \qquad \qquad \qquad \qquad \times \, \Big( \phi^{1 \
    (\pm)}_{(q,m)_n}- \phi^3_{(q-1,m+3)_{{n\pm 1}}}{}^\dag \, \phi^{2 \
  (\mp)}_{(q+1,m+3)_{n\pm 1}}{}^\dag\Big) \nonumber\\ 
&& - \,
\sqrt{\mbox{$\frac{(n\pm q+1\mp 1)\, (n-q+2)\,
        (n+q)}{2(n+1)}$}}\, \Big( \phi^{1 \
    (\pm)}_{(q,m)_n}- \phi^{2 \
  (\mp)}_{(q-1,m+3)_{n\pm 1}}{}^\dag \, \phi^{3}_{(q-2,m)_n}{}^\dag
\Big) \, \bigg]\bigg\} \ , \nonumber \\ [4pt]
\Fcal^{(q-1,m-3)_{n\pm1}\, (q,m)_n} &=&
\mbox{$\lambda_{k,l}^\mp(n\pm1,m-3)$}\,
  \bigg\{\s_2\,\sqrt{\mbox{$\frac{(n\mp q+1\pm 1)}{2(n+1\pm1)}$}} ~\widetilde\Th^\2  \wedge D \phi^{2 \
    (\pm)}_{(q,m)_n}  \nonumber
  \\ && +\, \frac{\s_3\, \s_1}{2}\, \widetilde\Th^3\wedge \widetilde\Th^1 \,\bigg[\, 
  \sqrt{\mbox{$\frac{((n+1\pm1)^2-q^2)\, (n\pm
        q+1\pm1)}{2(n+1\pm1)}$}}
\label{Fcalkloffdiag3} \\ && \qquad \qquad \qquad \qquad \qquad \times \, \Big( \phi^{2 \
    (\pm)}_{(q,m)_n}- \phi^{1 \
  (\mp)}_{(q-1,m-3)_{n\pm 1}}{}^\dag \, \phi^{3}_{(q-2,m)_{n}}{}^\dag \Big) \nonumber\\ && - \,
\sqrt{\mbox{$\frac{(n\pm q +1 \mp 1) \, (n-q+2)\, (n+q)}{2(n+1\pm 1)}$}}\, \Big( \phi^{2 \
    (\pm)}_{(q,m)_n}- \phi^{3}_{(q-1,m-3)_{n\pm 1}}{}^\dag\, \phi^{1 \
  (\mp)}_{(q+1,m-3)_{n\pm 1}}{}^\dag
\Big) \, \bigg]\bigg\}  \ , \nonumber \\[4pt]
\Fcal^{(q+2,m)_n\, (q,m)_{n}} &=& \frac{\s_3}{2}\, \sqrt{\mbox{$(n-q)\,
      (n+q+2)$}} ~ \widetilde\Th^\3 \wedge D\phi^{3}_{(q,m)_n} 
\nonumber\\ && \qquad +\, \mbox{$\frac{\s_1\, \s_2}{2}$}\,  \sqrt{(n+1)^2-(q+1)^2}\,
\widetilde\Th^1\wedge\widetilde\Th^2 
 \nonumber \\ && \qquad \qquad  \times
\sum_\pm\, \bigg[\, \lambda_{k,l}^\pm(n\mp 1,m-3)^2\, \Big(\phi^{3}_{(q,m)_n}- \phi^{2 \
  (\mp)}_{(q+2,m)_n}{}^\dag \,\phi^{1 \
  (\pm)}_{(q+1,m-3)_{n\mp 1}}{}^\dag  \Big) \nonumber \\ && 
\qquad\qquad -\,
\lambda^\pm_{k,l}(n,m)^2 \, \Big( \phi^{3}_{(q,m)_n}-  \phi^{1 \
  (\pm)}_{(q+2,m)_n}{}^\dag \, \phi^{2 \
  (\mp)}_{(q+1,m+3)_{n\pm 1}}{}^\dag \Big) \, \bigg]\ , \nonumber
\eea
and
\bea
\Fcal^{(q+2,m)_{n}\, (q-1,m+3)_{n\pm1}} &=& \mbox{$\frac{\s_1\,
    \s_3}{2}\, \sqrt{\frac{(n-q)\, (n+q+2)\, (n\mp
      q+1\pm1)}{2(n+1)}}$}\, \lambda_{k,l}^\pm(n,m)\,
\widetilde\Th^1\wedge \widetilde\Th^\3 \nonumber \\ && \times\, \Big(\phi^{3}_{(q,m)_{n}} \, \phi^{1 \ (\pm)}_{(q,m)_n}{}^\dag -
\phi^{1 \  (\pm)}_{(q+2,m)_{n}}{}^\dag \, \phi^{3}_{(q-1,m+3)_{n\pm 1}} \Big) \ ,
\nonumber \\[4pt]
\Fcal^{(q+2,m)_{n}\, (q-1,m-3)_{n\pm 1}} &=& \mbox{$\frac{\s_2\,\s_3}{2}\, \sqrt{\frac{(n-q)(n+q+2)\, (n\mp q+1\pm 1)}{2(n+1\pm1)}}$}\, \lambda_{k,l}^\mp(n\pm1,m-3)\, 
\widetilde\Th^2\wedge\widetilde\Th^\3   \nonumber\\ 
&&\times\, \Big(\phi^{3}_{(q,m)_{n}} \, \phi^{2 \
  (\pm)}_{(q,m)_{n}}{}^\dag - \phi^{2 \
  (\pm)}_{(q+2,m)_{n}}{}^\dag \, \phi^{3}_{(q-1,m-3)_{n\pm 1}} \Big) \ , 
\label{Fcalkloffdiag23}\\[4pt]
\Fcal^{(q+1,m-3)_{n\mp1}\, (q+1,m+3)_{n\pm1}} &=& \mbox{$\frac{\s_1\, \s_2}{2}\, \sqrt{\frac{((n+1)^2-(q+1)^2)}{(n+1\mp1)(n+1)}}$}\, \lambda_{k,l}^\pm(n,m)\, \lambda_{k,l}^\pm(n\mp1,m-3)\, \widetilde\Th^1\wedge \widetilde\Th^\2 \nonumber \\ && \times\, \Big(
\phi^{2 \
  (\mp)}_{(q+2,m)_{n}} \, \phi^{1 \
  (\pm)}_{(q+2,m)_{n}}{}^\dag 
- \phi^{1 \ (\pm)}_{(q+1,m-3)_{n\mp1}}{}^\dag \, \phi^{2 \
  (\mp)}_{(q+1,m+3)_{n\pm1}} \Big) \ , \nonumber
\eea
plus their hermitean conjugates $\Fcal^{(q',m'\, )_{n'}\,
  (q,m)_n}=-\big( \Fcal^{(q,m)_n\, (q',m'\, )_{n'}}\big)^\dag$ for $(q',m'\,)_{n'} \neq (q,m)_n$. Here
\bea
D\phi^{1 \
    (\pm)}_{(q,m)_n} &=& \diff \phi^{1 \
    (\pm)}_{(q,m)_n}+ A^{(q-1,m+3)_{n\pm1}} \,\phi^{1 \
    (\pm)}_{(q,m)_n}- \phi^{1 \
    (\pm)}_{(q,m)_n}\, A^{(q,m)_n} \ , \nonumber \\[4pt]
D \phi^{2\ (\pm)}_{(q,m)_n} &=& \diff\phi^{2\ (\pm)}_{(q,m)_n}+A^{(q-1,m-3)_{n\pm1}} \, \phi^{2\ (\pm)}_{(q,m)_n}-\phi^{2\ (\pm)}_{(q,m)_n}\, A^{(q,m)_n} \ , \nonumber \\[4pt]
D \phi^{3 \
    }_{(q,m)_n} &=& \diff \phi^{3 \
    }_{(q,m)_n}+ A^{(q+2,m)_n} \,\phi^{3 \
    }_{(q,m)_n}- \phi^{3 \
    }_{(q,m)_n}\, A^{(q,m)_n}\ .
\label{Qklbifund3}\eea
are bifundamental covariant derivatives of the Higgs fields on
$M$.


\bigskip

\begin{thebibliography}{99}

\bibitem{Agricola}
I.~Agricola, 
``Connections on naturally reductive spaces, their Dirac operator
and homogeneous models in string theory,''
Commun. Math. Phys. {\bf 232} (2003) 535--563
[arXiv:math.DG/0202094].

\bibitem{Bismut}
J.-M.~Bismut, 
``A local index theorem for non-K\"ahler manifolds,''
Math. Ann. {\bf 284}
(1989) 681--699.

\bibitem{BottTu}
  R.~Bott and L.W.~Tu,
  {\it Differential Forms in Algebraic Topology} (Springer, 1982).

\bibitem{Chatzistavrakidis:2009mh}
  A.~Chatzistavrakidis and G.~Zoupanos,
  ``Dimensional reduction of the heterotic string over nearly K\"ahler manifolds,''
  JHEP {\bf 09} (2009) 077
  [arXiv:0905.2398 [hep-th]].

\bibitem{CS}
  S.~Chiossi and S.M.~Salamon,
  ``The intrinsic torsion of SU(3) and G$_2$ structures,'' in: {\sl
    Differential Geometry, Valencia 2001}, eds. O.~Gil-Medrano and
  V.~Miquel (World Scientific, 2002) 115--133
  [arXiv:math.DG/0202282].

\bibitem{Dolan}
  B.P.~Dolan,
  ``The spectrum of the Dirac operator on coset spaces with
  homogeneous gauge fields,''
  JHEP {\bf 05} (2003) 018
  [arXiv:hep-th/0304037].

\bibitem{Dolan:2009ie}
  B.P.~Dolan and R.J.~Szabo,
  ``Dimensional reduction, monopoles and dynamical symmetry breaking,''
  JHEP {\bf 0903} (2009) 059
  [arXiv:0901.2491 [hep-th]].

\bibitem{Dolan:2009nz}
  B.P.~Dolan and R.J.~Szabo,
  ``Dimensional reduction and vacuum structure of quiver gauge theory,''
  JHEP {\bf 08} (2009) 038
  [arXiv:0905.4899 [hep-th]].

\bibitem{FI}
T.~Friedrich and S.~Ivanov, 
``Parallel spinors and connections with skew-symmetric torsion in
string theory,''
Asian J. Math. {\bf 6} (2002) 303--336
[arXiv:math.DG/0102142].

\bibitem{HKWY}
T.~Houri, D.~Kubiz\v{n}\'ak, C.~Warnick and Y.~Yasui, 
``Symmetries of the Dirac operator with skew-symmetric torsion,''
Class.\ Quant.\ Grav.\ {\bf 27} (2010) 185019
  [arXiv:1002.3616 [hep-th]].

\bibitem{Irges:2011de}
  N.~Irges and G.~Zoupanos,
  ``Reduction of $\mathcal{N}=1$ E$_8$ SYM over $\sut/\uo\times
    \uo\times \Z_3$ and its four-dimensional effective action,''
  Phys.\ Lett.\ B {\bf 698} (2011) 146--151
  [arXiv:1102.2220 [hep-ph]].

\bibitem{Kapetanakis:1992hf}
  D.~Kapetanakis and G.~Zoupanos,
  ``Coset space dimensional reduction of gauge theories,''
  Phys.\ Rept.\  {\bf 219} (1992) 1--76.

\bibitem{Kimura:2007a}
  T.~Kimura,
  ``Index theorems on torsional geometries,''
  JHEP {\bf 08} (2007) 048
  [arXiv:0704.2111 [hep-th]].

\bibitem{Kimura:2006af}
  T.~Kimura and P.~Yi,
  ``Comments on heterotic flux compactifications,''
  JHEP {\bf 07} (2006) 030
  [arXiv:hep-th/0605247].

\bibitem{Klaput:2011mz}
  M.~Klaput, A.~Lukas and C.~Matti,
  ``Bundles over nearly K\"ahler homogeneous spaces in heterotic string theory,''
  JHEP {\bf 09} (2011) 100
  [arXiv:1107.3573 [hep-th]].

\bibitem{Landweber}
G.D.~Landweber,
``Harmonic spinors on homogeneous spaces,''
Repr. Theory {\bf 4} (2000) 466--473
[arXiv:math.DG/0005056].

\bibitem{Lust}
D.~L\"ust,  
 ``Compactification of ten-dimensional superstring theories over Ricci flat coset spaces,''
   Nucl. Phys.  B {\bf 276} (1986) 220--240.

\bibitem{LNP}
 O.~Lechtenfeld, C.~N\"olle and A.D.~Popov,
  ``Heterotic compactifications on nearly K\"ahler manifolds,''
   JHEP {\bf 09} (2010) 074
   [arXiv:1007.0236 [hep-th]].

\bibitem{Lechtenfeld:2007st}
  O.~Lechtenfeld, A.D.~Popov and R.J.~Szabo,
  ``Quiver gauge theory and noncommutative vortices,''
  Prog.\ Theor.\ Phys.\ Suppl.\  {\bf 171} (2007) 258--268
  [arXiv:0706.0979 [hep-th]];
  B.P.~Dolan and R.J.~Szabo,
  ``Equivariant dimensional reduction and quiver gauge theories,''
  Gen.\ Rel.\ Grav.\  {\bf 43} (2011) 2453--2466
  [arXiv:1001.2429 [hep-th]].

\bibitem{LPS3}
  O.~Lechtenfeld, A.D.~Popov and R.J.~Szabo,
  ``SU(3)-equivariant quiver gauge theories and nonabelian
  vortices,'' 
  JHEP {\bf 08} (2008) 093
  [arXiv:0806.2791 [hep-th]].

\bibitem{LopesCardoso:2002hd}
  G.~Lopes Cardoso, G.~Curio, G.~Dall'Agata, D.~L\"ust, P.~Manousselis and G.~Zoupanos,
  ``Non-K\"ahler string backgrounds and their five torsion classes,''
  Nucl.\ Phys.\ B {\bf 652} (2003) 5--34
  [arXiv:hep-th/0211118].

\bibitem{Popov:2010rf}
  A.D.~Popov and R.J.~Szabo,
  ``Double quiver gauge theory and nearly K\"ahler flux compactifications,''
  JHEP {\bf 02} (2012) 033 [arXiv:1009.3208 [hep-th]].

\bibitem{Rajaraman}
R. Rajaraman, 
\emph{Solitons and Instantons}
(North Holland,
1982).

\bibitem{Scherk:1979zr}
  J.~Scherk and J.H.~Schwarz,
  ``How to get masses from extra dimensions,''
  Nucl.\ Phys.\ B {\bf 153} (1979) 61--88.

\bibitem{Strominger:1986uh}
  A.~Strominger,
  ``Superstrings with torsion,''
  Nucl.\ Phys.\ B {\bf 274} (1986) 253--284.

\end{thebibliography}
\end{document}